\documentclass[a4paper,fleqn]{cas-sc}

\usepackage[authoryear]{natbib}

\usepackage{dsfont}

\usepackage{graphicx}%
\usepackage{multirow}%
\usepackage{amsmath,amssymb,amsfonts}%
\usepackage{amsthm}%
\usepackage{mathrsfs}%
\usepackage[title]{appendix}%
\usepackage{xcolor}%
\usepackage{textcomp}%
\usepackage{manyfoot}%
\usepackage{booktabs}%
\usepackage{algorithm}%
\usepackage{algorithmicx}%
\usepackage{algpseudocode}%
\usepackage{listings}%
\usepackage{comment}%
\usepackage{multirow}%
\usepackage{rotating}%
\usepackage{tabularx}%
\usepackage{siunitx}%
\usepackage{rotating}%

\usepackage{lineno}

\usepackage{float}

\usepackage{enumitem}

\newcommand{\diff}{\mathrm{d}}

\def\tsc#1{\csdef{#1}{\textsc{\lowercase{#1}}\xspace}}
\tsc{WGM}
\tsc{QE}
\tsc{EP}
\tsc{PMS}
\tsc{BEC}
\tsc{DE}

\begin{document}
    
\let\WriteBookmarks\relax
\def\floatpagepagefraction{1}
\def\textpagefraction{.001}
\shorttitle{}
\shortauthors{}

\title [mode = title]{
Steadily moving semi-infinite fracture in plane poroelasticity
}

\author[1]{Evgenii Kanin}
\author[1]{Andreas Möri}[]
\author[2]{Dmitry Garagash}
\author[1]{Brice Lecampion}[]
\cormark[1]
\ead{brice.lecampion@epfl.ch}

\address[1]{Geo-Energy Laboratory-Gaznat Chair, Institute of Civil Engineering, École Polytechnique Fédérale de Lausanne (EPFL), Lausanne, Switzerland}
\address[2]{Department of Civil and Resource Engineering, Dalhousie University, Halifax, Canada}

\begin{abstract}
We present a boundary integral formulation for steadily propagating semi-infinite plane strain tensile and shear fractures in poroelastic media. By combining fundamental solutions of plane strain poroelasticity for an instantaneous fluid source and instantaneous edge dislocations (normal and slip modes) with temporal and spatial superposition principles, we derive boundary integral equations for steadily moving fractures under the adopted hydraulic boundary conditions. These equations relate the tractions (normal and shear stresses) and the pore fluid pressure on the fracture surfaces to the fracture opening, slip, and the fluid displacement function. Assuming prescribed traction and pore fluid pressure profiles, we develop a numerical methodology to solve the governing equations for fracture opening, slip, and the fluid displacement function. The formulation is systematically verified on several relevant problems, including a tensile fracture with exponential normal loading, a stress-free tensile fracture with an imposed exponential pore fluid pressure, and a shear fracture under uniform shear loading over a finite region, demonstrating excellent agreement with analytical and semi-analytical solutions. The resulting boundary integral framework provides an accurate and efficient tool for analyzing semi-infinite steadily propagating cracks in permeable poroelastic media. By supplementing the formulation with appropriate closure relations and additional physics, such as lubrication flow in hydraulic fractures or frictional strength evolution in shear fractures, it can be used to investigate a broad range of coupled fracture-fluid problems. The approach may also be adapted to other classes of elasto-diffusive problems by modifying the underlying physical parameters.
\end{abstract}


\maketitle

\section{Introduction}

Fracture propagation in poroelastic media is a fundamental process underlying numerous natural and engineered phenomena, including tensile fractures and shear faulting in the Earth's crust, slope instability, hydraulic and chemically induced cracking, and fracture processes in soft hydrophilic polymers \citep{rubin1995propagation, EcNo00,economides2002unified, tsai2010model,rivalta2015review, YuLa18, LeTi19, YuLa20}. In these systems, solid deformation and diffusion processes are intrinsically coupled, producing time-dependent stress and diffusion fields that govern the evolution of fracture opening, slip, and mass exchange. Accurately capturing this coupling poses challenges in modeling tensile (Mode I) and shear (Mode II) fractures \citep{boone1990numerical, sarris2012modeling, zhou2018phase, kamali2020role}.

Fully coupled poroelastic fracture problems can be addressed using the finite element method, which provides a general framework for capturing the two-way coupling between solid deformation and diffusion (see \cite{carrier2012numerical, BoLa15, salimzadeh2017three, baykin2018application, golovin2018influence, YuBo20} among others). However, such simulations require meshing the entire computational domain and applying substantial mesh refinement in the vicinity of the fracture to adequately resolve the steep gradients in stresses, displacements, and pore pressure. When fracture propagation is coupled with additional physical processes, such as cohesive zone models, lubrication flow within the fracture, or other nonlinear effects, the resulting system becomes strongly nonlinear and increasingly stiff. Moreover, elasto-diffusive processes often require small time steps to maintain stability and temporal resolution. Fully coupled finite element simulations remain computationally expensive for problems spanning large spatial and temporal scales.

Alongside numerical developments, mathematical formulations have also been proposed for hydraulic fractures in poroelastic media involving contact, cohesion, and non-penetration constraints between fracture faces. \cite{itou2022asymptotic} derived asymptotic solutions in the form of power series for a plane poroelastic model with a non-penetrating fluid-driven crack and obtained the associated stress intensity factors. More recently, \cite{itou2025nonlinear} formulated a nonlinear porous body model for a fluid-driven fracture with cohesion contact conditions, leak-off, and fluid volume control, and established well-posedness of the resulting variational problem.

Boundary integral methods provide an efficient and rigorous framework for analyzing fracture problems in elastic media \citep{crouch1983boundary, weertman1996dislocation, hills2013solution}, and have been extended to poroelasticity (see \cite{cheng1998singular, Chen16} for a review). In this approach, fractures are represented as continuous distributions of fundamental solutions corresponding to edge dislocations (opening and slip modes) and fluid sources embedded in an infinite domain. By superposing these solutions, the tractions and pore fluid pressure on the fracture surfaces are expressed through integral relations posed directly on the fracture geometry, thereby reducing the governing system of partial differential equations to boundary integral equations \citep{detournay1987poroelastic, vandamme1989two}. These formulations are particularly attractive because they exactly satisfy the governing equations in the surrounding medium.


When fracture problems in poroelastic media are examined using boundary integral formulations, a principal difficulty arises from the presence of space-time convolution integrals that relate displacement and flux discontinuities to the evolving pore pressure and stress fields \citep{detournay1987poroelastic, vandamme1989two}. These convolution integrals significantly increase computational cost and, in most cases, preclude closed-form analytical solutions. Spectral boundary integral formulations have also been used for fault problems in poroelastic solids with different types of hydraulic boundary conditions on the fault plane \citep{heimisson2022spectral,noda2022dynamic,heimisson2024analytical}. These studies provide formulations in the spatial Fourier wavenumber--time domain; however, for non-steady problems, temporal convolution remains necessary. A major simplification occurs when a steadily propagating semi-infinite fracture is considered. In a coordinate system moving with the fracture tip, the governing equations become time-independent and contain single spatial convolutions, with fracture characteristics depending solely on the distance from the tip \citep{Clea78, cheng1984boundary}. Consequently, the steadily propagating semi-infinite fracture has emerged as a canonical configuration, serving both as a fundamental problem in its own right and as a local representation of fracture fronts in more complex geometries \citep{PeDe08, peirce2015modeling, Deto16, dontsov2017multiscale, ZiLe20}. Moreover, it enables so-called ``equation of motion''-type approximate solutions for finite fracture problems with idealized geometries \citep{dontsov2016approximate, dontsov2017approximate, garagash2019cohesive, kanin2021radial, garagash2025propagation}.

Fundamental solutions of plane strain poroelasticity have been developed in a series of classical works. \cite{RiCl76} derived solutions for point forces and edge dislocations in both slip and opening modes, based on the decomposition introduced by \cite{McGi60}. Fundamental solutions for an instantaneous fluid source were subsequently obtained by \cite{Clea76, Clea77}. These building blocks were employed by \cite{RiSi76} to analyze steadily propagating shear cracks under imposed loading, while \cite{Clea76a} investigated the formation of shear bands in softening poroelastic media. Using complex potential methods, \cite{ruina1978} studied a steadily moving tensile fracture subjected to uniform normal loading over a finite region adjacent to the fracture tip. This approach was later extended by \cite{AtCr91} to exponentially decreasing normal loading and generalized to non-steadily propagating fractures.

A significant advance was made by \cite{Clea78}, who derived fundamental solutions for steadily moving fluid source and edge dislocation in poroelastic media and identified their potential as building blocks for fracture problems. A limitation of the solutions presented in \citep{Clea78} for a steadily moving edge dislocation is that they were not reduced to closed-form analytical expressions, thereby restricting their practical applicability. Subsequent studies employed subsets of these solutions to address specific problems, usually neglecting some coupling terms. \cite{DeGa03} utilized the solution for pore pressure generated by a steadily moving fluid source to analyze the pore fluid circulation zone near the tip of a fluid-driven fracture. \cite{Kova10} applied the solutions for pore pressure and normal stress on the fracture surfaces associated with a moving fluid source to investigate two-dimensional fluid exchange and the resulting poroelastic backstress (the stress induced by deformation driven by pore pressure diffusion that resists fracture opening) in the model of a semi-infinite steadily propagating hydraulic fracture. However, the analysis in \cite{Kova10} neglected the poroelastic contribution of rock deformation to the normal stress on the fracture surfaces, considering only the elastic contribution, and deformation-induced pore pressure perturbations.

Within the steadily moving semi-infinite setting, existing studies of cracks in poroelastic media have relied on uncoupled or partially coupled formulations. In particular, a boundary integral formulation that accounts for the full poroelastic response associated with the adopted hydraulic boundary conditions has not, to the best of our knowledge, been developed previously. The present work addresses this gap by deriving a boundary integral formulation for steadily propagating semi-infinite plane strain tensile (Mode I) and shear (Mode II) fractures in poroelastic media. In addition, we develop a numerical methodology for solving the resulting system of equations.

The main contribution of this work is therefore an analytical-numerical boundary integral framework for steadily propagating semi-infinite cracks in poroelastic media. We first derive analytical solutions for the stress and pore pressure fields generated by a steadily moving fluid source and by steadily moving edge dislocations in slip and normal modes. These solutions are obtained from the corresponding instantaneous fundamental solutions by applying the temporal superposition principle. We then represent the fracture as a continuous distribution of fluid sources and edge dislocations and, using spatial superposition, formulate the boundary integral equations for tensile and shear cracks. The resulting system is solved using an asymptotics-informed collocation method. The formulation provides a boundary operator that can be coupled, in future applications, with additional closure relations, such as lubrication flow in hydraulic fractures or frictional strength evolution in shear fractures.

The paper is organized as follows. Section~\ref{sec:fundamental_solutions} introduces plane strain poroelastic fundamental solutions for pore pressure and stress components induced by instantaneous fluid source and edge dislocations (normal and slip modes). Section~\ref{sec:steadily_moving_loadings} extends these solutions to steadily moving loadings via temporal superposition and characterizes their singularities and far-field behavior. Section~\ref{sec:boundary_integral_equations} uses spatial superposition to derive boundary integral equations governing the distributions of tractions and pore fluid pressure on the surfaces of a semi-infinite steadily propagating fracture. Section~\ref{sec:dimensionless_equations} presents the normalized form of these equations, while Section~\ref{sec:numerical_method} details the numerical algorithm for their solution. Section~\ref{sec:verification_tests} verifies the formulation on benchmark problems by comparing numerical results with available analytical and semi-analytical solutions.

\section{Fundamental solutions in plane strain poroelasticity}
\label{sec:fundamental_solutions}

We begin by recalling the governing equations for a fluid-saturated, linear, isotropic poroelastic medium as first outlined by \cite{Biot41}. We then present plane strain fundamental solutions for the pore pressure and stress fields induced by an instantaneous fluid source and by an instantaneous edge dislocation (both normal and slip modes) in an infinite domain. This section closely follows the presentation available in existing references on small-strain linear poroelasticity \citep{Biot41,RiCl76,DeCh93,Chen16}.

\subsection{Governing equations}

\subsubsection{Constitutive relations}

The mechanics of a fluid-saturated porous medium is described by the kinematics of a solid obeying the small strain hypothesis, namely the solid displacement vector $u_i$ and the infinitesimal strain tensor $\varepsilon_{ij} = (u_{i,j} + u_{j, i})/2$. In addition, the evolution of the amount of fluid in the representative element is tracked by the variation of fluid content $\zeta$ defined as $\zeta = (m_f-m_f^0)/\rho_f^0$, i.e., the variation of fluid mass with respect to the reference state divided by the fluid density in the reference state.

In parallel to these kinematic quantities $\varepsilon_{ij}$, $\zeta$, the total stress tensor of the porous medium $\sigma_{ij}$ and the pore fluid pressure $p$ describe the current state of the representative element. The linear isotropic poroelastic constitutive relations (in the absence of pre-stress and pore pressure in the reference state) can be expressed as:
\begin{flalign}
    & \sigma_{ij} = 2G \varepsilon_{ij} + \frac{2 G \nu }{1-2\nu } \varepsilon_v \delta_{ij} -\alpha p \delta_{ij}, \label{eq:poro_constitutive_1} \\
    & \zeta = \alpha \varepsilon_v +\frac{p}{M}, \label{eq:poro_constitutive_2}
\end{flalign}
where $G$ is the shear modulus, $\nu$ is the drained Poisson's ratio, $\alpha$ and $M$ are the Biot coefficient and Biot modulus, respectively; $\varepsilon_v = \varepsilon_{kk}$ is the volumetric strain, $\delta_{ij}$ is the Kronecker delta. Note that both equations use the summation of repeated indices.

In the drained limit, the fluid pore pressure variation is zero, such that the variation of fluid content is directly related to the volumetric strain as $\zeta = \alpha \varepsilon_v$. The solid behaves elastically with a drained Poisson's ratio $\nu$.

The undrained limit corresponds to the case in which the fluid has no time to flow, so that the variation of fluid content vanishes, implying $p=-\alpha M \varepsilon_v$. In that limit, the total stress can be solely related to strain via the introduction of an undrained bulk modulus $K_u = K + \alpha^2 M$, where $K=2G(1+\nu)/[3(1-2\nu)]$ is the drained bulk modulus, or alternatively an undrained Poisson's ratio: 
\begin{equation*}
    \nu_u =\frac{1}{2} \left(1-\frac{G (1-2 \nu )}{G + \alpha ^2 M (1-2 \nu )}\right).
\end{equation*}

Thermodynamic constraints on elastic materials require Poisson's ratio to satisfy $0 \leq \nu \leq 1/2$, implying $0 \leq \nu \leq \nu_u \leq 1/2$.

\subsubsection{Field equations and Darcy's law}
The static balance of momentum of the porous solid and the conservation of fluid mass in the pore space are the fundamental conservation laws to be satisfied in a poroelastic domain. Accounting for a total body force $F_i$ and fluid source $\gamma$, they read respectively:
\begin{flalign}
    & \sigma_{ij,j} + F_i = 0, \label{eq:momentum_conservation} \\
    & \frac{\partial \zeta }{\partial t} + q_{i,i} = \gamma. \label{eq:mass_conservation}
\end{flalign}
The fluid discharge vector $q_i$ corresponds to the fluid velocity (with respect to the solid) averaged over the pore space. It is related to the pore fluid pressure through Darcy's law, under the classical assumption of laminar flow in a porous medium:
\begin{equation}
    q_i = - \kappa (p_{,i} - f_i),
    \label{eq:darcy_law}
\end{equation}
where $\kappa=k/\mu$ is the fluid mobility coefficient (with $k$ the intrinsic permeability and $\mu$ the fluid viscosity), $f_i$ is the fluid body force. 

Combining Eqs.~\eqref{eq:poro_constitutive_1} -- \eqref{eq:momentum_conservation} yields the elasticity (Navier) equation with a fluid coupling term:
\begin{equation}
    G u_{i, jj} + \frac{G}{1-2\nu_u} u_{j, ji} - \alpha M \zeta_{, i} + F_i = 0.
    \label{eq:navier_with_coupling}
\end{equation}

Using Eqs.~\eqref{eq:poro_constitutive_2}, \eqref{eq:mass_conservation}, \eqref{eq:darcy_law}, and \eqref{eq:navier_with_coupling}, we obtain the diffusion equation:
\begin{equation}
    \frac{\partial \zeta }{\partial t} - c \nabla^2 \zeta = \frac{\eta c}{G}F_{i, i} - \kappa f_{i, i} + \gamma,
    \label{eq:diffusion}
\end{equation}
where  
\begin{equation*}
   c = \kappa/S
\end{equation*}
is the hydraulic diffusivity coefficient (also referred to as the generalized consolidation coefficient \citep{RiCl76}), while $S$ is the uniaxial strain / constant stress storage coefficient \citep{Chen16}, defined as:
\begin{equation*}
    S=\frac{1}{M}+\frac{\alpha^2}{K+4G/3}=\frac{\alpha^2(1-2\nu)^2(1-\nu_u)}{2G(1-\nu)(\nu_u-\nu)}.
\end{equation*}
In Eq.~\eqref{eq:diffusion}, we also introduced the poroelastic stress coefficient \citep{DeCh93}:
\begin{equation}
    \eta = \frac{\alpha (1-2\nu)}{2(1-\nu)},
    \label{eq:poroelastic_stress_coefficient}
\end{equation}
varying within the range $0 \leq \eta \leq 0.5$. The lower bound corresponds to either the uncoupled case ($\alpha = 0$) or the incompressible solid limit ($\nu = 0.5$), while the upper bound represents a poroelastic medium with $\alpha = 1$ and $\nu = 0$.

In addition to the constitutive and field equations, a well-posed set of boundary and initial conditions completes the formulation of a poroelastic initial-boundary value problem \citep{Chen16}.

\subsection{Some fundamental solutions}

In Eqs.~\eqref{eq:navier_with_coupling} and \eqref{eq:diffusion}, the body force and source terms can be used to introduce field
singularities into the solution \citep{detournay1987poroelastic, cheng1998singular, Chen16}. Fundamental solutions for a point source and an edge dislocation are derived using the decomposition technique introduced by \cite{biot1956general}:
\begin{equation}
    u_i = u_i^0 + \frac{\eta}{GS}\Phi_{,i}. 
    \label{eq:biot_decomposition}
\end{equation}
We require $u_i^0$ to satisfy the undrained Navier equation of elasticity:
\begin{equation}
    G u_{i, jj}^0 + \frac{G}{1-2\nu_u} u_{j, ji}^0 + F_i = 0,
    \label{eq:navier_without_coupling}
\end{equation}
such that the substitution of Eqs.~\eqref{eq:biot_decomposition} and \eqref{eq:navier_without_coupling} into Eq.~\eqref{eq:navier_with_coupling} provides the definition of the potential $\Phi$ as:
\begin{equation*}
    \zeta = \nabla^2 \Phi.
\end{equation*}

Substituting the preceding expression into Eq.~\eqref{eq:diffusion} and relaxing a Laplacian, we obtain the diffusion equation for the potential $\Phi$:
\begin{equation}
    \frac{\partial \Phi }{\partial t} - c \nabla^2 \Phi = g_1 + g_2 + g_3,
    \label{eq:diffusion_potential}
\end{equation}
where the functions $g_1$, $g_2$, and $g_3$ satisfy the following equations:
\begin{equation*}    
    \nabla^2 g_1 = \frac{\eta c}{G} F_{i, i}, ~~~ \nabla^2 g_2 = - \kappa f_{i, i}, ~~~ \nabla^2 g_3 = \gamma.
\end{equation*}
It is important to note that the solution for $\Phi$ is not unique, as it is defined only up to an arbitrary harmonic function.
Once the potential $\Phi$ is determined in the solution process, the quantities $\zeta$, $q_i$, $p$ can be computed without using $u_i^0$. The stress components, however, require both $u_i^0$ and $\Phi$. 

In summary, the displacement field can be decomposed into an undrained part, $u_i^0$, which satisfies an elasticity equation with undrained elastic moduli, and an irrotational part derived from the potential $\Phi$ governed by a diffusion equation. Although these components are generally coupled through boundary conditions, in infinite domains the undrained part becomes time-independent, and all time-dependence is contained in the irrotational part. This makes the Biot decomposition particularly effective for deriving fundamental solutions, reducing the problem to solving two uncoupled singular equations, \eqref{eq:navier_without_coupling} and \eqref{eq:diffusion_potential} \citep{detournay1987poroelastic, Chen16}.

\begin{figure}[]
    \centering
    \includegraphics[width=0.9\textwidth]{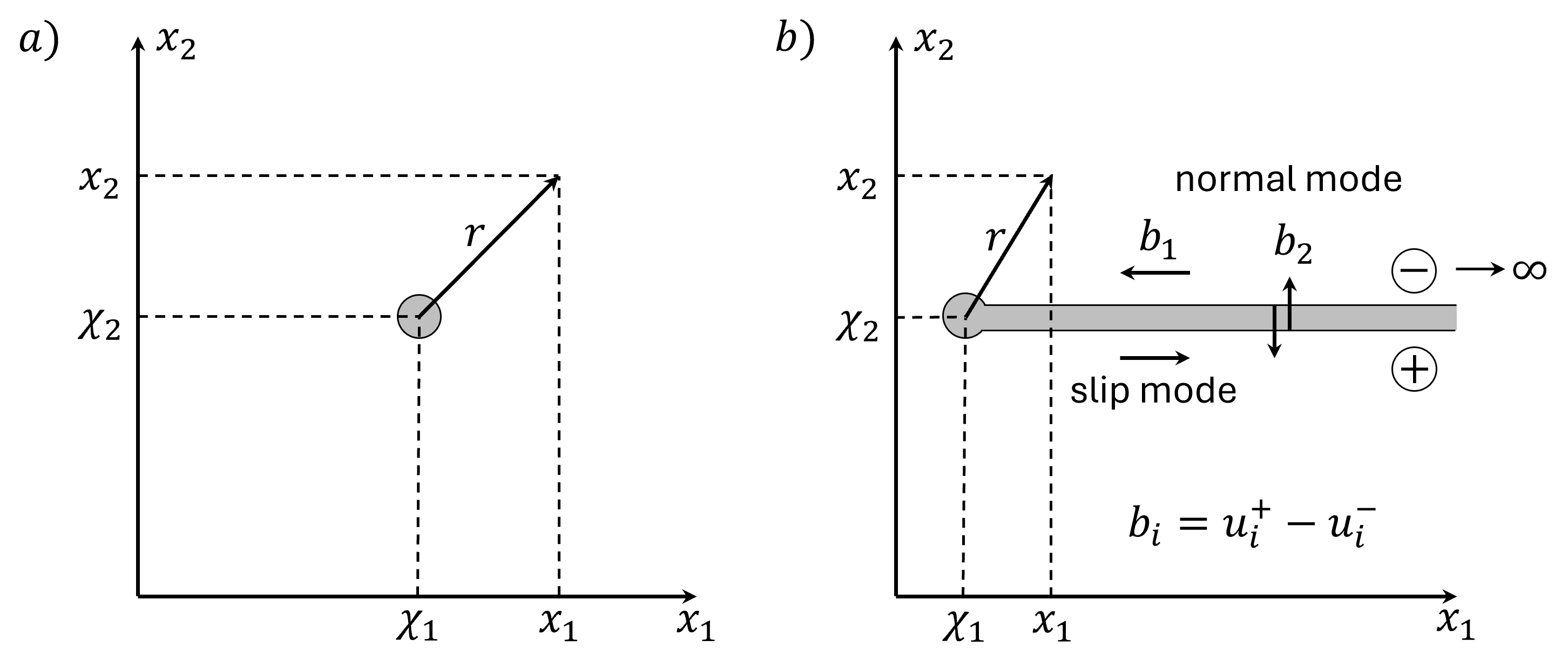}
    \caption{Fundamental loadings: (a) a fluid source and (b) an edge dislocation located at $\vec{\chi} = (\chi_1, \chi_2)$. The observation point is denoted by $\vec{x} = (x_1, x_2)$.}
    \label{fig:fundamanetal_fluid_source_edge_dislocation}
\end{figure}

\subsubsection{Instantaneous fluid source solutions}

A fluid source (Fig.~\ref{fig:fundamanetal_fluid_source_edge_dislocation}a) corresponds to the injection of fluid mass at a point within the medium. The instantaneous fluid source is modeled by the term $\gamma = \delta(\vec{x}-\vec{\chi})\delta(t - \tau)$, where $\delta(\cdot)$ denotes the Dirac delta function, $\vec{x} = (x_1, x_2)$ is the observation point, $t$ is the current time, and $\vec{\chi} = (\chi_1, \chi_2)$ specifies the location of the fluid source active at time $t = \tau$. 

The pore pressure and stress fields resulting from the instantaneous fluid source are given by the following expressions (see \cite{detournay1987poroelastic, Chen16} for more details):
\begin{flalign}
    & p^{si} = \frac{1}{\pi S} \frac{1}{r^2} \varsigma^2 e^{-\varsigma^2}, \label{eq:p_si} \\ 
    & \sigma^{si}_{ij} = \frac{\eta}{\pi S} \frac{1}{r^2} \left[\left(\delta_{ij} - 2r_{, i}r_{, j}\right)\left(1-e^{-\varsigma^2}\right) - 2\left(\delta_{ij} - r_{, i}r_{, j}\right)\varsigma^2 e^{-\varsigma^2}\right]. \label{eq:sigma_si} 
\end{flalign}
where we have introduced the following definitions:
\begin{equation*}
    \varsigma = \frac{r}{\sqrt{4 c (t - \tau)}}, \quad r = |\vec{x} - \vec{\chi}|, \quad  r_{, i} = \frac{\partial r}{\partial x_i} = \frac{x_i-\chi_i}{r}.
\end{equation*}
$\varsigma$ is the self-similar diffusion variable, $r$ is the Cartesian distance between the observation and source points, and $r_{,i}$ denotes the derivative of $r$ with respect to $x_i$.

\subsubsection{Instantaneous edge dislocation solutions}
\label{sec:instantaneous_edge_dislocation}


An edge dislocation (Fig.~\ref{fig:fundamanetal_fluid_source_edge_dislocation}b) corresponds to an imposed uniform displacement jump along a semi-infinite cut in an infinite medium. Two in-plane dislocation modes can be distinguished: a slip mode and a normal mode, corresponding to slip (shear/glide) and normal (climb) dislocations, respectively.

Edge dislocations can also be classified by their temporal evolution. A continuous edge dislocation is introduced at time $t = \tau$ and remains constant for $t > \tau$. In contrast, an instantaneous edge dislocation is applied at $t = \tau$ and is represented by a Dirac delta function in time. 

The pore pressure and stress fields induced by an edge dislocation depend on the flow boundary condition imposed on the dislocation plane. \cite{RiCl76, detournay1987poroelastic} derived solutions for a continuous edge dislocation subject to a zero pressure boundary condition along the dislocation line for the slip mode and a no-flow boundary condition for the normal mode. This choice of boundary conditions is natural. For a slip edge dislocation, continuity of pore pressure across the dislocation plane, together with the antisymmetry of the problem, requires zero pressure on the dislocation plane. For a normal edge dislocation, continuity of the normal derivative of pore pressure across the plane, combined with the symmetry of the problem, implies that no fluid flow can occur through the dislocation plane. It is important to note that this does not preclude fluid exchange between the fracture surfaces and the surrounding material, as modeled via the superposition of dislocations and fluid sources, as described later. However, the pressure field ahead of the fracture along its plane satisfies the corresponding boundary conditions: a no-flow (symmetry) condition for the normal mode, and zero induced pore pressure for the slip mode.

\cite{rudnicki1987plane} examined alternative boundary conditions on the dislocation plane: an impermeable plane for a slip edge dislocation and a permeable plane for a normal edge dislocation. For the slip mode, the dislocation plane may be effectively impermeable in some problems, such as when modeling a fault with clay gouge that inhibits fluid flow across the interface. In this case, the appropriate boundary condition is zero normal fluid flux, which results in a discontinuity in pore pressure across the plane. For the normal mode, a thin, highly permeable layer can be present along the dislocation plane. Such a layer can maintain the pore pressure at its ambient value, either because fluid flows easily within it or because opening is accompanied by fluid injection. This would lead to a boundary condition in which the pore pressure is prescribed on the plane and the normal flux is discontinuous. 

More general hydraulic descriptions of the dislocation plane are also possible and may be more appropriate for partially sealed faults, finite-thickness gouge layers, or interfaces with finite hydraulic resistance. For example, \cite{song2017plane} considered a slip edge dislocation on a leaky plane, for which the pressure jump across the interface is proportional to the normal fluid flux. This model introduces an interfacial hydraulic resistance and provides a continuous transition between the permeable limit, characterized by zero induced pore pressure on the slip plane, and the impermeable limit, characterized by zero normal pressure gradient on the interface.



In this study, we adopt the boundary conditions and associated solutions derived by \cite{detournay1987poroelastic}. These boundary conditions correspond to a specific classical hydraulic idealization rather than to a general interface law. Therefore, the kernels used below should be interpreted as fundamental solutions for the hydraulic limits embedded in these classical solutions. Alternative interface models, such as the leaky, partially sealed, or finite-thickness interface descriptions mentioned above, would require the use of the corresponding fundamental solutions and are not considered in the present work. However, the proposed methodology can be applied to such hydraulic interface models, provided that the required fundamental solutions are available.


\cite{detournay1987poroelastic} showed that the fields (pore pressure, displacement vector, and stress components) associated with a continuous edge dislocation can be decomposed into time-independent and time-dependent parts. The time-independent contribution corresponds to the undrained response of the poroelastic medium. 

Furthermore, the presence of a continuous edge dislocation oriented along the positive $x_1-$axis is equivalent to introducing a body force term $F_{ik, i} = 2G \mathrm{H}(t-\tau) e_{kj} \delta_{, j}(\vec{x} - \vec{\chi})$ on the right hand side of Eq.~\eqref{eq:diffusion}, where $k$ denotes the dislocation mode ($k = 1$ for slip mode, $k = 2$ for normal mode), $e_{ki}$ is the two-dimensional permutation symbol ($e_{11} = e_{22} = 0$, $e_{12} = -e_{21} = 1$), and $\mathrm{H}(\cdot)$ is the Heaviside function. 

Since this body force term involves the spatial derivative $\delta_{, j}(\vec{x} - \vec{\chi})$, the time-dependent part of the solution for a continuous edge dislocation is governed by a continuous fluid dipole: a spatially distributed pair of equal-and-opposite fluid sources whose potential $\Phi$ is equivalent to the directional derivative of the fluid source potential taken along the line connecting the source and sink. In particular, for a continuous edge dislocation located along the positive $x_1-$axis, the equivalent continuous dipole has strength $\mp 2\eta c$ ($-$ for slip mode, $+$ for normal mode) and is oriented $\pi/2$ clockwise relative to the Burgers vector $b_i = u_i^+ - u_i^-$, where $u_i^{\pm}$ denote the displacements on the positive (bottom) and negative (top) sides of the dislocation, respectively. 

To obtain the pore pressure and stress fields induced by an instantaneous edge dislocation oriented along the positive $x_1-$axis, we take the time derivative, $\partial/\partial t$, of the corresponding fields for a continuous edge dislocation reported by \cite{detournay1987poroelastic}, leading to the following expressions:
\begin{flalign}
    & p^{ei}_k = (p^{ei}_k)^0 + \Delta p^{ei}_k, \label{eq:p_ei} \\
    & (p^{ei}_k)^0 = \frac{\eta}{\pi S} e_{ki} \frac{r_{, i}}{r} \delta(t-\tau), ~~~ \Delta p^{ei}_k = -\frac{4 c \eta}{\pi S} e_{ki} \frac{r_{, i}}{r^3} \varsigma^4 e^{-\varsigma ^2}; \nonumber \\
    & \sigma^{ei}_{ijk} = (\sigma^{ei}_{ijk})^0 + \Delta \sigma^{ei}_{ijk}, \label{eq:sigma_ei} \\
    & (\sigma^{ei}_{ijk})^0 = \frac{G}{2\pi (1-\nu_u)} e_{kl} \frac{1}{r} \left(\delta_{il}r_{,j} + \delta_{jl}r_{,i} - \delta_{ij}r_{,l} - 2 r_{, i} r_{, j} r_{, l}\right) \delta(t-\tau), \nonumber \\
    & \Delta \sigma^{ei}_{ijk} = -\frac{4c\eta^2}{\pi S} e_{kl} \frac{1}{r^3} \bigg\{\left(\delta_{il}r_{,j} + \delta_{jl}r_{,i} + \delta_{ij}r_{,l} - 4 r_{, i} r_{, j} r_{, l}\right) \left[1 - (1+\varsigma^2)e^{-\varsigma ^2}\right] + 2 \left(r_{, i} r_{, j} r_{, l} - \delta_{ij} r_{, l}\right) \varsigma ^4 e^{-\varsigma ^2} \bigg\}. \nonumber
\end{flalign}
where $k$ represents the dislocation mode ($k = 1$ for slip mode, $k = 2$ for normal mode).





\section{Steadily moving fluid source and edge dislocation solutions}
\label{sec:steadily_moving_loadings}

We now derive the distributions of pore pressure and stress components produced by a steadily moving (i) fluid source and (ii) edge dislocation (normal and slip modes). These fields are obtained in a coordinate system moving with the respective loadings and will serve as the foundation for developing boundary integral equations for problems involving steadily propagating cracks.

We denote the fixed coordinate system by $(X, Y)$ and the moving system by $(x, y)$. The loading travels along the $X-$axis in the negative direction at constant velocity $V$. The moving coordinates are defined by $x = X + V t$, $y = Y$, so that the solution is steady in the moving frame $(x, y)$. Consequently, time derivatives in the fixed frame convert into spatial derivatives in the moving coordinates, and the transient problem reduces to a steady one in the co-moving frame.

The solution for the field $f^m = f^m(x, y)$ in the moving coordinate system, corresponding to a moving fluid source (or a moving edge dislocation), can be obtained using the temporal superposition principle. In this approach, the field is constructed by integrating the fundamental solution $f^i$, associated with an instantaneous fluid source (or an instantaneous edge dislocation). The instantaneous fundamental solution depends on the variables $f^i = f^i(X - \chi_1, Y - \chi_2, t - \tau)$, where $\vec{\chi} = (\chi_1, \chi_2) = (-V\tau, 0)$ denotes the position of the instantaneous loading at time $\tau$. Using these considerations, we obtain the general form of the field $f^m$ from the following integration:
\begin{equation}
    f^m (x, y) = \int_{-\infty}^{t} f^i \left[x - V (t-\tau), y, t-\tau\right] \diff \tau = \int_0^{\infty} f^i \left(x - V u, y, u\right) \diff u.
    \label{eq:field_moving_loading}
\end{equation}
The second equality in Eq.~\eqref{eq:field_moving_loading} follows from the change of integration variable $\tau \to u = t - \tau$, which removes the explicit dependence of $f^m$ on time $t$ and yields a steady field in the moving coordinate system.

We note that, for an edge dislocation, only the time-dependent components of the pore pressure ($\Delta p^{ei}_k$) and stresses ($\Delta \sigma^{ei}_{ijk}$) are substituted into Eq.~\eqref{eq:field_moving_loading}. The remaining terms, $(p^{ei}_k)^0$ and $(\sigma^{ei}_{ijk})^0$, correspond to the undrained response and provide the instantaneous fields evaluated at $\tau = t$; application of temporal superposition is unnecessary for these contributions.

\begin{figure}[]
    \centering
    \includegraphics[width=0.9\textwidth]{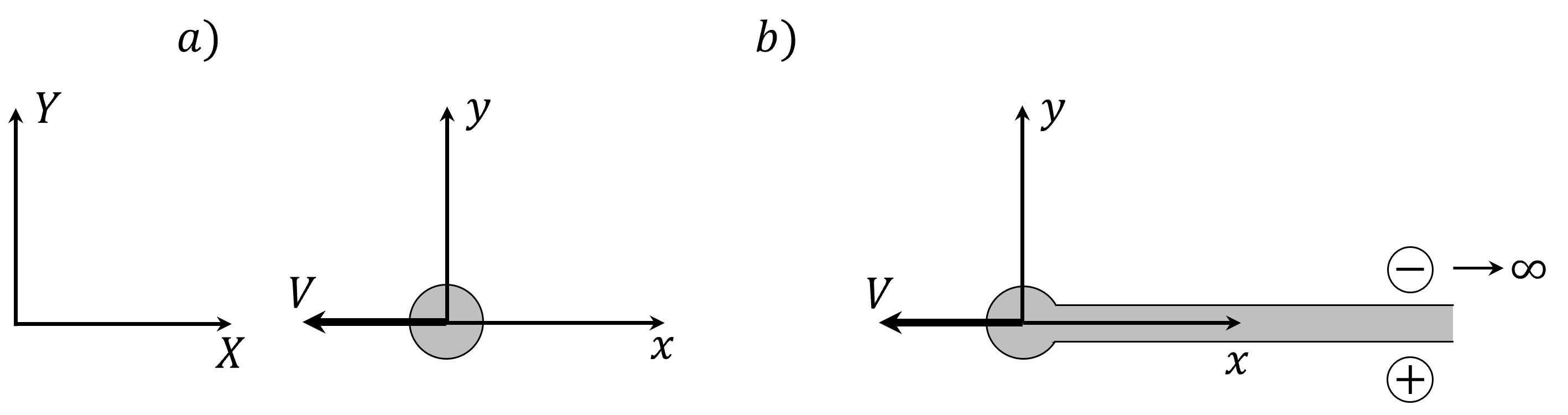}
    \caption{Fundamental loadings moving at constant velocity $V$: (a) a fluid source and (b) an edge dislocation. Both the fixed $(X, Y)$ and moving $(x, y)$ coordinate systems are indicated.}
    \label{fig:moving_fluid_source_edge_dislocation}
\end{figure}

Below, we present the analytical solutions for the pore pressure and the stress components along the $x-$axis induced by a steadily moving fluid source (Fig.~\ref{fig:moving_fluid_source_edge_dislocation}a) and by a steadily moving edge dislocation (Fig.~\ref{fig:moving_fluid_source_edge_dislocation}b). Detailed derivations are provided in Appendix~\ref{sec:derivations_moving_loadings}.

\subsection{Steadily moving fluid source}
\label{sec:solutions_moving_fluid_source}

We summarize the results of Appendix~\ref{sec:derivation_fluid_source} and present the pore pressure and stress distributions generated by a steadily moving fluid source, evaluated at $y=0$.

Pressure field:
\begin{equation}
    p^{sm} (x) = \frac{1}{2\pi\kappa} \mathcal{P}^{sm} \left(\frac{V x}{2 c}\right), ~~~~ \mathcal{P}^{sm}(\xi) = e^\xi \mathrm{K}_0 (|\xi|).
    \label{eq:p_sm}
\end{equation}

Stress tensor components:
\begin{equation}
    \sigma^{sm}_{ij} (x) = 
    \begin{pmatrix}
        \sigma^{sm}_{xx} (x)  \\ 
        \sigma^{sm}_{yy} (x)  \\ 
        \sigma^{sm}_{xy}(x) 
    \end{pmatrix} 
    = \frac{\eta}{2 \pi \kappa} \mathcal{S}^{sm}_{ij}\left(\frac{V x}{2 c}\right), ~~~~
    \mathcal{S}^{sm}_{ij}(\xi) =
    \begin{pmatrix}
        \mathcal{S}^{sm}_{xx} (\xi)  \\ 
        \mathcal{S}^{sm}_{yy} (\xi)  \\ 
        \mathcal{S}^{sm}_{xy}(\xi) 
    \end{pmatrix} = 
    \begin{pmatrix}
    \frac{1}{\xi} - e^\xi \left[\mathrm{K}_0(|\xi|) \pm  \mathrm{K}_1(|\xi|)\right] \\ 
    -\frac{1}{\xi} - e^\xi \big[\mathrm{K}_0(|\xi|) \mp \mathrm{K}_1(|\xi|)\big] \\ 
    0
    \end{pmatrix},
    \label{eq:sigma_sm}
\end{equation}
where the upper signs correspond to $\xi > 0$, and the lower signs to $\xi < 0$.

Based on the stress tensor components (Eq.~\eqref{eq:sigma_sm}), the mean and deviatoric stresses are defined as:
\begin{flalign*}
    & \sigma^{sm}_m (x) = \frac{\sigma^{sm}_{xx}(x) + \sigma^{sm}_{yy}(x)}{2} = \frac{\eta}{2 \pi \kappa} \mathcal{S}^{sm}_m \left(\frac{V x}{2 c}\right), \quad \mathcal{S}^{sm}_m(\xi) = -e^\xi \mathrm{K}_0 (|\xi|); \\
    & \sigma^{sm}_d (x) = \frac{\sigma^{sm}_{xx}(x) - \sigma^{sm}_{yy}(x)}{2} = \frac{\eta}{2 \pi \kappa} \mathcal{S}^{sm}_d \left(\frac{V x}{2 c}\right), \quad \mathcal{S}^{sm}_d (\xi) = \frac{1}{\xi} \mp e^\xi \mathrm{K}_1(|\xi|).
\end{flalign*}
Since $\mathcal{P}^{sm}(\xi) = -\mathcal{S}^{sm}_m(\xi)$, the pore pressure is related to the mean stress via: $p^{sm} (x) = -\sigma^{sm}_m (x) / \eta$.

An expression for the temperature field analogous to Eq.~\eqref{eq:p_sm} is given by \cite{CaJa59} for the equivalent problem in heat conduction. Using a different analytical approach, \cite{Clea78} considered a steadily moving fluid source and derived expressions for the associated pore pressure and stresses, which are identical to Eqs.~\eqref{eq:p_sm} and \eqref{eq:sigma_sm}, through a considerably more elaborate derivation. Eq.~\eqref{eq:p_sm} was employed by \cite{DeGa03} in a near-tip model of a fluid-driven fracture to investigate the pore fluid circulation zone, within which two-dimensional fluid exchange between the fracture and the surrounding formation occurs. Finally, \cite{Kova10} used Eqs.~\eqref{eq:p_sm} and \eqref{eq:sigma_sm} to account for the two-dimensional fluid exchange and poroelastic backstress in the fracture tip model.

\subsection{Steadily moving edge dislocation}
\label{sec:solutions_moving_edge_dislocation}

This section reports the pore pressure and stress components associated with a steadily moving edge dislocation, as derived in Appendix~\ref{sec:derivation_edge_dislocation}, evaluated at $y=0$.

Pressure field:
\begin{itemize}
    \item slip edge dislocation: 
    \begin{equation}
        p_1^{em} (x) = 0;
        \label{eq:p_em_1}
    \end{equation}
    \item normal edge dislocation:
    \begin{equation}
        p_2^{em}(x) = \frac{\eta V}{2 \pi \kappa} \mathcal{P}^{em}_2\left(\frac{V x}{2c}\right), ~~~~ \mathcal{P}^{em}_2(\xi) = -\frac{1}{\xi} - e^\xi \left[\mathrm{K}_0(|\xi|) \mp \mathrm{K}_1(|\xi|)\right].
        \label{eq:p_em_2}
    \end{equation}
\end{itemize}
Eq.~\eqref{eq:p_em_1} should be interpreted in the context of the adopted hydraulic boundary condition for a slip edge dislocation. It does not imply that shear slip in a poroelastic medium cannot generate pore pressure perturbations in general. Here, we adopted the assumption of pore pressure continuity across the dislocation plane, which, together with the antisymmetry of the problem, requires the induced pore pressure to vanish on the dislocation plane. In formulations with an impermeable, leaky, partially sealed, or finite-thickness shear zone, slip-induced pore pressure changes may appear on or across the interface and influence the effective normal stress and shear resistance \citep{rudnicki1987plane, song2017plane}. Such effects are outside the scope of the present work; however, the fundamental solutions presented by \cite{rudnicki1987plane} and \cite{song2017plane} could be extended to the steadily moving setting.

Stress tensor components:
\begin{itemize}
    \item slip edge dislocation: 
    \begin{equation}
        \sigma^{em}_{ij1}(x) = 
        \begin{pmatrix}
            \sigma^{em}_{xx1} (x) \\ 
            \sigma^{em}_{yy1} (x) \\ 
            \sigma^{em}_{xy1} (x)
        \end{pmatrix} = \frac{G}{2\pi (1-\nu_u)} ~ \frac{1}{x} ~ \mathcal{S}^{em}_{ij1}\left(\frac{V x}{2c}\right), ~~~~ 
        \mathcal{S}^{em}_{ij1}(\xi) = 
        \begin{pmatrix}
           \mathcal{S}^{em}_{xx1} (\xi) \\ 
            \mathcal{S}^{em}_{yy1} (\xi) \\ 
            \mathcal{S}^{em}_{xy1} (\xi)
        \end{pmatrix} = 
        \begin{pmatrix}
        0 \\ 0 \\ 1 + \beta \left[\frac{1}{\xi} \mp e^\xi \mathrm{K}_1(|\xi|)\right]
        \end{pmatrix};
        \label{eq:sigma_em_1}
    \end{equation}
    \item normal edge dislocation:
    \begin{flalign}
        & \sigma^{em}_{ij2}(x) = 
        \begin{pmatrix}
            \sigma^{em}_{xx2}(x) \\ 
            \sigma^{em}_{yy2}(x) \\ 
            \sigma^{em}_{xy2}(x)
        \end{pmatrix} = \frac{G}{2\pi (1-\nu_u)} ~ \frac{1}{x} ~ \mathcal{S}^{em}_{ij2} \left(\frac{V x}{2c}\right), \nonumber \\
        & \mathcal{S}^{em}_{ij2}(\xi) = 
        \begin{pmatrix}
            \mathcal{S}^{em}_{xx2}(\xi) \\ 
            \mathcal{S}^{em}_{yy2}(\xi) \\ 
            \mathcal{S}^{em}_{xy2}(\xi)
        \end{pmatrix} =
        \begin{pmatrix}
        1 + \beta \left[\frac{1}{\xi} \mp e^\xi \mathrm{K}_1(|\xi|)\right] \\ 
        1 - \beta \left(\frac{1}{\xi} \mp e^\xi \mathrm{K}_1(|\xi|) - 2 \xi e^\xi\bigg[\mathrm{K}_0(|\xi|) \mp  \mathrm{K}_1(|\xi|)\bigg]\right) \\ 
        0
        \end{pmatrix}.
        \label{eq:sigma_em_2}
    \end{flalign}
\end{itemize}

In Eqs.~\eqref{eq:p_em_2} -- \eqref{eq:sigma_em_2}, the upper signs correspond to $\xi > 0$, and the lower signs to $\xi < 0$. In the expressions for the stress components (Eqs.~\eqref{eq:sigma_em_1} and \eqref{eq:sigma_em_2}), the Cauchy-type singularity $1/x$ is explicitly factored out, and the poroelastic parameter $\beta$ is introduced, defined as:
\begin{equation}
    \beta = \frac{\nu_u - \nu}{1 - \nu}.
    \label{eq:poissons_contrast}
\end{equation}
It represents the normalized contrast between the undrained and drained Poisson's ratios. The parameter $\beta$ ranges from $0 \leq \beta \leq 0.5$, where the lower bound corresponds to the uncoupled case ($\alpha = 0$, resulting in $\nu_u = \nu$), and the upper bound reflects the maximum possible difference, attained for $\nu = 0$ and $\nu_u = 0.5$.

It is interesting to note that $\mathcal{P}^{em}_2(\xi) \equiv \mathcal{S}^{sm}_{yy} (\xi)$ and $\mathcal{S}^{em}_{xx2}(\xi) \equiv \mathcal{S}^{em}_{xy1} (\xi)$. Similar relations among Green's functions associated with different fields, including pore fluid pressure, stresses, and fluid displacement, for both continuous and instantaneous singularity types have been reported by \cite{cheng1998singular} (see also \cite{Chen16}).

\cite{Clea78} derived expressions for the pore pressure and stresses induced by a steadily moving edge dislocation using a distinct and substantially more involved analytical approach. Notably, the solutions in \citep{Clea78} were retained in integral form and were not reduced to closed-form expressions, in contrast to the explicit formulas obtained in the present work. These integral representations of the steadily moving solutions can be used in practice only if the corresponding integrals are precomputed and tabulated prior to application, which makes their use cumbersome and potentially prone to inaccuracies compared with the analytical expressions obtained here.

To the best of our knowledge, the solutions presented in the current section have not been applied in the literature to model a steadily moving semi-infinite fracture. Instead, existing models typically consider only the elastic contribution of solid deformation to the normal stress acting on the fracture surfaces, i.e., Eqs.~\eqref{eq:sigma_em_1} and \eqref{eq:sigma_em_2} with $\nu_u = \nu$, resulting in $\beta = 0$, while deformation-induced pore pressure perturbations (Eq.~\eqref{eq:p_em_2}) are often entirely neglected.

\subsection{Asymptotic behavior in the near-field and far-field}

It is useful to examine the near-field ($x \to 0$) and far-field ($x \to \pm \infty$) asymptotes of the pore pressure and stresses generated by a steadily moving fluid source (Section~\ref{sec:solutions_moving_fluid_source}) and edge dislocation (Section~\ref{sec:solutions_moving_edge_dislocation}). 

We present the characteristic shapes of the asymptotic fields as functions of the dimensionless coordinate:
\begin{equation}
    \xi = \frac{V x}{2c},
    \label{eq:dimensionless_coordinate_xi}
\end{equation}
retaining the sign and spatial dependence of each field while omitting constant dimensional prefactors.

We begin by considering the limit $\xi \to 0$, where the non-trivial fields exhibit singular behavior. For a fluid source, both the pore pressure and the non-zero stresses possess a logarithmic singularity of the form:
\begin{flalign*}
    & \mathcal{P}^{sm} (\xi \to 0) \sim - \gamma_E -\ln{\left(\frac{|\xi|}{2}\right)}, \quad \mathcal{S}^{sm}_{xx}(\xi \to 0) \sim -1 + \gamma_E + \ln{\left(\frac{|\xi|}{2}\right)}, \quad \mathcal{S}^{sm}_{yy}(\xi \to 0) \sim 1 + \gamma_E + \ln{\left(\frac{|\xi|}{2}\right)}, 
\end{flalign*}
where $\gamma_E$ is the Euler's constant. It is worth noting that the mean stress is characterized by a logarithmic singularity, whereas the deviatoric stress remains non-singular:   
\begin{equation*}
    \mathcal{S}^{sm}_m (\xi \to 0) \sim \gamma_E +\ln{\left(\frac{|\xi|}{2}\right)}, \quad \mathcal{S}^{sm}_d (\xi \to 0) \sim -1.
\end{equation*}

For a normal edge dislocation, the pore pressure field also exhibits a logarithmic singularity:
\begin{equation*}
    \mathcal{P}^{em}_2 (\xi \to 0) \sim 1 + \gamma_E + \ln{\left(\frac{|\xi|}{2}\right)}.
\end{equation*}

The non-zero stresses, in turn, display a combination of Cauchy-type and logarithmic singularities. Since the Cauchy-type singularity is explicitly factored out in Eq.~\eqref{eq:sigma_em_2} for the stress components $\sigma_{ij2}^{em}$, we present their characteristic shapes in the following form: 
\begin{flalign*}
    & \frac{\mathcal{S}^{em}_{xx2}(\xi)}{\xi}\bigg|_{\xi \to 0} \sim \frac{1-\beta }{\xi}-\frac{\beta}{4} \left[1 + 2 \gamma_E  + 2 \ln{\left(\frac{|\xi|}{2}\right)}\right], \quad \frac{\mathcal{S}^{em}_{yy2}(\xi)}{\xi}\bigg|_{\xi \to 0} \sim \frac{1-\beta }{\xi}-\frac{\beta}{4} \left[7 + 6 \gamma_E  + 6 \ln{\left(\frac{|\xi|}{2}\right)}\right],
\end{flalign*}

For a slip edge dislocation, the only non-zero field is the shear stress, which exhibits a logarithmic singularity:
\begin{equation*}
    \frac{\mathcal{S}^{em}_{xy1}(\xi)}{\xi}\bigg|_{\xi \to 0} \sim \frac{1-\beta }{\xi}-\frac{\beta}{4} \left[1 + 2 \gamma_E  + 2 \ln{\left(\frac{|\xi|}{2}\right)}\right].
\end{equation*}
Here, similarly to the previous equation, we account for the Cauchy-type singularity factored out in Eq.~\eqref{eq:sigma_em_1} for the stress components $\sigma_{ij1}^{em}$.

Next, we examine the far-field asymptotic behavior. In the domains ahead ($\xi \to -\infty$) and behind ($\xi \to + \infty$) the fluid source and the edge dislocation, the pore pressure and stresses behave as follows:
\begin{itemize}
    \item fluid source:
        \begin{align*}
            \mathcal{P}^{sm} (\xi \to -\infty) &\sim e^{2\xi} \sqrt{\frac{\pi}{2|\xi|}}, 
            & \mathcal{P}^{sm} (\xi \to +\infty) &\sim \sqrt{\frac{\pi}{2 \xi}} - \frac{1}{8} \sqrt{\frac{\pi}{2}} \frac{1}{\xi^{3/2}}; \\
            \mathcal{S}^{sm}_{xx}(\xi \to -\infty) &\sim \frac{1}{\xi} + \frac{e^{2\xi}}{2}\sqrt{\frac{\pi}{2}}\frac{1}{|\xi|^{3/2}}, 
            & \mathcal{S}^{sm}_{xx}(\xi \to +\infty) & \sim -\sqrt{\frac{2\pi}{\xi}} +\frac{1}{\xi}; \\
            \mathcal{S}^{sm}_{yy}(\xi \to -\infty) & \sim -\frac{1}{\xi} - e^{2\xi} \sqrt{\frac{2\pi}{|\xi|}}, 
            & \mathcal{S}^{sm}_{yy}(\xi \to +\infty) & \sim -\frac{1}{\xi} + \frac{1}{2} \sqrt{\frac{\pi}{2}} \frac{1}{\xi^{3/2}};
        \end{align*}
        
    \item normal edge dislocation:
        \begin{align*}
            \mathcal{P}^{em}_2 (\xi \to -\infty) & \sim -\frac{1}{\xi} - e^{2\xi} \sqrt{\frac{2\pi}{|\xi|}}, 
            & \mathcal{P}^{em}_2 (\xi \to +\infty) & \sim -\frac{1}{\xi} + \frac{1}{2} \sqrt{\frac{\pi}{2}} \frac{1}{\xi^{3/2}}; \\
            \frac{\mathcal{S}^{em}_{xx2} (\xi)}{\xi}\bigg|_{\xi \to -\infty} & \sim \frac{1}{\xi} + \frac{\beta}{\xi^2} - \beta e^{2\xi} \sqrt{\frac{\pi}{2}} \frac{1}{|\xi|^{3/2}}, 
            & \frac{\mathcal{S}^{em}_{xx2} (\xi)}{\xi} \bigg|_{\xi \to +\infty} & \sim \frac{1}{\xi} - \beta  \sqrt{\frac{\pi }{2}} \frac{1}{\xi^{3/2}} + \frac{\beta}{\xi^2}; \\
            \frac{\mathcal{S}^{em}_{yy2} (\xi)}{\xi}\bigg|_{\xi \to -\infty} & \sim \frac{1}{\xi} - \frac{\beta }{\xi^2} + 2 \beta e^{2\xi} \sqrt{\frac{2\pi}{|\xi|}}, 
            & \frac{\mathcal{S}^{em}_{yy2} (\xi)}{\xi}\bigg|_{\xi \to +\infty} & \sim \frac{1}{\xi} - \frac{\beta }{\xi^2} + \frac{3 \beta}{4} \sqrt{\frac{\pi}{2}} \frac{1}{\xi^{5/2}};
        \end{align*}
    \item slip edge dislocation:
        \begin{align*}
            \frac{\mathcal{S}^{em}_{xy1} (\xi)}{\xi}\bigg|_{\xi \to -\infty} & \sim \frac{1}{\xi} + \frac{\beta}{\xi^2} - \beta e^{2\xi} \sqrt{\frac{\pi}{2}} \frac{1}{|\xi|^{3/2}}, 
            & \frac{\mathcal{S}^{em}_{xy1} (\xi)}{\xi}\bigg|_{\xi \to +\infty} & \sim \frac{1}{\xi} - \beta  \sqrt{\frac{\pi }{2}} \frac{1}{\xi^{3/2}} + \frac{\beta}{\xi^2}.
        \end{align*}
\end{itemize}

\begin{figure}[]
    \centering
    \includegraphics[width=1\textwidth]{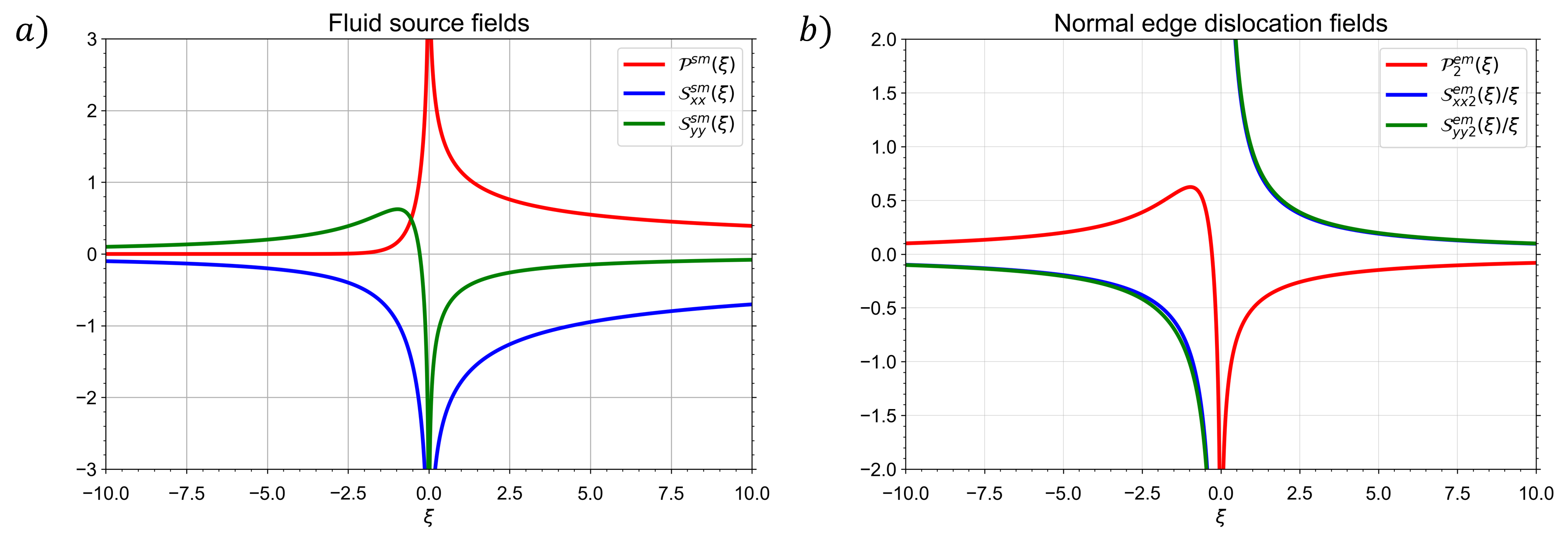}
    \caption{
    Characteristic shapes of the pore pressure and stresses induced by a steadily moving fluid source (a) and normal edge dislocation (b), plotted versus the dimensionless coordinate $\xi$ (Eq.~\eqref{eq:dimensionless_coordinate_xi}). The normalized undrained-drained Poisson's ratio contrast is taken as $\beta = 0.125$.
    }
    \label{fig:kernels}
\end{figure}

Fig.~\ref{fig:kernels} depicts the characteristic shapes of the pore pressure and stresses. Fig.~\ref{fig:kernels}a shows the fields induced by a fluid source, whereas Fig.~\ref{fig:kernels}b presents the distributions generated by a normal edge dislocation. The case of a slip edge dislocation is not included, since the shape of the shear stress $\mathcal{S}_{xy1}^{em}(\xi)/\xi$ (the only non-zero field) coincides with $\mathcal{S}_{xx2}^{em}(\xi)/\xi$ (see Fig.~\ref{fig:kernels}b).

\begin{figure}[]
    \centering
    \includegraphics[width=0.6\textwidth]{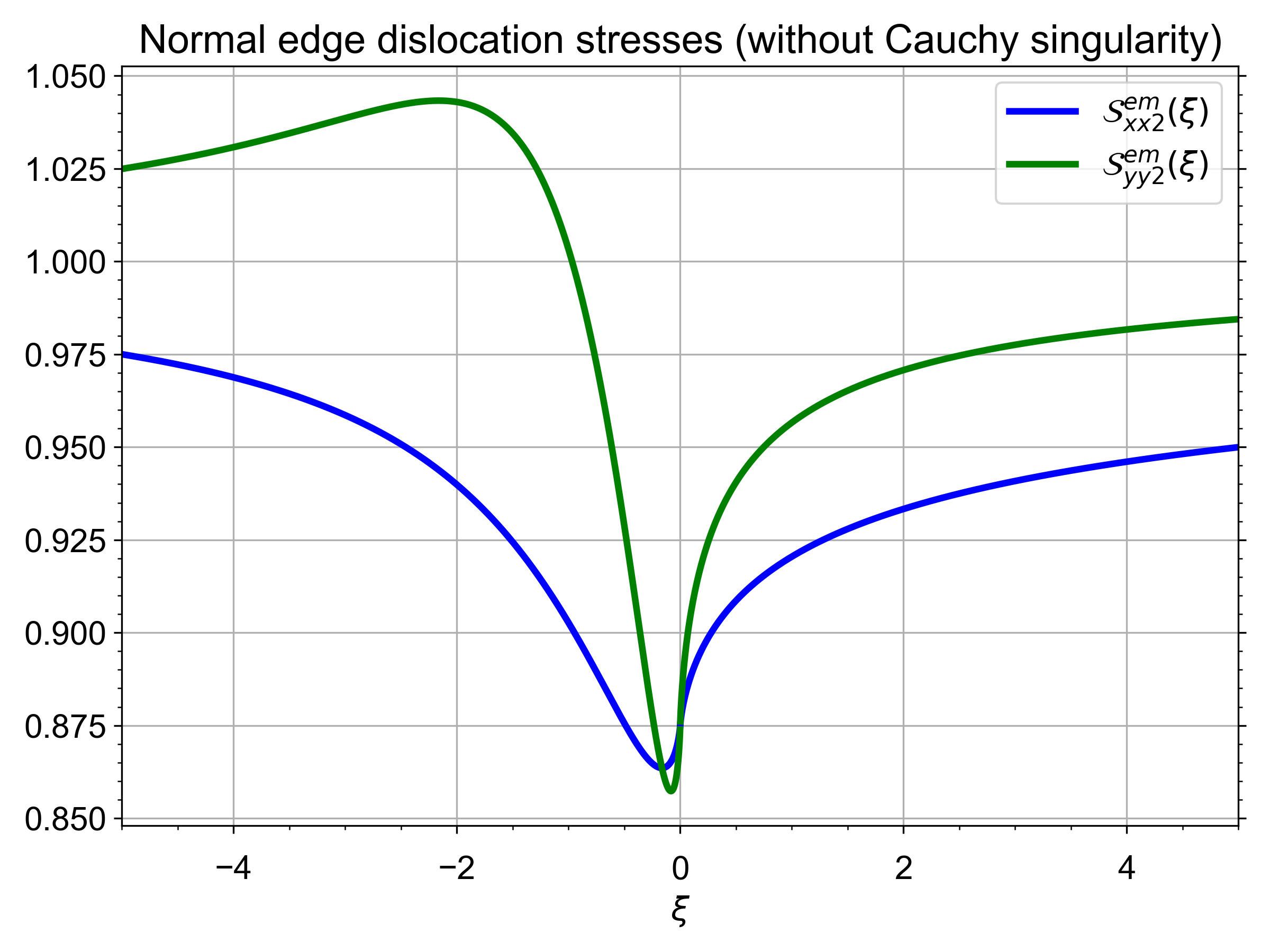}
    \caption{
    Characteristic shapes of the stresses induced by a steadily moving normal edge dislocation, with the Cauchy singularity removed by multiplying the governing equations by the dimensionless coordinate $\xi$ (Eq.~\eqref{eq:dimensionless_coordinate_xi}). The resulting distributions are plotted as functions of $\xi$. The value of the normalized undrained-drained Poisson's ratio contrast is $\beta = 0.125$. 
    }
    \label{fig:kernels_2}
\end{figure}

Since the dominant singularity of the stresses induced by an edge dislocation is of the Cauchy-type, we additionally examine the characteristic shapes of these fields after removing this singular contribution. The resulting regular (non-singular) functions are shown in Fig.~\ref{fig:kernels_2}.

The functions in Figs.~\ref{fig:kernels} and \ref{fig:kernels_2} exhibit strong, highly localized variations in the vicinity of $\xi = 0$, even after multiplying by $\xi$ to remove the Cauchy singularity (Fig.~\ref{fig:kernels_2}). Such complex behavior precludes the use of quadrature-based integral techniques (as employed in \cite{viesca2018numerical}) within the numerical framework described in Section~\ref{sec:numerical_method}, because Gauss-Chebyshev quadrature schemes cannot adequately resolve the sharp features of the integrand, leading to inaccurate results.

\section{Boundary integral equations for steadily moving fractures}
\label{sec:boundary_integral_equations}

We now formulate the boundary integral equations for steadily propagating semi-infinite cracks under the hydraulic boundary conditions embedded in the fundamental solutions described in Section~\ref{sec:instantaneous_edge_dislocation}. Tensile and shear cracks are considered separately. The steadily moving crack is represented as a superposition of moving edge dislocations and fluid sources. Hence, the solutions for steadily moving loadings, presented in Section~\ref{sec:steadily_moving_loadings}, can be used to formulate the boundary integral equations that relate the fracture opening or slip and the fluid exchange rate to the distributions of traction, i.e., normal or shear stress, and pore fluid pressure on the fracture surfaces. 

\begin{figure}[]
    \centering
    \includegraphics[width=1.0\textwidth]{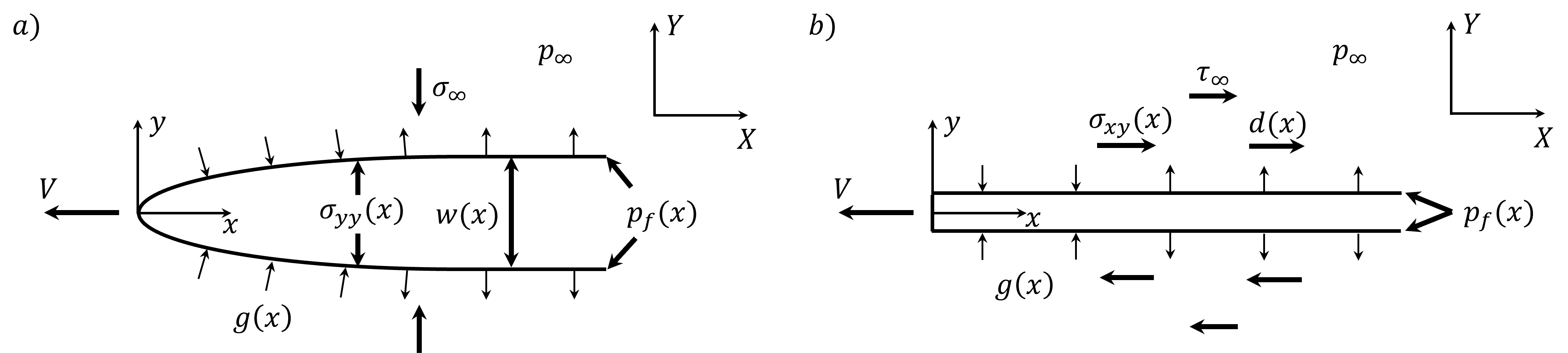}
    \caption{
    Schematic of a semi-infinite (a) tensile and (b) shear fractures propagating steadily at a constant velocity $V$ along the negative $X-$axis. The fixed $(X,Y)$ and moving $(x,y)$ coordinate systems are shown. The distributions of the pore fluid pressure $p_f(x)$, the normal stress $\sigma_{yy}$ (a), and the shear stress $\sigma_{xy}(x)$ (b) on the fracture surfaces are depicted. The fracture opening $w(x)$ (a) and slip $d(x)$ (b) are indicated. The reservoir is initially subjected to pore pressure $p_\infty$, uniform compressive stress $\sigma_\infty$ (a), and uniform shear stress $\tau_\infty$ (b). The local rate of fluid exchange between the fracture and the surrounding porous medium is denoted by $g(x)$.
    }
    \label{fig:semi_inf_poroelastic}
\end{figure}

We consider a semi-infinite crack propagating along the $X-$axis of the fixed coordinate system $(X, Y)$ in the negative direction at constant velocity $V$ (Fig.~\ref{fig:semi_inf_poroelastic}). We introduce a coordinate system $(x, y)$ moving together with a semi-infinite fracture. The fixed and moving coordinate systems are related to each other through the transformation: $x = X + Vt, ~ y = Y$. In the moving frame, the problem is stationary, and all fracture characteristics are functions of the distance from the moving tip $x$. These include the fracture opening profile $w(x)$ for the tensile fracture, the slip profile $d(x)$ for the shear fracture, and the fluid exchange rate profile $g(x)$, as well as the distributions on the fracture surfaces of the pore fluid pressure $p_f(x)$, the normal stress $\sigma_{yy}(x)$ for the tensile fracture, and the shear stress $\sigma_{xy}(x)$ for the shear fracture.

First, we discuss a tensile fracture (Fig.~\ref{fig:semi_inf_poroelastic}a). An arbitrary opening profile $w(x)$ can be represented as a superposition (pile-up) of infinitesimal normal edge dislocations distributed along the crack. The dislocation strength is characterized by the Burgers vector component $b_2 = u_2^+ - u_2^-$, where $u_2^{\pm}$ denote the displacements along the $y-$axis on the positive (bottom) and negative (top) sides of the dislocation, respectively. Under this convention, $b_2 < 0$ corresponds to an opening dislocation. At a point $x$, the dislocation strength is related to the opening profile as:
\begin{equation*}
    b_2(x) = -\diff w(x) = -\left[w(x + \diff x) - w(x)\right] = -\frac{\diff w}{\diff x}\diff x.
\end{equation*}
Using the principle of spatial superposition, the contribution of the distributed climb dislocations to the normal stress (positive in tension) and the pore fluid pressure on the fracture surfaces can be expressed as:
\begin{equation}
    \sigma_{yy}^e(x) = -\int_0^{\infty} \frac{\diff w}{\diff s} \sigma_{yy2}^{em}(x-s) \diff s, ~~~~ p^e(x) = -\int_0^{\infty} \frac{\diff w}{\diff s} p_2^{em}(x-s) \diff s,
    \label{eq:tensile_crack_dislocation_contribution}
\end{equation}
where the kernels $\sigma_{yy2}^{em}(x)$, $p_2^{em}(x)$ (given by Eqs.~\eqref{eq:sigma_em_2} and \eqref{eq:p_em_2}, respectively) correspond to the normal stress and the pore fluid pressure induced by a unit ($b_2 = 1$) normal steadily moving edge dislocation. It should be noted that a normal edge dislocation does not induce any shear stress on the fracture surfaces (see Eq.~\eqref{eq:sigma_em_2}).

Next, we consider a shear fracture (Fig.~\ref{fig:semi_inf_poroelastic}b). An arbitrary slip profile $d(x)$ can be described as a continuous distribution of infinitesimal slip edge dislocations positioned along the fracture. The strength of each dislocation is characterized by the Burgers vector component $b_1 = u_1^+ - u_1^-$, where $u_1^{\pm}$ denote the displacements along the $x-$axis on the positive (bottom) and negative (top) sides of the dislocation, respectively. Under this convention, the dislocation strength is linked with the slip profile as:
\begin{equation*}
    b_1(x) = -\diff d(x) = -\left[d(x + \diff x) - d(x)\right] = -\frac{\diff d}{\diff x}\diff x,
\end{equation*}
since, by definition, $d(x) = u_1(x, y=0^+) - u_1(x, y=0^-)$, where $u_1(x, y=0^{\pm})$ are the displacements along the $x-$axis on the top and bottom sides of a shear fracture. By the principle of superposition, the contribution of the distributed slip edge dislocations to the shear stress on the fracture surfaces can be written as:
\begin{equation}
    \sigma_{xy}^e(x) = -\int_0^{\infty} \frac{\diff d}{\diff s} \sigma_{xy1}^{em}(x-s) \diff s,
    \label{eq:shear_crack_dislocation_contribution}
\end{equation}
where the kernel $\sigma_{xy1}^{em}(x)$ (given by Eq.~\eqref{eq:sigma_em_1}) describes the shear stress produced by a unit ($b_1 = 1$) slip steadily moving edge dislocation. We should note that a slip edge dislocation does not contribute to the normal stress and the pore fluid pressure on the fracture surfaces (see Eq.~\eqref{eq:sigma_em_1}).

We also need to account for the impact of fluid exchange between the fracture and the surrounding permeable porous solid. This process is quantified by the fluid exchange rate $g(x)$ along the fracture, where a positive value corresponds to leak-off (fluid leaving the fracture) and a negative value to leak-in (fluid entering the fracture). Following \citep{cheng1998singular, Kova10, Chen16}, we introduce the fluid displacement function $v(x)$, which is related to the fluid exchange rate as:
\begin{equation*}
    v(x) = \frac{1}{V}\int_0^x g(s) \diff s  ~~~~ \to ~~~~  g(x) = V \frac{\diff v}{\diff x}.
\end{equation*} 
By representing the fracture as a continuous distribution of fluid sources and applying the principle of superposition, the contribution of the fluid sources to the normal stress and the pore fluid pressure on the fracture surfaces can be expressed as: 
\begin{equation}
    \sigma_{yy}^s(x) = V \int_0^{\infty} \frac{\diff v}{\diff s} \sigma_{yy}^{sm}(x-s) \diff s, ~~~~ p^s(x) = V \int_0^{\infty} \frac{\diff v}{\diff s} p^{sm}(x-s) \diff s,
    \label{eq:source_contribution}
\end{equation}
where the kernels $\sigma_{yy}^{sm}(x)$, $p^{sm}(x)$ (given by Eqs.~\eqref{eq:sigma_sm}, \eqref{eq:p_sm}, respectively) correspond to the normal stress and the pore fluid pressure induced by a unit ($g = 1$) steadily moving fluid source. One can note that a fluid source does not induce any shear stress on the fracture surfaces (see Eq.~\eqref{eq:sigma_sm}).

Combining Eqs.~\eqref{eq:tensile_crack_dislocation_contribution}, \eqref{eq:shear_crack_dislocation_contribution}, and \eqref{eq:source_contribution}, we obtain the boundary integral equations governing the distributions of the normal (tensile fracture) and shear (shear fracture) stresses, as well as the pore fluid pressure on the fracture surfaces:
\begin{itemize}
    \item tensile fracture:

    \begin{itemize}[label=$\circ$]
        \item normal stress:
            \begin{equation}
                \sigma_{yy}(x) = \sigma_\infty + \sigma_{yy}^e(x) + \sigma_{yy}^s(x) = \sigma_\infty - \int_0^{\infty} \frac{\diff w}{\diff s} \sigma_{yy2}^{em}(x-s) \diff s + V \int_0^{\infty} \frac{\diff v}{\diff s} \sigma_{yy}^{sm}(x-s) \diff s;
                \label{eq:tensile_crack_normal_stress}
            \end{equation}
        \item pore fluid pressure:
            \begin{equation}
                p_f(x) = p_\infty + p^e(x) + p^s(x) = p_\infty + V \int_0^{\infty} \frac{\diff v}{\diff s} p^{sm}(x-s) \diff s - \int_0^{\infty} \frac{\diff w}{\diff s} p_2^{em}(x-s) \diff s;
                \label{eq:fluid_pressure}
            \end{equation}
    \end{itemize}

    \item shear fracture:

    \begin{itemize}[label=$\circ$]
        \item shear stress:
            \begin{equation}
                \sigma_{xy}(x) = \tau_\infty + \sigma_{xy}^e(x) = \tau_\infty - \int_0^{\infty} \frac{\diff d}{\diff s} \sigma_{xy1}^{em}(x-s) \diff s;
                \label{eq:shear_crack_shear_stress}
            \end{equation}
        \item pore fluid pressure:
            \begin{equation}
                p_f(x) = p_\infty + p^s(x) = p_\infty + V \int_0^{\infty} \frac{\diff v}{\diff s} p^{sm}(x-s) \diff s.
                \label{eq:fluid_pressure_shear}
            \end{equation}
    \end{itemize}
\end{itemize}

In the system of equations presented above, we have accounted for the presence of a non-zero initial state of stress and pore pressure. In this initial configuration, in the absence of a crack, the medium is subjected to a uniform compressive stress $\sigma_\infty < 0$ acting normal to the crack plane in the case of a tensile fracture, a uniform shear stress $\tau_\infty < 0$ acting parallel to the crack plane in the case of a shear fracture, and a uniform pore fluid pressure $p_\infty$. 

Eqs.~\eqref{eq:shear_crack_shear_stress} and \eqref{eq:fluid_pressure_shear} are decoupled as a consequence of the hydraulic boundary condition adopted for the slip edge dislocation, namely the permeable slip-plane limit. Accordingly, the shear fracture formulation retains the poroelastic modification of the shear stress kernel but it does not include slip-induced pore pressure perturbations on the fracture plane. Such perturbations arise under other hydraulic conditions as previously discussed, including impermeable, leaky, partially sealed, or finite-thickness shear-zone models. Incorporating these effects into the present framework requires the use of fundamental solutions satisfying the corresponding interface boundary conditions \citep{rudnicki1987plane,song2017plane}.

On the one hand, the system of boundary integral equations allows the determination of the normal or shear stress and pore fluid pressure distributions when the dislocation density ($\diff w/\diff x$ or $\diff d / \diff x$) and the derivative of the fluid displacement function ($\diff v / \diff x$) are prescribed. On the other hand, if the stress and pore fluid pressure are specified on the fracture surfaces, the system can be solved for the dislocation density and the fluid exchange rate. In this sense, the present formulation defines a poroelastic boundary operator. To address multiphysics problems such as hydraulic fracture growth or frictional rupture propagation, the boundary integral equations must be supplemented with additional closure relations.

In the case of a hydraulic fracture, assuming permeable fracture surfaces and identical pore and fracturing fluids, the pore fluid pressure $p_f(x)$ can be identified with the pressure of the fracturing fluid within the fracture channel (in contrast to impermeable fracture surfaces, where pore fluid pressure and fracturing fluid pressure decouple). For a tensile (opening mode) fracture, the normal stress on the fracture surfaces is related to the fracturing fluid pressure inside the fracture as: $\sigma_{yy}(x) = -p_f(x)$. For a frictional shear fracture, the shear stress equals the frictional strength: $\sigma_{xy}(x) = -f\left[\sigma_{yy}(x) + p_f(x)\right]$, where $f$ denotes the friction coefficient. In addition, more general frictional models may additionally include rate-and-state effects, dilatancy, 
compaction, healing, or other shear-zone processes.


\section{Normalized form of the governing equations for steadily moving fractures}
\label{sec:dimensionless_equations}

This section presents the system of governing equations for tensile and shear fractures in dimensionless form. The normalized formulation reduces the number of governing parameters, highlighting the essential dimensionless groups. It is therefore convenient for computing general numerical solutions and systematically analyzing how the solution depends on the governing parameters.

\subsection{Steadily propagating semi-infinite tensile fracture}

The governing system for a tensile fracture (Fig.~\ref{fig:semi_inf_poroelastic}a) consists of Eqs.~\eqref{eq:tensile_crack_normal_stress} and \eqref{eq:fluid_pressure} which can be made explicit further as:
\begin{flalign}
    -(\sigma_{yy}(x) - \sigma_\infty) &= \frac{G}{2\pi (1-\nu_u)} \int_0^\infty \frac{\diff w}{\diff s}\mathcal{S}^{em}_{yy2} \left(\frac{V (x-s)}{2c}\right) \frac{\diff s}{x-s} - \frac{\eta V}{2 \pi \kappa} \int_0^{\infty} \frac{\diff v}{\diff s} \mathcal{S}^{sm}_{yy}\left(\frac{V (x-s)}{2c}\right) \diff s, \label{eq:tensile_crack_normal_stress_1} \\
    p_f(x) - p_\infty &= \frac{V}{2\pi\kappa}\int_0^{\infty} \frac{\diff v}{\diff s}  \mathcal{P}^{sm} \left(\frac{V (x-s)}{2c}\right) \diff s - \frac{\eta V}{2 \pi \kappa} \int_0^{\infty} \frac{\diff w}{\diff s} \mathcal{P}^{em}_2 \left(\frac{V (x-s)}{2c}\right) \diff s.
    \label{eq:fluid_pressure_1}
\end{flalign}
The distributions of the normal stress $\sigma_{yy}(x)$ and the pore fluid pressure $p_f(x)$ are assumed to be known, and the problem is solved for the fracture opening profile $w(x)$ and the fluid displacement function $v(x)$.

The uncoupled case, in which pore fluid flow does not induce solid deformation and, in turn, solid deformation does not generate pore pressure perturbations, corresponds to $\alpha = 0$. It is obtained from the system of Eqs.~\eqref{eq:tensile_crack_normal_stress_1} and \eqref{eq:fluid_pressure_1} by setting $\nu_u = \nu$ and $\beta = \eta = 0$.  

In the case of a fracture with impermeable surfaces, there is no fluid exchange between the fracture and the surrounding porous medium, such that $v(x) = 0$. The problem thus reduces to Eq.~\eqref{eq:tensile_crack_normal_stress_1} for the fracture opening profile, while Eq.~\eqref{eq:fluid_pressure_1} can be used to determine the pore pressure distribution on the fracture surfaces. 

The near-tip fracture opening is governed by the LEFM asymptote \citep{irwin1957analysis}:
\begin{equation}
    w(x) = 2\sqrt{\frac{2}{\pi}} \frac{K_I (1-\nu)}{G}\sqrt{x}, ~~~~ x \to 0,
    \label{eq:tip_asymptote}
\end{equation}
where $K_I$ is the mode I stress intensity factor, which is part of the solution and can be extracted from the fracture opening profile.

We introduce the dimensionless coordinates, fracture opening, fluid displacement function, stress intensity factor, net normal stress, and net pore fluid pressure:
\begin{equation}
    \xi = \frac{x}{\ell_*}, ~~~~ \zeta = \frac{y}{\ell_*}, ~~~~ \Omega = \frac{w}{w_*}, ~~~~ \Upsilon = \frac{v}{v_*}, ~~~~ \mathcal{K} = \frac{K_I}{K_*}, ~~~~ \Sigma_{yy} = -\frac{\sigma_{yy}-\sigma_\infty}{\sigma_*}, ~~~~ \Pi = \frac{p_f-p_\infty}{p_*},
    \label{eq:dimensionless_variables}
\end{equation}
where the characteristic scales are chosen as:
\begin{equation}
    \ell_* = \frac{2c}{V}, ~~~~ w_* = v_* = \frac{\sigma_* c (1-\nu)}{G V}, ~~~~ K_*= \frac{\sigma_*}{4}\sqrt{\frac{\pi c}{V}}, ~~~~ p_* = \sigma_*,
    \label{eq:characteristic_scales}
\end{equation}
and the stress (pressure) scale $\sigma_*$ is to be chosen based on the problem under consideration. Please note that the symbol $\zeta$ denotes the variation of fluid content only in Section~\ref{sec:fundamental_solutions}. In all subsequent sections, $\zeta$ refers to the dimensionless $y-$coordinate.

As a result, we obtain the normalized forms of Eqs.~\eqref{eq:tensile_crack_normal_stress_1}, \eqref{eq:fluid_pressure_1}, and \eqref{eq:tip_asymptote}:
\begin{flalign}
    & \Sigma_{yy}(\xi) = \frac{1}{4\pi (1-\beta)} \int_0^\infty \frac{\diff \Omega}{\diff s} \mathcal{S}^{em}_{yy2} \left(\xi-s\right) \frac{\diff s}{\xi-s} - \frac{\eta}{2 \pi \mathcal{S}}  \int_0^{\infty} \frac{\diff \Upsilon}{\diff s} \mathcal{S}^{sm}_{yy}\left(\xi-s\right) \diff s, \label{eq:tensile_crack_normal_stress_dimensionless} \\
    & \Pi(\xi) = \frac{1}{2\pi\mathcal{S}}\int_0^{\infty} \frac{\diff \Upsilon}{\diff s}  \mathcal{P}^{sm} \left(\xi-s\right) \diff s - \frac{\eta}{2 \pi \mathcal{S}} \int_0^{\infty} \frac{\diff \Omega}{\diff s} \mathcal{P}^{em}_2 \left(\xi-s\right) \diff s,
    \label{eq:fluid_pressure_dimensionless} \\
    & \Omega(\xi) = \mathcal{K} \sqrt{\xi}, ~~~~ \xi \to 0,
    \label{eq:tip_asymptote_dimensionless}
\end{flalign}
where $\mathcal{S}$ is a dimensionless storage coefficient \citep{Kova10}, which can be expressed in terms of $\beta$ and $\eta$ as:
\begin{equation}
    \mathcal{S} = \frac{G S}{1-\nu} = \frac{2 (1 - \beta)\eta^2}{\beta}.
    \label{eq:dimensionless_storage}
\end{equation}

\subsection{Steadily propagating semi-infinite shear fracture}

The governing system for a shear fracture (Fig.~\ref{fig:semi_inf_poroelastic}b) consists of Eqs.~\eqref{eq:shear_crack_shear_stress} and \eqref{eq:fluid_pressure_shear}:
\begin{flalign}
    -(\sigma_{xy}(x) - \tau_\infty) & = \frac{G}{2\pi(1-\nu_u)} \int_0^{\infty} \frac{\diff d}{\diff s} \mathcal{S}_{xy1}^{em} \left(\frac{V (x-s)}{2c}\right) \frac{\diff s}{x - s}. \label{eq:shear_crack_shear_stress_1} \\
    p_f(x) - p_\infty &= \frac{V}{2\pi\kappa}\int_0^{\infty} \frac{\diff v}{\diff s}  \mathcal{P}^{sm} \left(\frac{V (x-s)}{2c}\right) \diff s.
    \label{eq:fluid_pressure_shear_1}
\end{flalign}
The distributions of the shear stress $\sigma_{xy}(x)$ and the pore fluid pressure $p_f(x)$ are assumed to be known, and the problem is solved for the slip profile $d(x)$ and the fluid displacement function $v(x)$.

The uncoupled case follows from Eqs.~\eqref{eq:shear_crack_shear_stress_1} and \eqref{eq:fluid_pressure_shear_1} when $\nu_u = \nu$ and $\beta = 0$.

In the case of a fracture with impermeable surfaces, the problem reduces to Eq.~\eqref{eq:shear_crack_shear_stress_1} for the slip profile, while Eq.~\eqref{eq:fluid_pressure_shear_1} is omitted.

The near-tip slip profile takes the following form, in accordance with LEFM:
\begin{equation}
    d(x) = 2\sqrt{\frac{2}{\pi}} \frac{K_{II} (1-\nu)}{G}\sqrt{x}, ~~~~ x \to 0,
    \label{eq:tip_asymptote_shear_crack}
\end{equation}
where $K_{II}$ is the mode II stress intensity factor, which is part of the solution.

The dimensionless coordinates, slip along the fracture plane, fluid displacement function, stress intensity factor, net shear stress, and net pore fluid pressure are defined as:
\begin{equation}
    \xi = \frac{x}{\ell_*}, ~~~~ \zeta = \frac{y}{\ell_*}, ~~~~ \Delta = \frac{d}{d_*}, ~~~~ \Upsilon = \frac{v}{v_*}, ~~~~ \mathcal{K} = \frac{K_{II}}{K_*}, ~~~~ \Sigma_{xy} = -\frac{\sigma_{xy} - \tau_{\infty}}{\sigma_*}, ~~~~ \Pi = \frac{p_f-p_\infty}{p_*},
    \label{eq:dimensionless_variables_shear_crack}
\end{equation}
where the characteristic scales are similarly chosen as:
\begin{equation}
    \ell_* = \frac{2c}{V}, ~~~~ d_* = v_* = \frac{\sigma_* c (1-\nu)}{G V}, ~~~~ K_*= \frac{\sigma_*}{4}\sqrt{\frac{\pi c}{V}}, ~~~~ p_* = \sigma_*,
    \label{eq:characteristic_scales_shear_crack}
\end{equation}
and the stress (pressure) scale $\sigma_*$ is to be chosen based on the problem under consideration. 

Consequently, we obtain the normalized forms of Eqs.~\eqref{eq:shear_crack_shear_stress_1}, \eqref{eq:fluid_pressure_shear_1}, and \eqref{eq:tip_asymptote_shear_crack}:
\begin{flalign}
    & \Sigma_{xy}(\xi) = \frac{1}{4\pi(1-\beta)}\int_0^{\infty} \frac{\diff \Delta}{\diff s} \mathcal{S}_{xy1}^{em} \left(\xi - s\right) \frac{\diff s}{\xi - s}, 
    \label{eq:shear_crack_shear_stress_dimensionless} \\
    & \Pi(\xi) = \frac{1}{2\pi\mathcal{S}}\int_0^{\infty} \frac{\diff \Upsilon}{\diff s}  \mathcal{P}^{sm} \left(\xi-s\right) \diff s,
    \label{eq:fluid_pressure_shear_dimensionless} \\
    & \Delta(\xi) = \mathcal{K}\sqrt{\xi}, ~~~~ \xi \to 0,
    \label{eq:tip_asymptote_shear_crack_dimensionless}
\end{flalign}

\section{Numerical discretization}
\label{sec:numerical_method}

We outline the numerical approach used to solve the system of boundary integral equations, focusing on the tensile fracture case. The same discretization strategy can be applied to shear fractures. This numerical scheme 
draws inspiration from the methodology first presented in \citep{GaDe00, GaDe11}, where known near-field and far-field asymptotic behavior is embedded in the numerical approximation.

We develop a collocation method for the system of boundary integral equations (Eqs.~\eqref{eq:tensile_crack_normal_stress_dimensionless} and \eqref{eq:fluid_pressure_dimensionless}). The dislocation density $\omega(\xi) = \diff \Omega/\diff \xi$ and the derivative of the fluid displacement function $\upsilon(\xi) = \diff \Upsilon/\diff \xi$ are treated as nodal unknowns, while the integral equations are enforced at the midpoints of the computational grid. The unknown fields are interpolated over each spatial segment as a superposition of two power-law functions consistent with their near-field ($\xi \to 0$) and far-field ($\xi \to \infty$) behavior. Finally, the algorithm computes the dimensionless profiles of the fracture opening $\Omega(\xi)$ and the fluid displacement function $\Upsilon(\xi)$.

We define a computational domain $\xi \in [\xi_1, \xi_N]$ and discretize it into $N-1$ subintervals $[\xi_i, \xi_{i+1}]$, $i = 1, \ldots, N-1$, using a logarithmic grid. The functions $\omega(\xi)$ and $\upsilon(\xi)$ are evaluated at the mesh nodes and assembled into the vectors of unknowns $\boldsymbol{\omega} = \{\omega_i\}_{i=1}^N$ and $\boldsymbol{\upsilon} = \{\upsilon_i\}_{i=1}^N$, where $\omega_i = \omega(\xi_i)$ and $\upsilon_i = \upsilon(\xi_i)$, $i = 1, \ldots, N$.

The variation of each unknown function $f(\xi) \in \left\{\omega(\xi), \upsilon(\xi)\right\}$, both within and beyond the computational domain, is approximated as:
\begin{equation}
    f(\xi) = 
    \begin{cases}
        f_0^*(\xi), & \xi \leq \xi_1, \\
        A_i f_0(\xi) + B_i f_\infty(\xi), & \xi \in [\xi_i, \xi_{i+1}], ~ i = 1, \ldots, N-1, \\
        f_\infty^*(\xi), & \xi \geq \xi_N.
    \end{cases}
    \label{eq:omega_v_approximation}
\end{equation}
The functions $f_0^*(\xi)$ and $f_\infty^*(\xi)$ represent the near-field $f(\xi \to 0)$ and far-field $f(\xi \to \infty)$ asymptotic behavior of $f(\xi)$, respectively. The shape functions $f_0(\xi) = \xi^a$ and $f_\infty(\xi) = \xi^b$ are defined in such a way that they reproduce the near-field and far-field power-law behavior of $f(\xi)$, i.e., $f_0(\xi) \sim f_0^*(\xi)$ and $f_\infty(\xi) \sim f_\infty^*(\xi)$. The near-field and far-field asymptotes must be determined for the specific problem under consideration.

By imposing continuity of the approximation given in Eq.~\eqref{eq:omega_v_approximation} at the internal nodes $\xi_i$, $i = 2, \ldots, N - 1$, the coefficients $A_i$ and $B_i$ are expressed in terms of the unknown nodal values of the field, $f_i = f(\xi_i)$, at the endpoints of the $i-$th subinterval:
\begin{equation}
    A_i=\frac{f_\infty\left(\xi_i\right)f_{i+1}-f_\infty\left(\xi_{i+1}\right)f_i}{f_\infty\left(\xi_i\right)f_0(\xi_{i+1})-f_\infty\left(\xi_{i+1}\right)f_0(\xi_i)}, ~~~~ B_i=-\frac{f_0\left(\xi_i\right)f_{i+1}-f_0\left(\xi_{i+1}\right)f_i}{f_\infty\left(\xi_i\right)f_0\left(\xi_{i+1}\right)-f_\infty\left(\xi_{i+1}\right)f_0\left(\xi_i\right)}.
    \label{eq:coefficients_approximation}
\end{equation}

The field $f(\xi)$ continuity condition at the endpoints of the computational domain, $\xi_1$ and $\xi_N$, requires: $f_1 = f_0^*(\xi_1), ~~ f_N = f_\infty^*(\xi_N)$.

We numerically evaluate each integral in the system of Eqs.~\eqref{eq:tensile_crack_normal_stress_dimensionless} and \eqref{eq:fluid_pressure_dimensionless} at the midpoints of the computational mesh, $\xi_{i+1/2} = (\xi_i + \xi_{i+1})/2$, $i = 1, \ldots, N-1$, as follows:
\begin{flalign}
    & \mathcal{J}(\xi_{i+1/2}) = \int_0^\infty f(s) \mathcal{F}(\xi_{i+1/2} - s) \diff s = \underbrace{\int_0^{\xi_1}f_0^*(s) \mathcal{F}(\xi_{i+1/2} - s) \diff s}_{\mathcal{J}_1} + \nonumber \\
    & + \sum_{i = 1}^{N-1}\left[A_i \underbrace{\int_{\xi_i}^{\xi_{i+1}} f_0(s) \mathcal{F}(\xi_{i+1/2} - s) \diff s}_{\mathcal{J}_2} + B_i \underbrace{\int_{\xi_i}^{\xi_{i+1}} f_\infty(s) \mathcal{F}(\xi_{i+1/2} - s) \diff s}_{\mathcal{J}_3}\right] + \underbrace{\int_{\xi_N}^{\infty}f_\infty^*(s) \mathcal{F}(\xi_{i+1/2} - s) \diff s}_{\mathcal{J}_4},
    \label{eq:calculation_integral}
\end{flalign}
where the integrals $\mathcal{J}_1$  to $\mathcal{J}_4$ are precomputed numerically prior to solving the problem.

Based on Eqs.~\eqref{eq:coefficients_approximation}, \eqref{eq:calculation_integral}, we obtain the linear system of equations:
\begin{equation}
    \begin{pmatrix}
    \boldsymbol{\Sigma}_{yy} \\ \boldsymbol{\Pi}
    \end{pmatrix}
    = \begin{pmatrix}
    \frac{1}{4\pi (1-\beta)} \mathbb{S}^e & - \frac{\eta}{2 \pi \mathcal{S}}\mathbb{S}^s \\
     - \frac{\eta}{2 \pi \mathcal{S}}\mathbb{P}^e &  \frac{1}{2\pi\mathcal{S}} \mathbb{P}^s
    \end{pmatrix}
    \begin{pmatrix}
    \boldsymbol{\omega} \\ \boldsymbol{\upsilon}
    \end{pmatrix} + \mathbf{L},
    \label{eq:linear_system}
\end{equation}      
where $\boldsymbol{\Sigma}_{yy} = \left\{(\Sigma_{yy})_{i+1/2}\right\}_{i=1}^{N-1}$, $\boldsymbol{\Pi} = \left\{\Pi_{i + 1/2}\right\}_{i=1}^{N-1}$ are vectors of the dimensionless net normal stress $(\Sigma_{yy})_{i+1/2} = \Sigma_{yy}(\xi_{i+1/2})$ and net pore fluid pressure $\Pi_{i+1/2} = \Pi(\xi_{i+1/2})$ at the midpoints of the computational mesh. In turn, $\mathbb{S}^e$, $\mathbb{S}^s$, $\mathbb{P}^e$, and $\mathbb{P}^s$ denote matrices constructed from the integrals $\mathcal{J}_2$ and $\mathcal{J}_3$ (Eq.~\eqref{eq:calculation_integral}) combined with the coefficients given by Eq.~\eqref{eq:coefficients_approximation}. The function $\mathcal{F}(\xi)$ corresponds, respectively, to $\mathcal{S}^{em}_{yy2}(\xi) / \xi$, $\mathcal{S}^{sm}_{yy}(\xi)$, $\mathcal{P}^{em}_2(\xi)$, and $\mathcal{P}^{sm}(\xi)$ for $\mathbb{S}^e$, $\mathbb{S}^s$, $\mathbb{P}^e$, and $\mathbb{P}^s$. Finally, the vector $\mathbf{L}$ contains the constant values resulting from integrals $\mathcal{J}_1$ and $\mathcal{J}_4$ (Eq.~\eqref{eq:calculation_integral}).

We assume that the far-field asymptotes $\omega_\infty^*(\xi)$ and $\upsilon_\infty^*(\xi)$ are known, which prescribes the values of $\omega_N$ and $\upsilon_N$, and that each near-field asymptote, $\omega_0^*(\xi)$ and $\upsilon_0^*(\xi)$, contains a single unknown coefficient that depends linearly on the corresponding value at the first node, i.e., $\omega_1$ or $\upsilon_1$ (Eq.~\eqref{eq:tip_asymptote_dimensionless}). In this case, the contributions of the integral $\mathcal{J}_1$ (Eq.~\eqref{eq:calculation_integral}) are incorporated into the blocks of the system matrix (Eq.~\eqref{eq:linear_system}). Consequently, the resulting system consists of $2N - 2$ equations for $2N - 2$ unknowns, $\omega_i$ and $\upsilon_i$, with $i = 1, \ldots, N - 1$, which can be solved using a standard direct solver.

As a final step, we analytically integrate $\omega(\xi)$ and $\upsilon(\xi)$ using their representations (Eq.~\eqref{eq:omega_v_approximation}), once all coefficients have been obtained from the solution of the linear system (Eq.~\eqref{eq:linear_system}). This yields the fracture opening and the fluid displacement function at the computational nodes, $\Omega_i = \Omega(\xi_i)$ and $\Upsilon_i = \Upsilon(\xi_i)$, $i = 1, \ldots, N$.

The accuracy of the proposed numerical method depends on the appropriate choice of the bounds of the computational domain, $\xi_1$ and $\xi_N$, beyond which the near-field ($\xi \leq \xi_1$) and far-field ($\xi \geq \xi_N$) asymptotic solutions are imposed. These bounds must be selected such that the asymptotes are well established in the corresponding domains. For each problem considered, the relevant near-tip and far-field asymptotes must be identified, and the computational interval must be checked for convergence with respect to the domain bounds, $\xi_1$ and $\xi_N$, and the number of collocation points. In the benchmark problems considered in Section~\ref{sec:verification_tests}, these choices are verified through convergence tests and by comparison with available analytical or semi-analytical solutions.

\section{Verification tests}
\label{sec:verification_tests}

We verify the boundary integral equations formulated for a steadily moving semi-infinite crack, which govern the tractions and the pore fluid pressure on the fracture surfaces (Section~\ref{sec:boundary_integral_equations}). These relations are applied to several benchmark problems. The numerical results show excellent agreement with analytical and semi-analytical solutions from the literature, thereby demonstrating the correctness of the derivations and the accuracy of the numerical implementation for the considered benchmark problems.


\subsection{Semi-infinite tensile fracture under exponential loading}

As a first test, we consider the problem studied by \cite{AtCr91}. It involves a steadily moving semi-infinite tensile crack (Fig.~\ref{fig:semi_inf_poroelastic}a) subjected to a prescribed normal loading. The poroelastic medium is initially stress-free ($\sigma_{\infty} = 0$) and has zero pore fluid pressure ($p_{\infty} = 0$). The normal traction applied on the fracture surfaces has an exponential spatial dependence, $\sigma_{yy}(x) = -\sigma_0 e^{-x/a}$, where $a$ is the loading distance. Accordingly, the characteristic stress (pressure) scale in this problem is $\sigma_* = \sigma_0$ (see Eqs.~\eqref{eq:dimensionless_variables} and \eqref{eq:characteristic_scales}). The fracture surfaces can be either impermeable or permeable. In the impermeable case, no fluid exchange occurs, $v(x) = 0$, and only Eq.~\eqref{eq:tensile_crack_normal_stress_1} should be solved for the fracture opening profile $w(x)$. In the permeable case, the pore fluid pressure on the fracture surfaces is prescribed as $p_f(x) = 0$, and the problem is governed by Eqs.~\eqref{eq:tensile_crack_normal_stress_1} and \eqref{eq:fluid_pressure_1}, which determine the opening $w(x)$ and the fluid displacement function $v(x)$. 

\cite{AtCr91} obtained analytical solutions for both the impermeable and permeable cases by applying a spatial Fourier transform and using the Wiener-Hopf technique to solve the resulting functional equations. Their approach is restricted to specific forms of traction and pore pressure on the fracture surfaces, for which the functional equations admit closed-form solutions. In contrast, the formulation presented in the current work (Section~\ref{sec:boundary_integral_equations}) expresses the coupled poroelastic tensile crack problem directly in the physical domain via boundary integral equations, providing a numerical framework for prescribed distributions of traction and pore fluid pressure on the fracture surfaces, while recovering the analytical solutions of \cite{AtCr91} as a special case corresponding to their specific exponential loading.

\subsubsection{Impermeable fracture surfaces}
\label{sec:atkinson_craster_impermeable}

For the case of impermeable fracture surfaces, the absence of fluid exchange between the fracture and the surrounding porous medium allows the crack opening profile $w(x)$ to be determined by solving Eq.~\eqref{eq:tensile_crack_normal_stress_1}. The dimensionless form of this equation is given in Eq.~\eqref{eq:tensile_crack_normal_stress_dimensionless}, which, for the present problem, reduces to:
\begin{equation}
    e^{-\xi/\mathcal{A}} = \frac{1}{4\pi (1-\beta)}\int_0^\infty \frac{\diff \Omega}{\diff s}\mathcal{S}^{em}_{yy2} \left(\xi - s\right) \frac{\diff s}{\xi-s},
    \label{eq:normal_stress_atkinson_impermeable_dimensionless}
\end{equation}
where $\mathcal{A} = a V / (2c)$ is the dimensionless loading distance. The opening profile $\Omega(\xi)$ is governed by two parameters: the normalized undrained-drained Poisson's ratio contrast $\beta$ (Eq.~\eqref{eq:poissons_contrast}) and the dimensionless loading distance $\mathcal{A}$.

First, we consider the uncoupled case, corresponding to $\beta = 0$. In this limit, Eq.~\eqref{eq:normal_stress_atkinson_impermeable_dimensionless} is given by:
\begin{equation*}
    e^{-\xi/\mathcal{A}} = \frac{1}{4\pi}\int_0^\infty \frac{\diff \Omega}{\diff s} \frac{\diff s}{\xi-s}.
\end{equation*}

Using Eq.~(A6) from \citep{GaDe00}, we obtain the expression for $\diff \Omega / \diff \xi$. Integrating the opening gradient gives the dimensionless elastic fracture opening profile:
\begin{equation}
    \Omega_e(\xi) = \frac{8\mathcal{A}}{\sqrt{\pi}} D\left(\sqrt{\frac{\xi}{\mathcal{A}}}\right),
    \label{eq:opening_atkinson_impermeable_uncoupled_dimensionless}
\end{equation}
where $D(x)$ denotes the Dawson's integral \citep{abramowitz1948handbook}.

Applying a Taylor expansion, we extract the leading-order term of the opening profile near the fracture tip, which takes the form of Eq.~\eqref{eq:tip_asymptote_dimensionless} with the specific value of the dimensionless stress intensity factor: 
\begin{equation}
    \mathcal{K}_e = 8\sqrt{\frac{\mathcal{A}}{\pi}}.
    \label{eq:elastic_K_I}
\end{equation}
This quantity is referred to as the dimensionless elastic stress intensity factor, and its expression coincides with the result reported by \cite{AtCr91}, their Eq.~(88).

From Eq.~\eqref{eq:opening_atkinson_impermeable_uncoupled_dimensionless}, we can also determine the far-field asymptotic behavior of the fracture opening. Expanding the opening profile for $\xi \to \infty$ using a Taylor series, we find:
\begin{equation}
    \Omega_e(\xi) = \frac{4\mathcal{A}^{3/2}}{\sqrt{\pi \xi}}, ~~~~ \xi\to\infty.
    \label{eq:far_field_opening_atkinson_impermeable_uncoupled_dimensionless}
\end{equation}

Next, we proceed to the general numerical solution of the problem, which is obtained using the algorithm presented in Section~\ref{sec:numerical_method}. The computational domain is defined by the endpoints $\xi_1 = 10^{-8}$ and $\xi_N = 10^{10}$, and it is divided into $299$ segments distributed logarithmically, i.e., $N = 300$. 

In the near-field, the LEFM asymptote given by Eq.~\eqref{eq:tip_asymptote_dimensionless} governs the fracture opening, giving $\omega_0^*(\xi) = \mathcal{K}\xi^{-1/2}/2$, where $\mathcal{K}$ is the (unknown) stress intensity factor that forms part of the solution, and $\omega_0(\xi) = \xi^{-1/2}$.

In the far field, we assume that the crack aperture follows the same power-law behavior as in the uncoupled case, see Eq.~\eqref{eq:far_field_opening_atkinson_impermeable_uncoupled_dimensionless}, providing $\omega_\infty^*(\xi) \sim \xi^{-3/2}$ and $\omega_\infty(\xi) = \xi^{-3/2}$. The far-field asymptote, together with the very large value of the upper bound of the computational domain, $\xi_N$, results in a negligible contribution from the integral $\mathcal{J}_4$ in Eq.~\eqref{eq:calculation_integral}. Consequently, we set $\omega_\infty^*(\xi) = 0$ and impose $\omega_N = 0$.

For impermeable fracture surfaces, the problem reduces to a linear system of $N - 1$ equations for $N - 1$ unknowns, namely $\boldsymbol{\omega} = \{\omega_i\}_{i=1}^{N-1}$:
\begin{equation*}
    \boldsymbol{\Sigma}_{yy} = \frac{1}{4\pi(1-\beta)} \mathbb{S}^e \boldsymbol{\omega}. 
\end{equation*}

\begin{figure}[]
    \centering
    \includegraphics[width=1.0\textwidth]{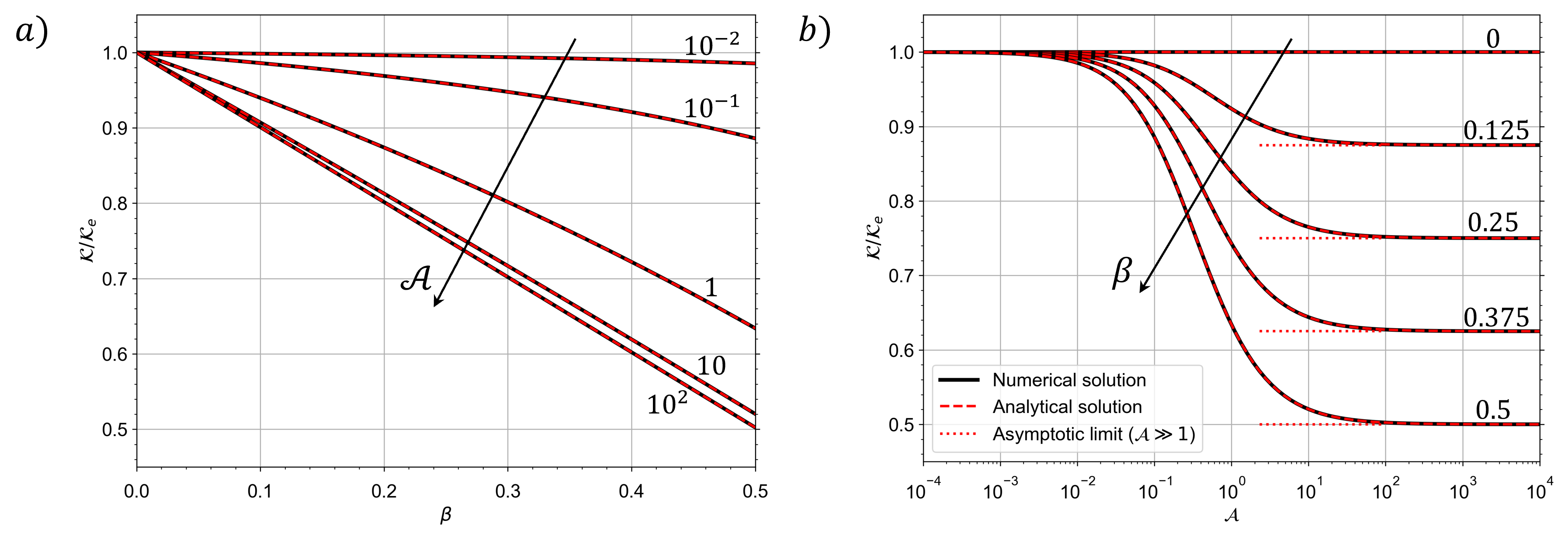}
    \caption{
    Dimensionless stress intensity factor for a semi-infinite tensile fracture with impermeable surfaces, loaded by an exponential normal stress, normalized by the dimensionless elastic stress intensity factor (Eq.~\eqref{eq:elastic_K_I}), as a function of (a) the normalized undrained-drained Poisson's ratio contrast $\beta$ and (b) the dimensionless loading distance $\mathcal{A}$. The numerical solution is shown by black lines, while the analytical solution (Eq.~\eqref{eq:K_I_imperm}) is represented by dashed red lines. In panel (b), the asymptotic limit $\mathcal{K}/\mathcal{K}_e \to 1-\beta$ for $\mathcal{A} \gg 1$ is indicated by dotted red lines. In panel (a), results are shown for $\mathcal{A} = \{10^{-2}, ~10^{-1}, ~1, ~10, ~10^2\}$, while in panel (b), results correspond to $\beta = \{0, ~0.125, ~0.25, ~0.375, ~0.5\}$.
    }
    \label{fig:atkinson_imperm_K_I}
\end{figure}

Fig.~\ref{fig:atkinson_imperm_K_I} shows the dependence of the normalized dimensionless stress intensity factor, $\mathcal{K}/\mathcal{K}_e$, on (a) the normalized undrained-drained Poisson's ratio contrast $\beta$ and (b) the dimensionless loading distance $\mathcal{A}$. The numerical solution is compared with the analytical solution reported by \cite{AtCr91}, their Eq.~(177), which, expressed in terms of the present scaling (Eqs.~\eqref{eq:dimensionless_variables} and \eqref{eq:characteristic_scales}), takes the following form:
\begin{equation}
    \mathcal{K} = \frac{8 \mathcal{A}^{3/2}(1-\beta) \sqrt{1+2\mathcal{A}}}{\sqrt{\pi} \left(\beta + (\mathcal{A}-\beta) \sqrt{1 + 2 \mathcal{A}} \right)},
    \label{eq:K_I_imperm}
\end{equation}
which reduces to Eq.~\eqref{eq:elastic_K_I}, when $\beta = 0$.

From Fig.~\ref{fig:atkinson_imperm_K_I}, the following trends can be observed: $\mathcal{K}/\mathcal{K}_e \to 1$ for $\mathcal{A} \ll 1$ and $\mathcal{K}/\mathcal{K}_e \to 1 - \beta$ for $\mathcal{A} \gg 1$.

\begin{figure}[]
    \centering
    \includegraphics[width=1.0\textwidth]{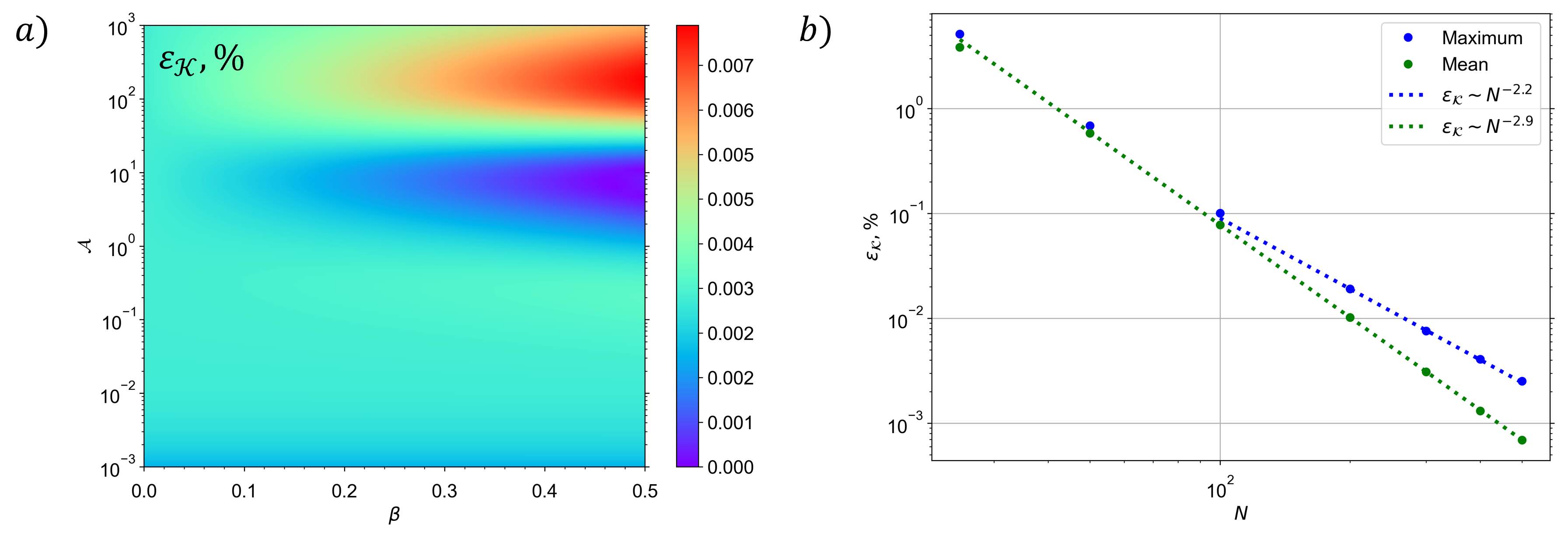}
    \caption{
    Relative difference, $\varepsilon_{\mathcal{K}}$ (in \%), between numerical and analytical (Eq.~\eqref{eq:K_I_imperm}) solutions for the dimensionless stress intensity factor $\mathcal{K}$ of a semi-infinite tensile fracture with impermeable surfaces subjected to exponential normal loading. (a) Relative difference as a function of the parameters $\beta$ and $\mathcal{A}$. (b) Convergence analysis showing the dependence of the maximum (blue circles) and mean (green circles) relative differences, evaluated over the selected ranges of $\beta$ and $\mathcal{A}$, on the number of grid nodes $N$ along the computational domain. Power-law fits are shown by blue and green dotted lines for the maximum and mean relative differences, respectively.
    }
    \label{fig:atkinson_imperm_K_I_error}
\end{figure}

The relative difference between the two solutions, $\varepsilon_{\mathcal{K}}$, as a function of the parameters $\beta$ and $\mathcal{A}$, is presented in Fig.~\ref{fig:atkinson_imperm_K_I_error}a, and does not exceed 0.008 \%. In addition, Fig.~\ref{fig:atkinson_imperm_K_I_error}b presents the results of the convergence analysis, illustrating how the maximum and mean relative differences between the numerical and analytical solutions, evaluated over selected ranges of $\beta$ and $\mathcal{A}$, depend on the number of nodes $N$ along the computational domain $\xi \in [\xi_1,\xi_N]$. Power-law fits indicate an approximately quadratic convergence rate for the maximum relative difference and a cubic convergence rate for the mean relative difference, demonstrating good numerical accuracy and consistency of the proposed discretization.

\begin{figure}[]
    \centering
    \includegraphics[width=1.0\textwidth]{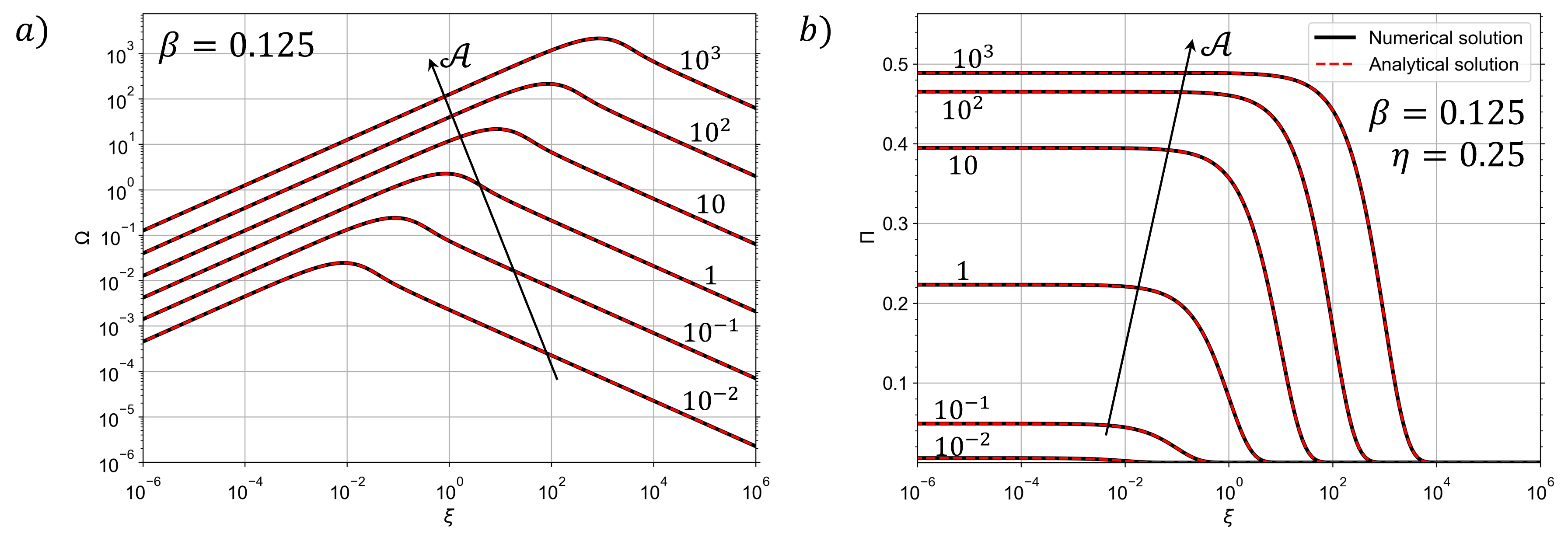}
    \caption{
    Dimensionless (a) opening and (b) pore fluid pressure profiles along a semi-infinite tensile fracture with impermeable surfaces, loaded by an exponential normal stress, shown for $\mathcal{A} = \{10^{-2}, ~10^{-1}, ~1, ~10, ~10^2, ~10^3\}$ with $\beta = 0.125$ and $\eta = 0.25$. The numerical results are shown by black lines, while the analytical solutions (Eq.~\eqref{eq:opening_atkinson_impermeable_dimensionless} with $\mathcal{K}$ given by Eq.~\eqref{eq:K_I_imperm} for the opening and Eq.~\eqref{eq:pressure_profile_impermeable_dimensionless} for the pressure) are indicated by dashed red lines.
    }
    \label{fig:atkinson_imperm_w_p}
\end{figure}

Fig.~\ref{fig:atkinson_imperm_w_p}a shows the dimensionless opening profiles $\Omega(\xi)$ for different values of the dimensionless loading distance $\mathcal{A}$ for $\beta = 0.125$. The analytical expression was reported by \cite{AtCr91}, their Eq.~(163), for the case of permeable fracture surfaces only. Following the structure of that solution, we derive an expression for the opening profile corresponding to impermeable fracture surfaces:
\begin{equation}
    \Omega(\xi) = \mathcal{K} \sqrt{\mathcal{A}} D\left(\sqrt{\xi/\mathcal{A}}\right),
    \label{eq:opening_atkinson_impermeable_dimensionless}
\end{equation}
with $\mathcal{K}$ given by Eq.~\eqref{eq:K_I_imperm}, which reduces to Eq.~\eqref{eq:opening_atkinson_impermeable_uncoupled_dimensionless} when $\beta = 0$.

The opening profiles shown in Fig.~\ref{fig:atkinson_imperm_w_p}a can be substituted into Eq.~\eqref{eq:fluid_pressure_dimensionless}, which, for the impermeable fracture surfaces, takes the form:
\begin{equation}
    \Pi(\xi) = - \frac{\eta}{2 \pi \mathcal{S}} \int_0^{\infty} \frac{\diff \Omega}{\diff s} \mathcal{P}^{em}_2 \left(\xi-s\right) \diff s ~~ \to ~~ \mathbf{\Pi} = - \frac{\eta}{2 \pi \mathcal{S}}\mathbb{P}^e \boldsymbol{\omega},
    \label{eq:pressure_profile_impermeable_dimensionless}
\end{equation}
to compute the pore fluid pressure distribution on the fracture surfaces. In the equation above, the dimensionless storage coefficient $\mathcal{S}$ is related to $\beta$ and $\eta$ through Eq.~\eqref{eq:dimensionless_storage}.

Fig.~\ref{fig:atkinson_imperm_w_p}b compares the numerical solution with the analytical solution reported by \cite{AtCr91} (the unnumbered equation between their Eqs.~(180) and (181)), which, when expressed in terms of the scaling given by Eq.~\eqref{eq:characteristic_scales}, takes the following form:
\begin{equation}
    \Pi(\xi) = \frac{\mathcal{A} \left(\sqrt{1+2 \mathcal{A}}-1\right) \beta }{\eta  \left(\beta  + (\mathcal{A}-\beta )\sqrt{1+2 \mathcal{A}}\right)} e^{-\xi/\mathcal{A}}.
    \label{eq:pressure_atkinson_impermeable_dimensionless}
\end{equation}

Next, we compute the pore pressure field around the fracture. Since Eq.~\eqref{eq:pressure_profile_impermeable_dimensionless} provides the pore pressure distribution only along $\zeta = 0$, we employ the general expression for the pore pressure field induced by a unit normal steadily moving edge dislocation (see Appendix~\ref{sec:derivation_edge_dislocation_pressure}) and apply the principle of spatial superposition (see Section~\ref{sec:boundary_integral_equations}):
\begin{equation}
    \Pi(\xi, \zeta) = -\frac{\eta}{2 \pi \mathcal{S}} \int_0^{\infty} \frac{\diff \Omega}{\diff s} \left(-\frac{\xi-s}{r^2} + e^{\xi - s} \left[\frac{\xi-s}{r} \mathrm{K}_1\left(r\right) - \mathrm{K}_0\left(r\right)\right] \right) \diff s,
    \label{eq:pressure_distribution_around_impermeable_dimensionless}
\end{equation}
where $r = \sqrt{(\xi-s)^2 + \zeta^2}$.

\begin{figure}[]
    \centering
    \includegraphics[width=1.0\textwidth]{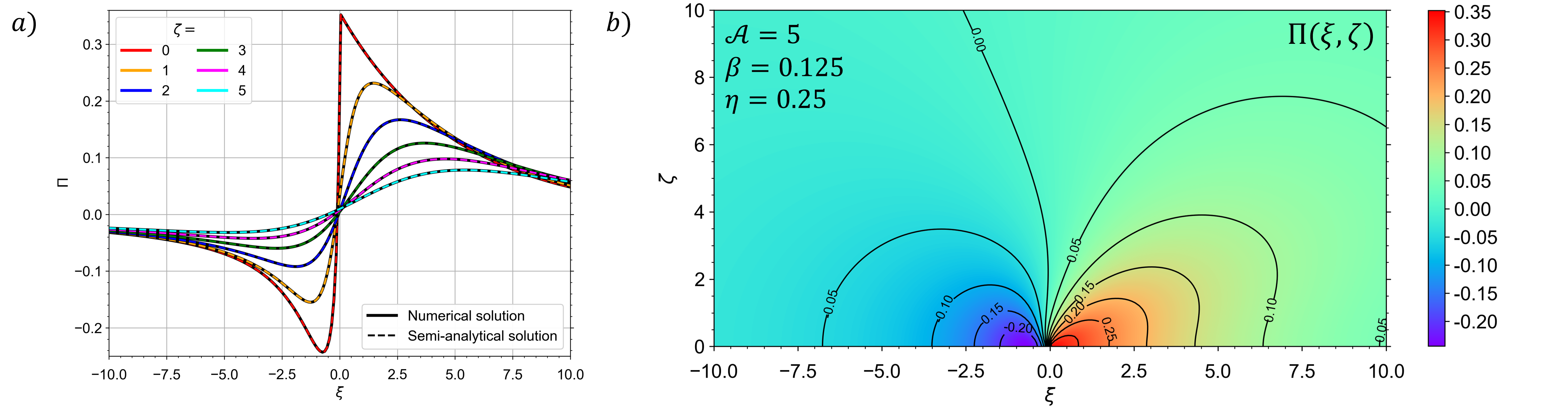}
    \caption{
    Dimensionless pore fluid pressure around a semi-infinite tensile fracture with impermeable surfaces, loaded by an exponential normal stress. (a) Profiles along lines parallel to the fracture at $\zeta = \{0, 1, 2, 3, 4, 5\}$ for $\mathcal{A} = 5$, $\beta = 0.125$, and $\eta = 0.25$. The numerical results obtained from Eq.~\eqref{eq:pressure_distribution_around_impermeable_dimensionless} are plotted as solid black lines, whereas the reference semi-analytical solution from Eq.~\eqref{eq:pressure_atkinson_impermeable_dimensionless_field} is shown as dashed colored lines. (b) Pore pressure field around the fracture tip for $\zeta > 0$, computed from Eq.~\eqref{eq:pressure_distribution_around_impermeable_dimensionless}. Isolines corresponding to $\Pi(\xi, \zeta) = \mathrm{const}$ are shown as solid black lines.
    }
    \label{fig:atkinson_imperm_pressure}
\end{figure}

Fig.~\ref{fig:atkinson_imperm_pressure}a compares the pore pressure distributions obtained using Eq.~\eqref{eq:pressure_distribution_around_impermeable_dimensionless} and the expression reported by \cite{AtCr91} (the unnumbered equation between their Eqs.~(180) and (181)). The results are shown along lines $\zeta = \mathrm{const}$ parallel to the fracture plane $\zeta = 0$ for different values of $\zeta$, for $\mathcal{A} = 5$, $\beta = 0.125$, and $\eta = 0.25$. In terms of our scaling (Eqs.~\eqref{eq:dimensionless_variables} and \eqref{eq:characteristic_scales}), the expression from \cite{AtCr91} can be written as follows:
\begin{flalign}
    \Pi(\xi, \zeta) = \frac{\mathcal{A}^{3/2} \beta \sqrt{1 + 2 \mathcal{A}} }{\sqrt{2\pi} \eta \left(\beta + (\mathcal{A}-\beta)\sqrt{1 + 2 \mathcal{A}} \right)} & \bigg[\frac{\sqrt{r(\xi) - \xi}}{r(\xi)} \left(e^{\xi-r(\xi)} - 1\right) - \nonumber \\
    & - \frac{e^{-\xi/\mathcal{A}}}{\mathcal{A}} \int_{-\infty}^\xi e^{s/\mathcal{A}}\frac{\sqrt{r(s) - s}}{r(s)} \left(e^{s-r(s)} - 1\right) \diff s\bigg],
    \label{eq:pressure_atkinson_impermeable_dimensionless_field}
\end{flalign}
where $r(\xi) = \sqrt{\xi^2 + \zeta^2}$. We note that the integral in the equation above can be evaluated analytically for $\zeta = 0$, yielding Eq.~\eqref{eq:pressure_atkinson_impermeable_dimensionless} when $\xi > 0$.

For a visual representation of the pressure distribution around the fracture tip, we also compute $\Pi(\xi, \zeta)$ for $\zeta > 0$, using Eq.~\eqref{eq:pressure_distribution_around_impermeable_dimensionless}, and present it in Fig.~\ref{fig:atkinson_imperm_pressure}b.

\subsubsection{Permeable fracture surfaces with prescribed zero pore fluid pressure}
\label{sec:atkinson_craster_permeable}

For the case of permeable fracture surfaces with prescribed pore fluid pressure $p_f(x) = 0$, the problem is formulated in terms of the crack opening profile $w(x)$ and the fluid displacement function $v(x)$. The governing system is given by Eqs.~\eqref{eq:tensile_crack_normal_stress_1} and \eqref{eq:fluid_pressure_1}. In dimensionless form, the system is expressed by Eqs.~\eqref{eq:tensile_crack_normal_stress_dimensionless} and \eqref{eq:fluid_pressure_dimensionless}, which, for the present problem, reduce to:
\begin{flalign}
    & e^{-\xi/\mathcal{A}} = \frac{1}{4\pi (1-\beta)}\int_0^\infty \frac{\diff \Omega}{\diff s}\mathcal{S}^{em}_{yy2} \left(\xi - s\right) \frac{\diff s}{\xi-s} - \frac{\eta}{2 \pi \mathcal{S}} \int_0^{\infty} \frac{\diff \Upsilon}{\diff s} \mathcal{S}^{sm}_{yy}(\xi-s) \diff s,
    \label{eq:normal_stress_atkinson_permeable_dimensionless} \\
    & 0 = \int_0^{\infty} \frac{\diff \Upsilon}{\diff s}  \mathcal{P}^{sm} (\xi-s) \diff s - \eta \int_0^{\infty} \frac{\diff \Omega}{\diff s} \mathcal{P}^{em}_2  (\xi-s) \diff s \label{eq:pressure_atkinson_permeable_dimensionless} .
\end{flalign}

The opening $\Omega(\xi)$ and fluid displacement function $\Upsilon(\xi)$ profiles are governed by three parameters: the normalized undrained-drained Poisson's ratio contrast $\beta$ (Eq.~\eqref{eq:poissons_contrast}), the dimensionless loading distance $\mathcal{A}$, and the poroelastic stress coefficient $\eta$ (Eq.~\eqref{eq:poroelastic_stress_coefficient}). The dimensionless storage coefficient $\mathcal{S}$ is related to $\beta$ and $\eta$ through Eq.~\eqref{eq:dimensionless_storage}.

The general numerical solution of the problem is obtained using the algorithm described in Section~\ref{sec:numerical_method}. The computational grid and all elements of the numerical scheme associated with the opening gradient $\omega(\xi)$ remain the same as in the case of impermeable fracture surfaces (Section~\ref{sec:atkinson_craster_impermeable}), due to the LEFM near-tip asymptote (Eq.~\eqref{eq:tip_asymptote_dimensionless}) and the shape of the far-field asymptote given by Eq.~\eqref{eq:far_field_opening_atkinson_impermeable_uncoupled_dimensionless}. In the following, we provide clarifications on how fluid exchange is incorporated.

We assume that the fluid displacement function $\Upsilon(\xi)$ exhibits the same asymptotic behavior as the fracture opening $\Omega(\xi)$ in both the near-field and far-field. Based on this assumption, we define the algorithmic elements associated with the gradient of the fluid displacement function $\upsilon(\xi)$ as: $\upsilon_0^*(\xi) = \mathcal{K}_Y\xi^{-1/2}/2$, $\upsilon_0(\xi) = \xi^{-1/2}$, $\upsilon_\infty^*(\xi) = 0$, $\upsilon_\infty(\xi) = \xi^{-3/2}$. As in Section~\ref{sec:atkinson_craster_impermeable}, the contribution of the integral $\mathcal{J}_4$ in Eq.~\eqref{eq:calculation_integral} can be neglected, allowing us to impose a vanishing far-field asymptote, $\upsilon_N = 0$. The coefficient $\mathcal{K}_\Upsilon$ is unknown and treated as part of the solution. 

Consequently, for permeable fracture surfaces, we solve a linear system of $2N - 2$ equations for $2N - 2$ unknowns, namely $\boldsymbol{\omega} = \{\omega_i\}_{i=1}^{N-1}$ and $\boldsymbol{\upsilon} = \{\upsilon_i\}_{i=1}^{N-1}$, using Eq.~\eqref{eq:linear_system}, with $\mathbf{L} = 0$.

\begin{figure}[]
    \centering
    \includegraphics[width=1.0\textwidth]{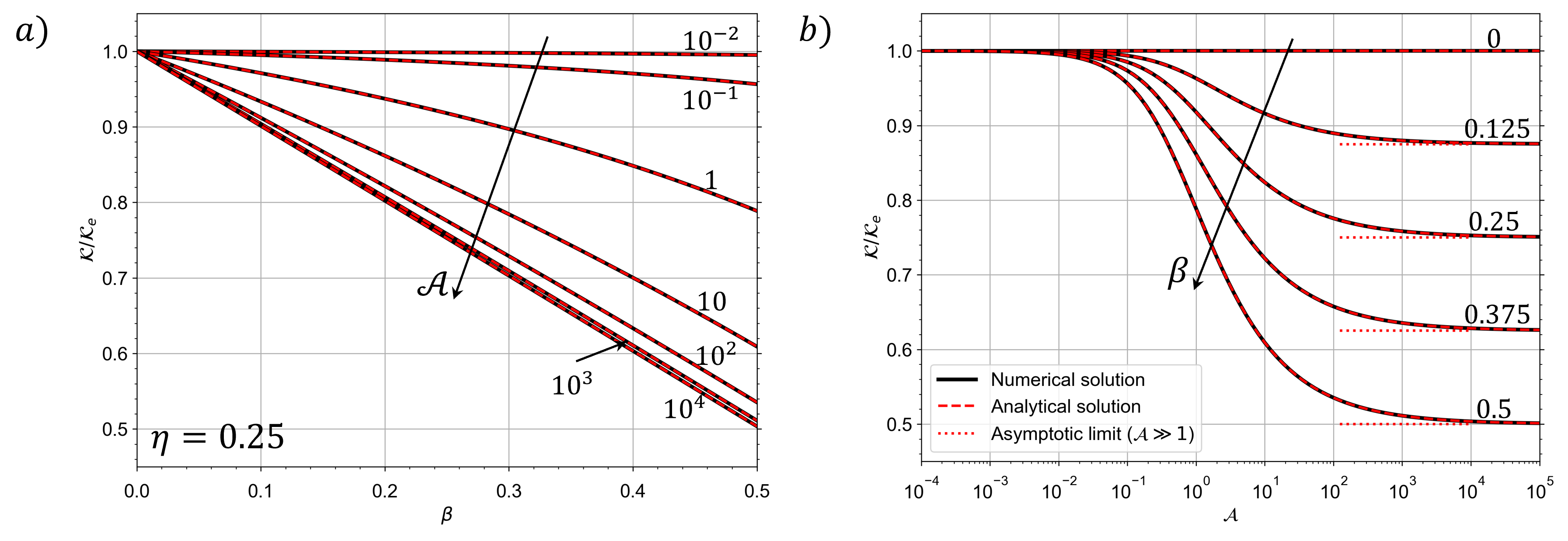}
    \caption{
    Dimensionless stress intensity factor for a semi-infinite tensile fracture with permeable surfaces, subjected to prescribed zero pore fluid pressure and an exponential normal stress, normalized by the dimensionless elastic stress intensity factor (Eq.~\eqref{eq:elastic_K_I}), as a function of (a) the normalized undrained-drained Poisson's ratio contrast $\beta$ and (b) the dimensionless loading distance $\mathcal{A}$. The poroelastic stress coefficient is set to $\eta = 0.25$. The numerical solution is shown by black lines, while the analytical solution (Eq.~\eqref{eq:K_I_perm}) is represented by dashed red lines. In panel (b), the asymptotic limit $\mathcal{K}/\mathcal{K}_e \to 1-\beta$ for $\mathcal{A} \gg 1$ is indicated by dotted red lines. In panel (a), results are shown for $\mathcal{A} = \{10^{-2}, ~10^{-1}, ~1, ~10, ~10^2, ~10^3, ~10^4\}$, while in panel (b), results correspond to $\beta = \{0, ~0.125, ~0.25, ~0.375, ~0.5\}$.
    }
    \label{fig:atkinson_perm_K_I}
\end{figure}

Fig.~\ref{fig:atkinson_perm_K_I} shows the dependence of the normalized dimensionless stress intensity factor, $\mathcal{K}/\mathcal{K}_e$, on (a) the normalized undrained-drained Poisson's ratio contrast $\beta$ and (b) the dimensionless loading distance $\mathcal{A}$. The poroelastic stress coefficient is fixed at $\eta = 0.25$. The numerical solution is compared with the analytical solution reported by \cite{AtCr91}, their Eq.~(174), which, expressed in terms of the present scaling (Eqs.~\eqref{eq:dimensionless_variables} and \eqref{eq:characteristic_scales}), takes the following form:
\begin{equation}
    \mathcal{K} = \frac{8 \mathcal{A}^{3/2}(1-\beta)}{\sqrt{\pi} \left(\mathcal{A} + \beta -\beta\sqrt{1+2\mathcal{A}} \right)},
    \label{eq:K_I_perm}
\end{equation}
which reduces to Eq.~\eqref{eq:elastic_K_I}, when $\beta = 0$.

The relative difference between the numerical and analytical solutions does not exceed 0.008 \%. The trends discussed in Section~\ref{sec:atkinson_craster_impermeable} are preserved in the current case: $\mathcal{K}/\mathcal{K}_e \to 1$ for $\mathcal{A} \ll 1$ and $\mathcal{K}/\mathcal{K}_e \to 1 - \beta$ for $\mathcal{A} \gg 1$. Furthermore, the numerical calculations indicate that, for fixed $\beta > 0$ and $\mathcal{A}$, the dimensionless opening profile $\Omega(\xi)$ and, consequently, the dimensionless stress intensity factor $\mathcal{K}$ are independent of the poroelastic stress coefficient $\eta > 0$. 

\begin{figure}[]
    \centering
    \includegraphics[width=1.0\textwidth]{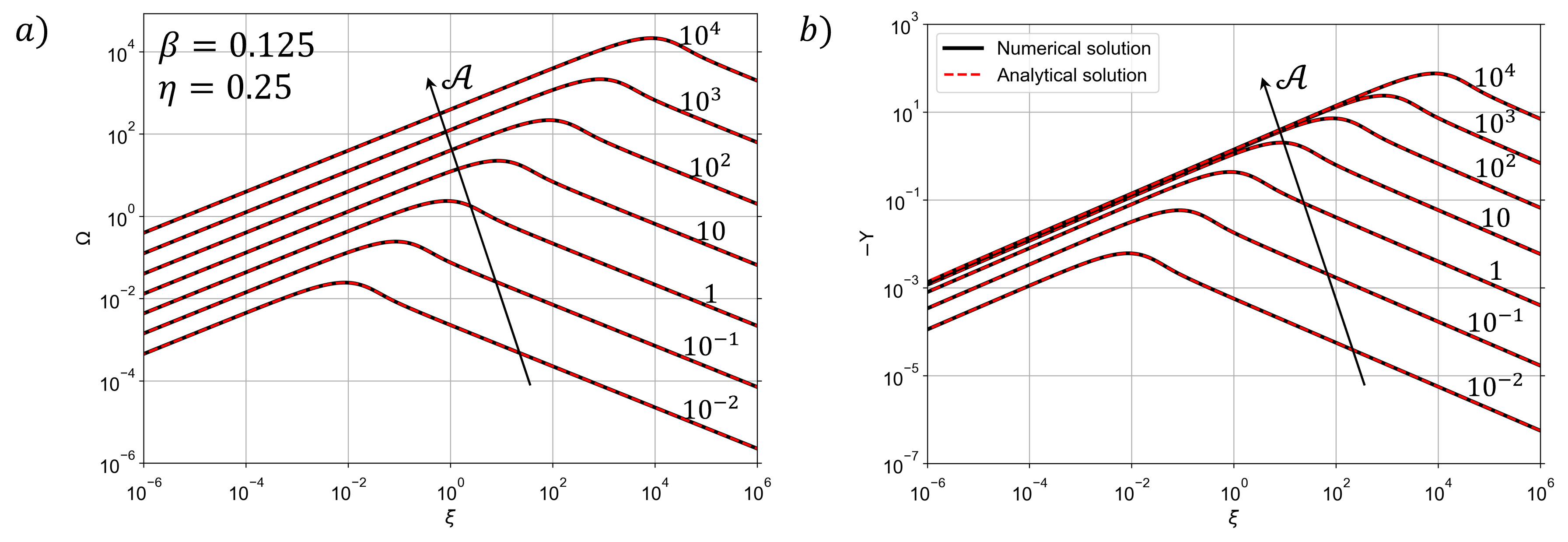}
    \caption{
    Dimensionless profiles of (a) the opening and (b) the fluid displacement function along a semi-infinite tensile fracture with permeable surfaces, subjected to prescribed zero pore fluid pressure and an exponential normal stress, shown for $\mathcal{A} = \{10^{-2}, ~10^{-1}, ~1, ~10, ~10^2, ~10^3, ~10^4\}$ with $\beta = 0.125$ and $\eta = 0.25$. The fluid displacement function is depicted with reversed sign. The numerical results are shown by black lines, while the analytical solutions (Eq.~\eqref{eq:opening_atkinson_impermeable_dimensionless} with $\mathcal{K}$ given by Eq.~\eqref{eq:K_I_perm} for the opening and Eq.~\eqref{eq:fluid_displacement_atkinson_permeable_dimensionless} for the fluid displacement function) are indicated by dashed red lines.
    }
    \label{fig:atkinson_perm_w_v}
\end{figure}

Fig.~\ref{fig:atkinson_perm_w_v}a presents the dimensionless opening profiles $\Omega(\xi)$ for different values of the dimensionless loading distance $\mathcal{A}$ at $\beta = 0.125$ and $\eta = 0.25$. \cite{AtCr91} reported the analytical solution in their Eq.~(163). Comparing this expression with their Eq.~(174) for the stress intensity factor suggests a typo in the numerical prefactor. In terms of our scaling (Eqs.~\eqref{eq:dimensionless_variables} and \eqref{eq:characteristic_scales}), the opening profile is given by Eq.~\eqref{eq:opening_atkinson_impermeable_dimensionless} with $\mathcal{K}$ defined by Eq.~\eqref{eq:K_I_perm}, which reduces to Eq.~\eqref{eq:opening_atkinson_impermeable_uncoupled_dimensionless}, when $\beta = 0$.

The profiles of the fluid displacement function are shown in Fig.~\ref{fig:atkinson_perm_w_v}b. Since $\Upsilon(\xi) < 0$, the distributions are plotted with the reversed sign. The analytical solution is obtained from the expression for the pore pressure distribution given by \cite{AtCr91}, their Eq.~(160). In terms of the scaling (Eq.~\eqref{eq:characteristic_scales}), the analytical solution for the fluid displacement function takes the following form: 
\begin{equation}
    \Upsilon(\xi) = -\frac{8 \mathcal{A} \left(\sqrt{1 + 2 \mathcal{A}}-1\right) (1-\beta) \eta }{\sqrt{\pi } \left(\mathcal{A} + \beta -\sqrt{1 + 2 \mathcal{A}} \beta\right)} D\left(\sqrt{\frac{\xi}{\mathcal{A}}}\right) = -\frac{\eta}{\mathcal{A}}\left(\sqrt{1 + 2 \mathcal{A}} - 1\right) \Omega(\xi),
    \label{eq:fluid_displacement_atkinson_permeable_dimensionless}
\end{equation}
where $\Omega(\xi)$ is the opening profile given by Eq.~\eqref{eq:opening_atkinson_impermeable_dimensionless}, evaluated with the appropriate dimensionless stress intensity factor, Eq.~\eqref{eq:K_I_perm}.

Using Eq.~\eqref{eq:fluid_displacement_atkinson_permeable_dimensionless} together with the identity $\mathcal{P}^{em}_2(\xi) \equiv \mathcal{S}^{sm}_{yy} (\xi)$, one can show that the system given by Eqs.~\eqref{eq:normal_stress_atkinson_permeable_dimensionless} and \eqref{eq:pressure_atkinson_permeable_dimensionless} reduces to a single equation expressed in terms of the opening gradient $\diff \Omega / \diff \xi$ and independent of the poroelastic stress coefficient $\eta$. As noted above, this behavior is confirmed by the numerical results.

Next, we compute the pore pressure field around the fracture. For an arbitrary $\zeta$ coordinate, the pore pressure distribution is governed by the following equation (see Section~\ref{sec:boundary_integral_equations} and Appendices~\ref{sec:derivation_fluid_source_pressure}, \ref{sec:derivation_edge_dislocation_pressure}):
\begin{equation}
    \Pi(\xi, \zeta) = \frac{1}{2 \pi \mathcal{S}}\int_0^\infty \frac{\diff \Upsilon}{\diff s} e^{\xi - s} \mathrm{K}_0(r) \diff s - \frac{\eta}{2 \pi \mathcal{S}} \int_0^{\infty} \frac{\diff \Omega}{\diff s} \left(-\frac{\xi-s}{r^2} + e^{\xi - s} \left[\frac{\xi-s}{r} \mathrm{K}_1\left(r\right) - \mathrm{K}_0\left(r\right)\right] \right) \diff s,
    \label{eq:pressure_distribution_around_permeable_dimensionless}
\end{equation}
where $r = \sqrt{(\xi-s)^2 + \zeta^2}$.

\begin{figure}[]
    \centering
    \includegraphics[width=1.0\textwidth]{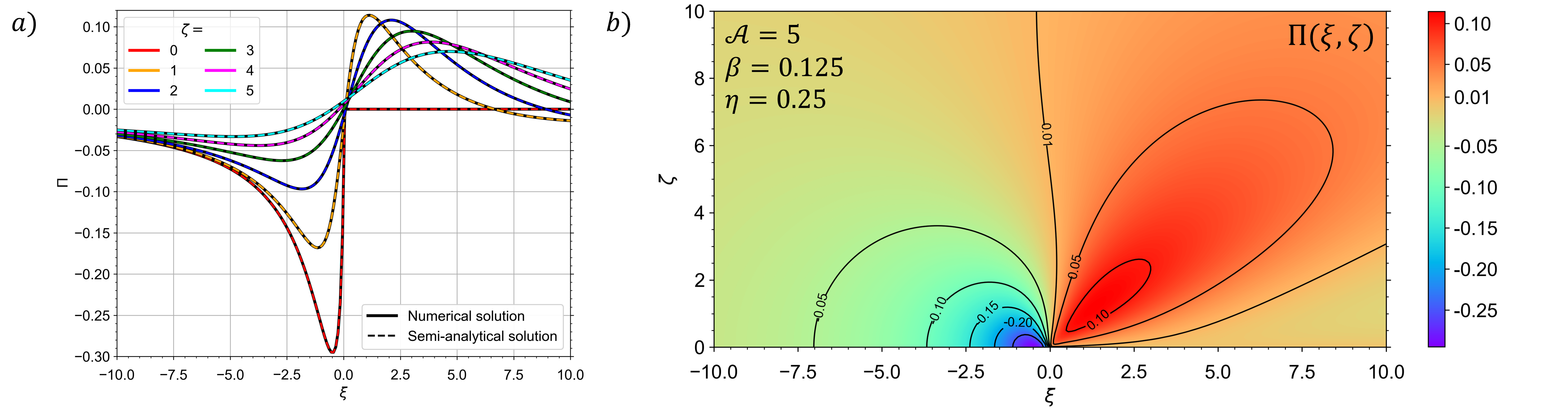}
    \caption{
    Dimensionless pore fluid pressure around a semi-infinite tensile fracture with permeable surfaces, subjected to prescribed zero pore fluid pressure and an exponential normal stress. (a) Profiles along lines parallel to the fracture at $\zeta = \{0, 1, 2, 3, 4, 5\}$ for $\mathcal{A} = 5$, $\beta = 0.125$, and $\eta = 0.25$. The numerical results obtained from Eq.~\eqref{eq:pressure_distribution_around_permeable_dimensionless} are plotted as solid black lines, whereas the reference semi-analytical solution from Eq.~\eqref{eq:pressure_atkinson_permeable_dimensionless_field} is shown as dashed colored lines. (b) Pore pressure field around the fracture tip for $\zeta > 0$, computed from Eq.~\eqref{eq:pressure_distribution_around_permeable_dimensionless}. Isolines corresponding to $\Pi(\xi, \zeta) = \mathrm{const}$ are shown as solid black lines.
    }
    \label{fig:atkinson_perm_pressure}
\end{figure}

Fig.~\ref{fig:atkinson_perm_pressure}a compares the pressure distributions along lines parallel to the fracture for different values of $\zeta$, for $\mathcal{A} = 5$, $\beta = 0.125$, and $\eta = 0.25$. The results are obtained using Eq.~\eqref{eq:pressure_distribution_around_permeable_dimensionless} and the expression reported by \cite{AtCr91}, their Eq.~(160). In terms of our scaling (Eqs.~\eqref{eq:dimensionless_variables} and \eqref{eq:characteristic_scales}), the solution from \cite{AtCr91} can be expressed as follows:
\begin{flalign}
    \Pi(\xi, \zeta) = \frac{\mathcal{A} \beta }{\sqrt{2 \pi } \eta  \left(\mathcal{A} + \beta -\sqrt{1 + 2 \mathcal{A}} \beta \right)} &  \bigg[\sqrt{\mathcal{A}}\frac{\sqrt{r(\xi) - \xi}}{r(\xi)} \left(e^{\xi-r(\xi)} - 1\right) - \nonumber \\
    & - \frac{e^{-\xi/\mathcal{A}}}{\sqrt{\mathcal{A}}} \int_{-\infty}^\xi e^{s/\mathcal{A}}\frac{\sqrt{r(s) - s}}{r(s)} \left(\sqrt{1 + 2\mathcal{A}} e^{s-r(s)} - 1 \right) \diff s\bigg],
    \label{eq:pressure_atkinson_permeable_dimensionless_field}
\end{flalign}
where $r(\xi) = \sqrt{\xi^2 + \zeta^2}$. The integral in the equation above can be evaluated analytically for $\zeta = 0$, yielding $\Pi(\xi) = 0$ for $\xi > 0$.
Fig.~\ref{fig:atkinson_perm_pressure}b shows the pore pressure field $\Pi(\xi, \zeta)$ for $\zeta > 0$, computed using Eq.~\eqref{eq:pressure_distribution_around_permeable_dimensionless} to visualize the distribution near the fracture tip.

\begin{figure}[]
    \centering
    \includegraphics[width=0.6\textwidth]{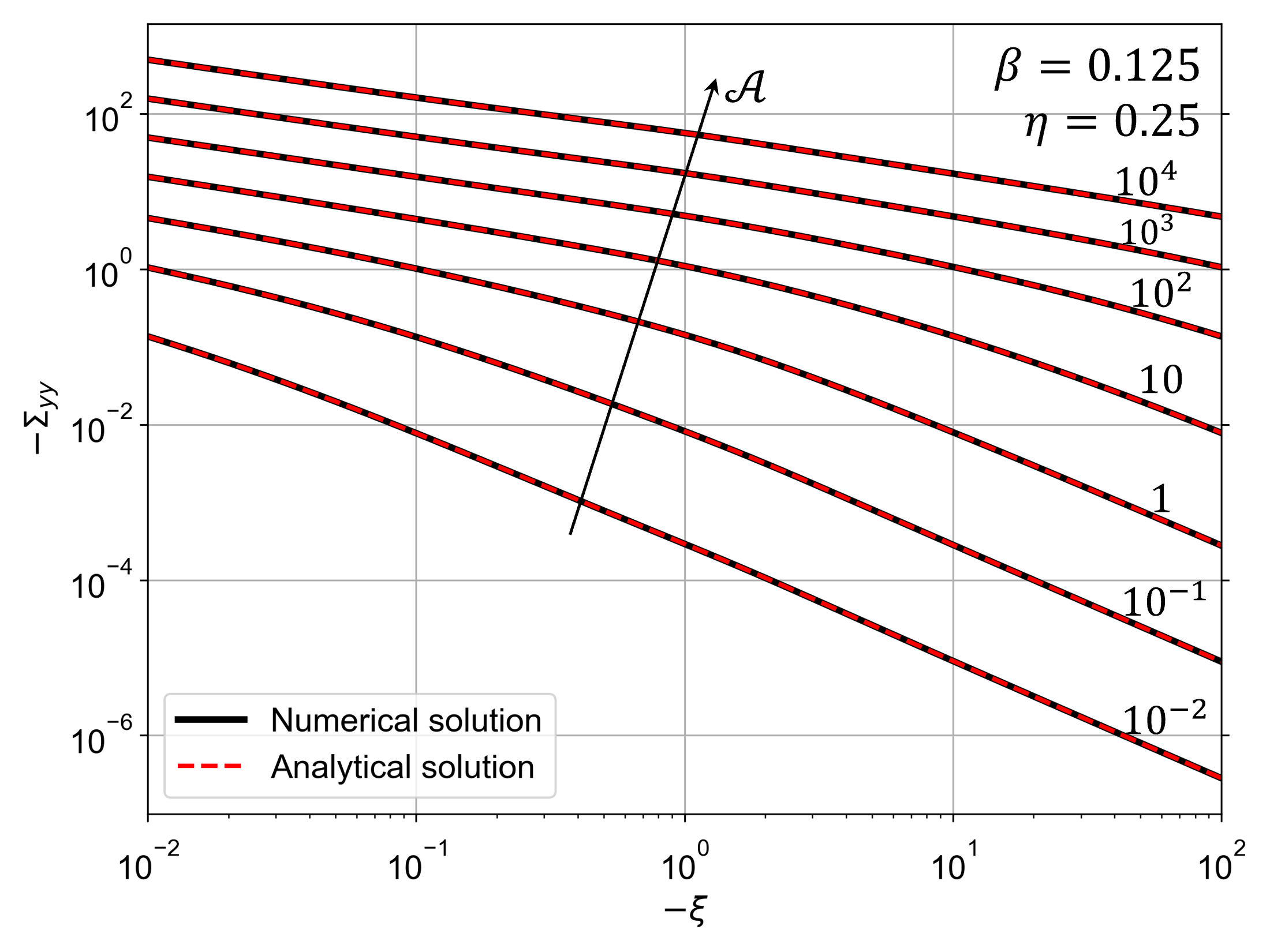}
    \caption{
    Dimensionless tensile stress profiles ahead ($\xi < 0$) of a semi-infinite tensile fracture with permeable surfaces, subjected to prescribed zero pore fluid pressure and an exponential normal stress. Results are shown for $\mathcal{A} = \{10^{-2}, ~10^{-1}, ~1, ~10, ~10^2, ~10^3, ~10^4\}$ with $\beta = 0.125$ and $\eta = 0.25$. Both the tensile stress and the $\xi-$coordinate are plotted with reversed sign. The numerical results are plotted by black lines, while the analytical solution (Eq.~\eqref{eq:stress_atkinson_permeable_dimensionless_field}) is indicated by dashed red lines.
    }
    \label{fig:atkinson_perm_sigma_yy}
\end{figure}

Finally, we evaluate the tensile stress distribution ahead of the fracture tip using Eq.~\eqref{eq:tensile_crack_normal_stress_dimensionless}, which provides $\Sigma_{yy}(\xi) = e^{-\xi/\mathcal{A}}$ along the fracture surface (see Eq.~\eqref{eq:normal_stress_atkinson_permeable_dimensionless}). In Fig.~\ref{fig:atkinson_perm_sigma_yy}, we demonstrate the tensile stress distributions for several values of the dimensionless loading distance $\mathcal{A}$, by taking $\beta = 0.125$ and $\eta = 0.25$.  Because $\Sigma_{yy}(\xi) < 0$ for $\xi < 0$, both the stress and the coordinate are plotted with reversed sign to facilitate visualization. The numerical results obtained from Eq.~\eqref{eq:tensile_crack_normal_stress_dimensionless} are shown in Fig.~\ref{fig:atkinson_perm_sigma_yy} together with the analytical solution of \cite{AtCr91} (their Eq.~(161)), which, expressed in terms of our scaling (Eqs.~\eqref{eq:dimensionless_variables} and \eqref{eq:characteristic_scales}), takes the following form:
\begin{flalign} 
    &\Sigma_{yy} (\xi) = -\frac{2 \mathcal{A} \beta }{\mathcal{A}+\beta - \beta \sqrt{1 + 2 \mathcal{A}}} 
    \Bigg\{\sqrt{\frac{2 \mathcal{A}}{\pi}} \Bigg(\frac{1-e^{2 \xi }}{4 \sqrt{2} (-\xi )^{3/2}} + \frac{\frac{1}{2 \beta }-e^{2 \xi }}{\sqrt{-2\xi} }\Bigg)-\frac{e^{-\xi / \mathcal{A}}}{2 \beta}\mathrm{Erfc}\left(\sqrt{-\frac{\xi }{\mathcal{A}}}\right) + \nonumber \\ & 
    + \sqrt{1 + \frac{1}{2 \mathcal{A}}} \Bigg(\frac{e^{2 \xi }-1}{\sqrt{-2 \pi \xi}} - \frac{e^{-\xi / \mathcal{A}}}{\sqrt{2 \mathcal{A}}} \Bigg[\frac{1}{\sqrt{1 + 2 \mathcal{A}}} \mathrm{Erfc}\left(\sqrt{-\frac{\xi}{\mathcal{A}}\left(1 + 2 \mathcal{A}\right)}\right) - \mathrm{Erfc}\left(\sqrt{-\frac{\xi }{\mathcal{A}}}\right)\Bigg]\Bigg)\Bigg\},
    \label{eq:stress_atkinson_permeable_dimensionless_field}
\end{flalign}
where $\mathrm{Erfc}(\cdot)$ is the complementary error function. 

\subsection{Stress-free semi-infinite tensile fracture with exponential pore pressure}

In this section, we examine a modification of the problem discussed in Section~\ref{sec:atkinson_craster_permeable}. Specifically, we consider a steadily propagating semi-infinite tensile crack (Fig.~\ref{fig:semi_inf_poroelastic}a) subject to a stress-free boundary condition, $\sigma_{yy}(x) = 0$, and an imposed pore fluid pressure distribution $p_f(x) = p_0 e^{-x/a}$. The poroelastic medium is initially stress-free ($\sigma_\infty = 0$) and at zero pore pressure ($p_\infty = 0$). Within the scaling defined by Eqs.~\eqref{eq:dimensionless_variables} and \eqref{eq:characteristic_scales}, we set $\sigma_* = p_0$.

We note that this problem formulation is somewhat artificial, as the prescribed loading is mechanically unphysical. In particular, imposing a pore pressure distribution in the absence of any external tensile stress induces volumetric expansion of the material surrounding the crack, which leads to crack closure; accordingly, the model predicts negative crack opening. Despite this limitation, the configuration remains useful as a benchmark for verifying the derived boundary integral equations.

The problem is formulated in terms of the crack opening profile $w(x)$ and the fluid displacement function $v(x)$, and is governed by Eqs.~\eqref{eq:tensile_crack_normal_stress_1} and \eqref{eq:fluid_pressure_1}. The dimensionless form of these equations is given by Eqs.~\eqref{eq:tensile_crack_normal_stress_dimensionless} and \eqref{eq:fluid_pressure_dimensionless}, which, for the present problem, reduce to:
\begin{flalign}
    & 0 = \frac{1}{4\pi (1-\beta)}\int_0^\infty \frac{\diff \Omega}{\diff s}\mathcal{S}^{em}_{yy2} \left(\xi - s\right) \frac{\diff s}{\xi-s} - \frac{\eta}{2 \pi \mathcal{S}} \int_0^{\infty} \frac{\diff \Upsilon}{\diff s} \mathcal{S}^{sm}_{yy}(\xi-s) \diff s,
    \label{eq:normal_stress_atkinson_pressure_bc_dimensionless} \\
    & e^{-\xi/\mathcal{A}} = \frac{1}{2\pi \mathcal{S}} \int_0^{\infty} \frac{\diff \Upsilon}{\diff s}  \mathcal{P}^{sm} (\xi-s) \diff s -  \frac{\eta}{2\pi \mathcal{S}} \int_0^{\infty} \frac{\diff \Omega}{\diff s} \mathcal{P}^{em}_2  (\xi-s) \diff s \label{eq:pressure_atkinson_pressure_bc_dimensionless} .
\end{flalign}

The opening $\Omega(\xi)$ and the fluid displacement function $\Upsilon(\xi)$ profiles are governed by the normalized undrained-drained Poisson's ratio contrast $\beta$ (Eq.~\eqref{eq:poissons_contrast}), the dimensionless loading distance $\mathcal{A}$, and the poroelastic stress coefficient $\eta$ (Eq.~\eqref{eq:poroelastic_stress_coefficient}). The numerical solution is obtained using the algorithm outlined in Section~\ref{sec:numerical_method}, with all problem-specific components being identical to those presented in Section~\ref{sec:atkinson_craster_permeable}.

\begin{figure}[]
    \centering
    \includegraphics[width=1.0\textwidth]{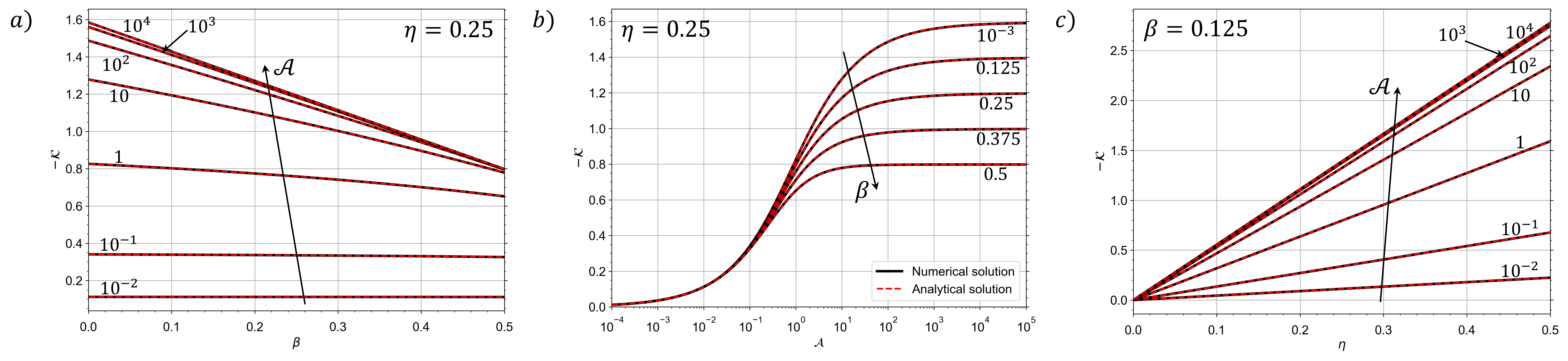}
    \caption{
    Dimensionless stress intensity factor for a semi-infinite tensile fracture with permeable surfaces, stress-free and subjected to an imposed exponential pore pressure, as a function of (a) the normalized undrained-drained Poisson's ratio contrast $\beta$, (b) the dimensionless loading distance $\mathcal{A}$, and (c) the poroelastic stress coefficient $\eta$. The dimensionless stress intensity factor is shown with reversed sign. The numerical solution is shown by black lines, while the analytical solution (Eq.~\eqref{eq:K_I_pressure_bc}) is represented by dashed red lines. In panel (a), results are shown for $\mathcal{A} = \{10^{-2}, ~10^{-1}, ~1, ~10, ~10^2, ~10^3, ~10^4\}$ with $\eta = 0.25$; in panel (b), results correspond to $\beta = \{10^{-3}, ~0.125, ~0.25, ~0.375, ~0.5\}$ with $\eta = 0.25$; in panel (c) results are given for $\mathcal{A} = \{10^{-2}, ~10^{-1}, ~1, ~10, ~10^2, ~10^3, ~10^4\}$ with $\beta = 0.125$.
    }
    \label{fig:atkinson_pressure_bc_K_I}
\end{figure}

Fig.~\ref{fig:atkinson_pressure_bc_K_I} shows the dependence of the dimensionless stress intensity factor, $\mathcal{K}$, on (a) the normalized undrained-drained Poisson's ratio contrast $\beta$, (b) the dimensionless loading distance $\mathcal{A}$, and (c) the poroelastic stress coefficient $\eta$. Since $\mathcal{K} < 0$ in this problem, we plot it with the reversed sign. The numerical solution is compared with the analytical solution reported by \cite{AtCr91}, their Eq.~(182), which, expressed in terms of the present scaling (Eqs.~\eqref{eq:dimensionless_variables} and \eqref{eq:characteristic_scales}), takes the following form:
\begin{equation}
    \mathcal{K} = -\frac{8 \sqrt{\mathcal{A}} \left(\sqrt{1 + 2 \mathcal{A}}-1\right) (1 - \beta) \eta }{\sqrt{\pi } \left(\mathcal{A} + \beta - \beta\sqrt{1 + 2 \mathcal{A}} \right)}.
    \label{eq:K_I_pressure_bc}
\end{equation}
It should be noted that the dimensionless stress intensity factor given in Eq.~\eqref{eq:K_I_pressure_bc} is twice as large as the value reported by \cite{AtCr91}. The origin of this discrepancy is not clear from their published expression; however, the value reported here is consistent with the results of the present boundary integral formulation and with the reconstructed profiles for the fracture opening and fluid displacement function.


The relative difference between the numerical and analytical solutions does not exceed 0.007 \%. Unlike the model discussed in Section~\ref{sec:atkinson_craster_permeable}, the dimensionless stress intensity factor depends on the poroelastic stress coefficient $\eta$. From Fig.~\ref{fig:atkinson_pressure_bc_K_I}c, one can observe that $\mathcal{K} \to 0$ as $\eta \to 0$, while for $\beta \to 0$, $\mathcal{K}$ varies as a function of both $\mathcal{A}$ and $\eta$ (see Fig.~\ref{fig:atkinson_pressure_bc_K_I}b).

\begin{figure}[]
    \centering
    \includegraphics[width=1.0\textwidth]{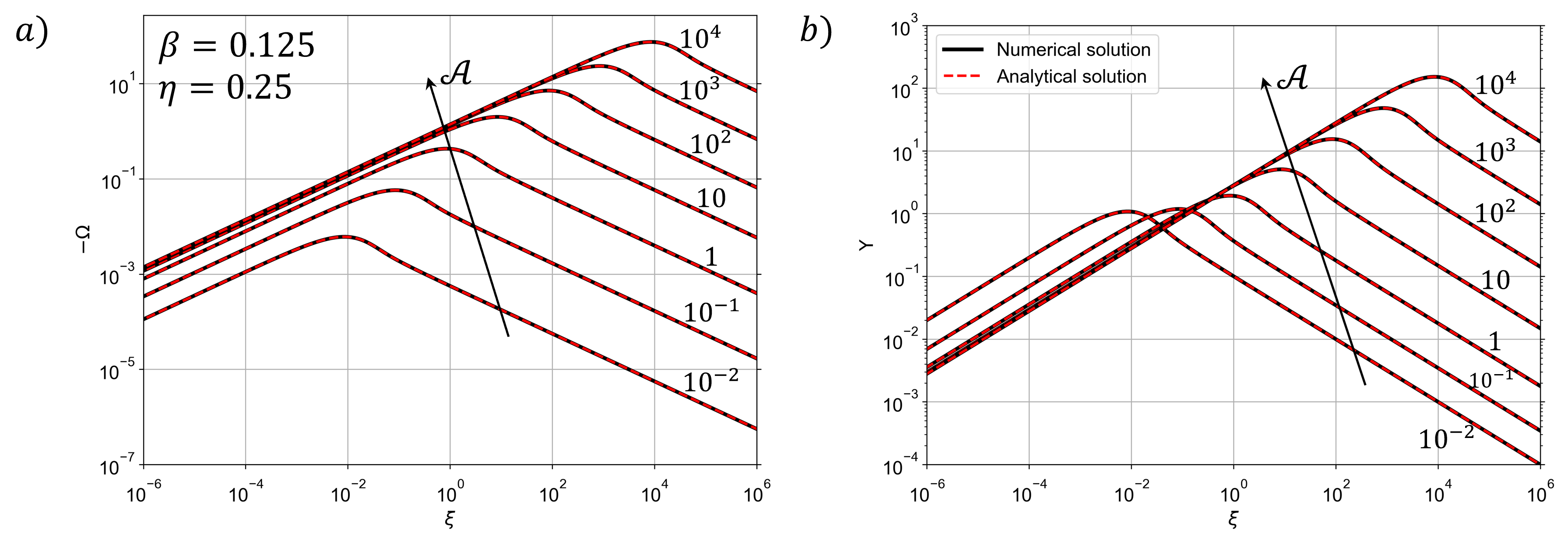}
    \caption{
    Dimensionless profiles of (a) the opening and (b) the fluid displacement function along a semi-infinite tensile fracture with permeable surfaces, stress-free and subjected to an imposed exponential pore pressure, shown for $\mathcal{A} = \{10^{-2}, ~10^{-1}, ~1, ~10, ~10^2, ~10^3, ~10^4\}$ with $\beta = 0.125$ and $\eta = 0.25$. The opening profile is shown with reversed sign. The numerical results are shown by black lines, while the analytical solutions (Eq.~\eqref{eq:opening_atkinson_impermeable_dimensionless} with $\mathcal{K}$ given by Eq.~\eqref{eq:K_I_pressure_bc} for the opening and Eq.~\eqref{eq:fluid_displacement_atkinson_pressure_bc_dimensionless} for the fluid displacement function) are indicated by dashed red lines.
    }
    \label{fig:atkinson_pressure_bc_w_v}
\end{figure}

Fig.~\ref{fig:atkinson_pressure_bc_w_v}a shows the dimensionless opening profiles $\Omega(\xi)$ for various values of the dimensionless loading distance $\mathcal{A}$ at $\beta = 0.125$ and $\eta = 0.25$. Since $\Omega(\xi) < 0$, the distributions are plotted with the reversed sign. Although the analytical solution was not explicitly provided in \citep{AtCr91}, it can be reconstructed based on the formulas presented in Sections~\ref{sec:atkinson_craster_impermeable} and \ref{sec:atkinson_craster_permeable}, yielding the opening profile given by Eq.~\eqref{eq:opening_atkinson_impermeable_dimensionless} with $\mathcal{K}$ defined by Eq.~\eqref{eq:K_I_pressure_bc}.

The profiles of the fluid displacement function are presented in Fig.~\ref{fig:atkinson_pressure_bc_w_v}b. The analytical solution is derived using the relations provided in Sections~\ref{sec:atkinson_craster_impermeable} and \ref{sec:atkinson_craster_permeable}:
\begin{equation}
    \Upsilon(\xi) = \frac{8 \left(\beta + \sqrt{1 + 2 \mathcal{A}} (\mathcal{A} - \beta)\right) (1 - \beta) \eta ^2}{\sqrt{\pi} \beta \left(\mathcal{A} + \beta - \beta\sqrt{1 + 2 \mathcal{A}} \right)} D\left(\sqrt{\frac{\xi}{\mathcal{A}}}\right).
    \label{eq:fluid_displacement_atkinson_pressure_bc_dimensionless}
\end{equation}

\subsection{Semi-infinite shear fracture under constant loading along a strip}

The next test is based on the problem examined by \cite{RiSi76}, which considers a steadily propagating semi-infinite shear crack (Fig.~\ref{fig:semi_inf_poroelastic}b) subjected to a uniform shear load applied over a finite region. The poroelastic medium is initially stress-free ($\tau_{\infty} = 0$) and has zero pore fluid pressure ($p_\infty = 0$). A uniform shear traction is applied on the fracture surfaces over a finite domain of length $a$ near the crack tip: $\sigma_{xy}(x) = -\tau_0 \mathrm{H}(a - x)$. Hence, the characteristic stress (pressure) scale in this problem is $\sigma_* = \tau_0$ (see Eqs.~\eqref{eq:dimensionless_variables} and \eqref{eq:characteristic_scales}). The fracture surfaces are permeable, with the pore fluid pressure prescribed as $p_f(x) = 0$. Consequently, Eq.~\eqref{eq:shear_crack_shear_stress_1} governs the slip profile $d(x)$, while the fluid displacement function is trivial, $v(x) = 0$. This benchmark verifies the boundary integral equation for the shear stress but does not test the equation for the pore fluid pressure. In the present shear fracture formulation, the pore fluid pressure is indeed decoupled from slip under the adopted permeable slip-plane hydraulic limit.


The dimensionless form of Eq.~\eqref{eq:shear_crack_shear_stress_1} is given by Eq.~\eqref{eq:shear_crack_shear_stress_dimensionless}, which, for the present problem, reduces to:
\begin{equation}
    \mathrm{H}(\mathcal{A} - \xi) = \frac{1}{4\pi(1-\beta)}\int_0^{\infty} \frac{\diff \Delta}{\diff s} \mathcal{S}_{xy1}^{em} \left(\xi - s\right) \frac{\diff s}{\xi - s}, 
    \label{eq:shear_stress_rice_dimensionless}
\end{equation}
where $\mathcal{A} = a V / (2c)$ is the dimensionless loading distance. The slip profile $\Delta (\xi)$ depends on the normalized undrained-drained Poisson's ratio contrast $\beta$ (Eq.~\eqref{eq:poissons_contrast}) and the dimensionless loading distance $\mathcal{A}$. 

First, we consider the uncoupled case, $\beta = 0$, in which Eq.~\eqref{eq:shear_stress_rice_dimensionless} reduces to:
\begin{equation*}
    \mathrm{H}(\mathcal{A} - \xi) = \frac{1}{4\pi}\int_0^\infty \frac{\diff \Delta}{\diff s} \frac{\diff s}{\xi-s}.
\end{equation*}
Following the approach described in Section~\ref{sec:atkinson_craster_impermeable}, we obtain the dimensionless elastic slip profile:
\begin{equation}
    \Delta_e(\xi) = \frac{4}{\pi}\left(2 \sqrt{\xi \mathcal{A} } + (\mathcal{A} - \xi) \ln \left|\frac{\sqrt{\xi} + \sqrt{\mathcal{A}}}{\sqrt{\xi} - \sqrt{\mathcal{A}}}\right| \right).  
    \label{eq:slip_rice_uncoupled_dimensionless}
\end{equation}
A Taylor expansion of Eq.~\eqref{eq:slip_rice_uncoupled_dimensionless} provides the leading-order term of the slip profile near the fracture tip, which takes the form of Eq.~\eqref{eq:tip_asymptote_shear_crack_dimensionless} with the specific value of the dimensionless stress intensity factor: 
\begin{equation}
    \mathcal{K}_e = \frac{16\sqrt{\mathcal{A}}}{\pi}.
    \label{eq:elastic_K_II}
\end{equation}
This quantity is referred to as the dimensionless elastic stress intensity factor, and its expression coincides with the result reported by \cite{RiSi76}, their Eq.~(6).

Moreover, using Eq.~\eqref{eq:slip_rice_uncoupled_dimensionless}, we determine the far-field asymptote of the slip profile:
\begin{equation}
    \Delta_e(\xi) = \frac{16 \mathcal{A}^{3/2}}{3 \pi \sqrt{\xi}}, ~~~~ \xi\to\infty.
    \label{eq:far_field_opening_rice_uncoupled_dimensionless}
\end{equation}

The general numerical solution of the problem is obtained using the methodology presented in Section~\ref{sec:numerical_method}. All problem-specific components are identical to those described in Section~\ref{sec:atkinson_craster_impermeable}, since the near-field asymptote (Eq.~\eqref{eq:tip_asymptote_shear_crack_dimensionless}) and the shape of the far-field asymptote (Eq.~\eqref{eq:far_field_opening_rice_uncoupled_dimensionless}) are identical to the problem discussed in Section~\ref{sec:atkinson_craster_impermeable}.

\begin{figure}[]
    \centering
    \includegraphics[width=1.0\textwidth]{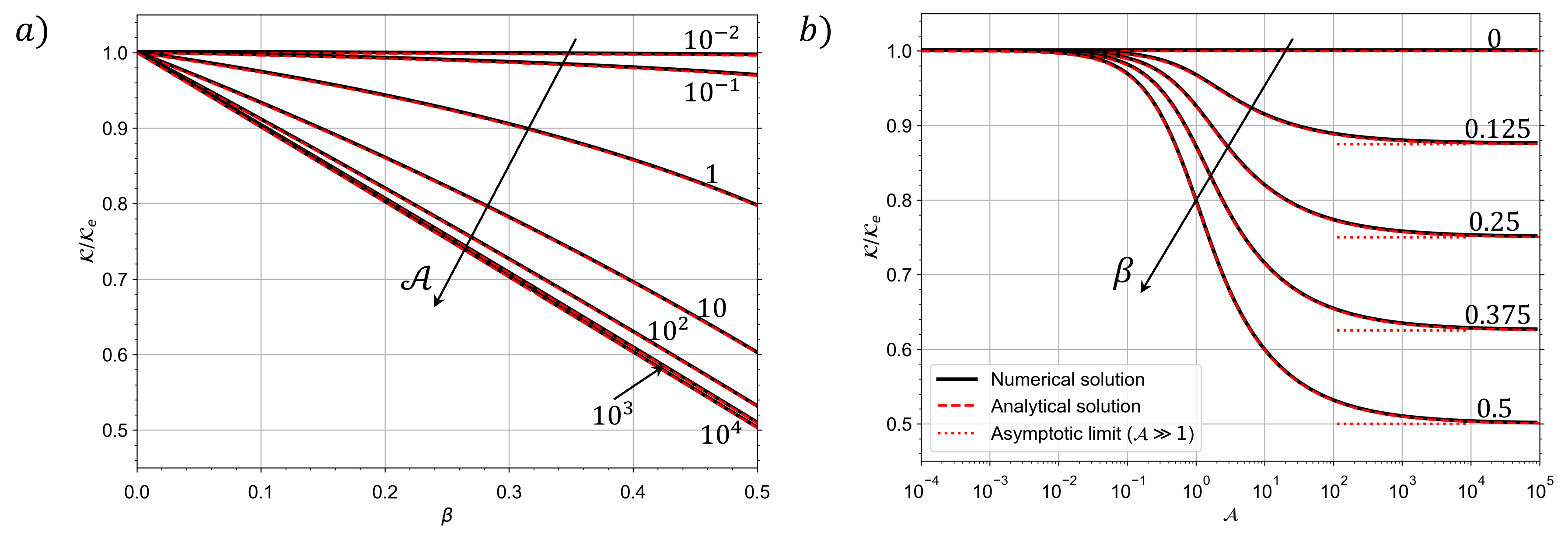}
    \caption{
    Dimensionless stress intensity factor for a semi-infinite shear fracture with permeable surfaces, subjected to prescribed zero pore fluid pressure and a uniform shear stress applied over a finite region, normalized by the dimensionless elastic stress intensity factor (Eq.~\eqref{eq:elastic_K_II}), as a function of (a) the normalized undrained-drained Poisson's ratio contrast $\beta$ and (b) the dimensionless loading distance $\mathcal{A}$. The numerical solution is shown by black lines, while the analytical solution (Eq.~\eqref{eq:K_II_rice}) is represented by dashed red lines. In panel (b), the asymptotic limit $\mathcal{K}/\mathcal{K}_e \to 1-\beta$ for $\mathcal{A} \gg 1$ is indicated by dotted red lines. In panel (a), results are shown for $\mathcal{A} = \{10^{-2}, ~10^{-1}, ~1, ~10, ~10^2, ~10^3, ~10^4\}$, while in panel (b), results correspond to $\beta = \{0, ~0.125, ~0.25, ~0.375, ~0.5\}$.
    }
    \label{fig:rice_K_II}
\end{figure}

Fig.~\ref{fig:rice_K_II} presents the normalized stress intensity factor, $\mathcal{K}/\mathcal{K}_e$, as a function of (a) the normalized undrained-drained Poisson's ratio contrast $\beta$ and (b) the dimensionless loading distance $\mathcal{A}$. The numerical results are compared with the analytical solution derived by \cite{RiSi76}, their Eq.~(37), using Fourier transform and Wiener-Hopf techniques. Expressed in terms of the present scaling (Eqs.~\eqref{eq:dimensionless_variables} and \eqref{eq:characteristic_scales}), the analytical solution reads:
\begin{flalign}
    & \mathcal{K} = -\mathcal{K}_e \frac{e^{-i\pi/4}}{4i\sqrt{\pi}} \int_{-\infty}^{\infty} \frac{1 - e^{is}}{s \hat{D}^-(s)\hat{m}^-(s)}\diff s, \label{eq:K_II_rice}\\
    & \hat{D}^-(s) = \frac{1}{1-\beta} + \frac{i s \beta}{\mathcal{A}(1-\beta)}\left[1 - \frac{\hat{n}^-(s)}{\hat{m}^-(s)}\right], ~~~~ \hat{n}^-(s) = \sqrt{s - 2 i \mathcal{A}}, ~~~~ \hat{m}^-(s) = \sqrt{s}, \nonumber
\end{flalign}
where $i$ is the imaginary unit, and the functions $\hat{n}^-(s)$ and $\hat{m}^-(s)$ possess branch cuts along the positive imaginary axis of the complex $s-$plane, chosen such that both functions have positive real parts for $s>0$. Eq.~\eqref{eq:K_II_rice} reduces to Eq.~\eqref{eq:elastic_K_II}, when $\beta = 0$.

The relative difference between the numerical and analytical solutions does not exceed 0.18 \%. The asymptotic behavior of $\mathcal{K}/\mathcal{K}_e$ is similar to that discussed in Sections~\ref{sec:atkinson_craster_impermeable} and~\ref{sec:atkinson_craster_permeable}: $\mathcal{K}/\mathcal{K}_e \to 1$ for $\mathcal{A} \ll 1$ and $\mathcal{K}/\mathcal{K}_e \to 1 - \beta$ for $\mathcal{A} \gg 1$.

\begin{figure}[]
    \centering
    \includegraphics[width=0.6\textwidth]{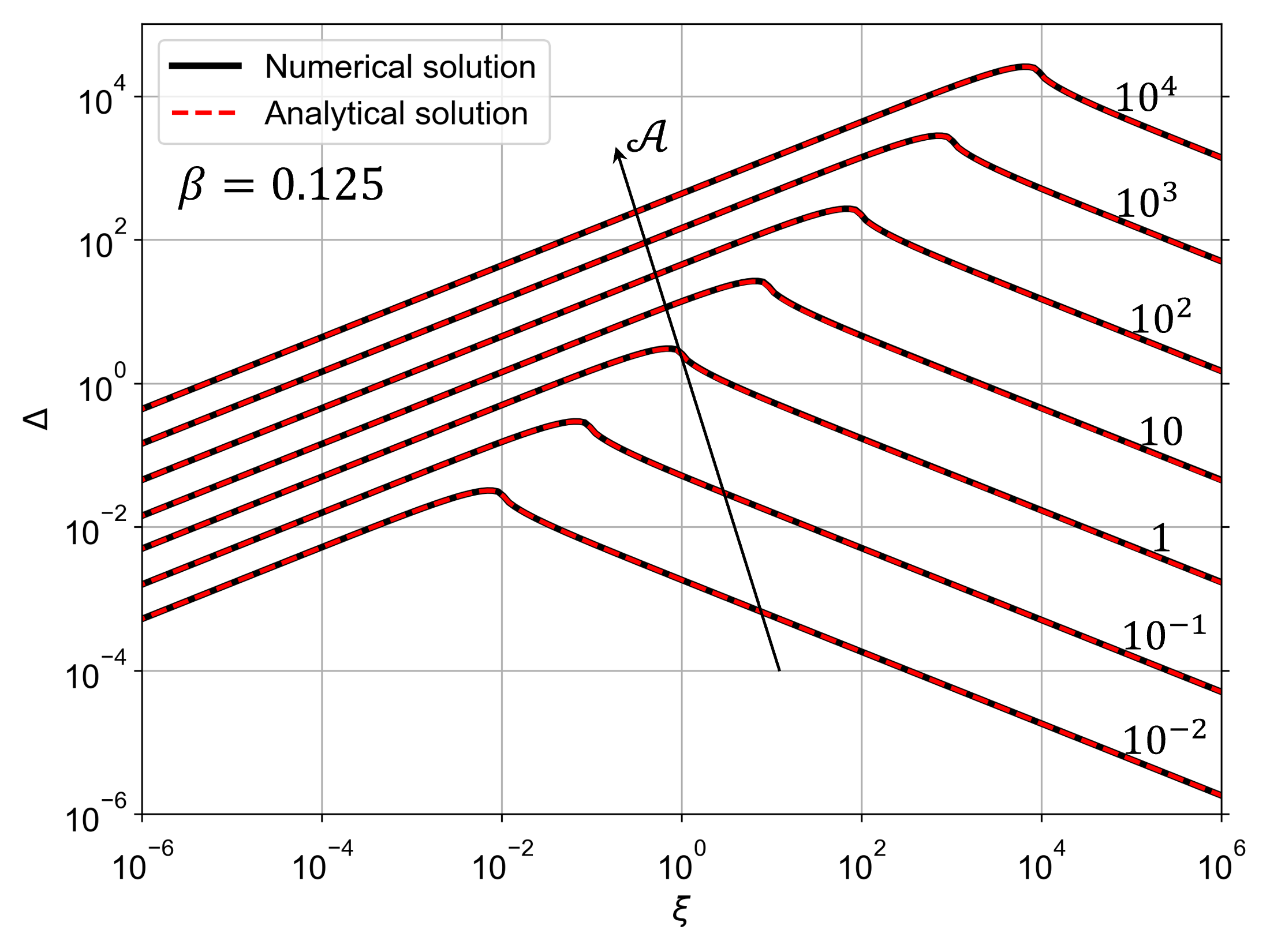}
    \caption{
    Dimensionless slip profiles along a semi-infinite shear fracture with permeable surfaces, subjected to prescribed zero pore fluid pressure and a uniform shear stress applied over a finite region, shown for $\mathcal{A} = \{10^{-2}, ~10^{-1}, ~1, ~10, ~10^2, ~10^3, ~ 10^4\}$ with $\beta = 0.125$. The numerical results are shown by black lines, while the analytical solution (Eq.~\eqref{eq:slip_rice_dimensionless}), are indicated by dashed red lines.
    }
    \label{fig:rice_d}
\end{figure}

The dimensionless slip profiles $\Delta(\xi)$ are shown in Fig.~\ref{fig:rice_d}, corresponding to different values of the dimensionless loading distance $\mathcal{A}$ at $\beta = 0.125$. \cite{RiSi76} did not provide an analytical expression for the slip profile. We therefore assume that the shape of the slip profile for $\beta > 0$ is identical to that in the uncoupled case, i.e., $\Delta_e(\xi)$ (Eq.~\eqref{eq:slip_rice_uncoupled_dimensionless}). On the basis of this assumption and Eq.~\eqref{eq:K_II_rice}, we reconstruct the slip distribution as:
\begin{equation}
    \Delta(\xi) = \frac{\mathcal{K}}{4\sqrt{\mathcal{A}}} \left(2 \sqrt{\xi \mathcal{A} } + (\mathcal{A} - \xi) \ln \left|\frac{\sqrt{\xi} + \sqrt{\mathcal{A}}}{\sqrt{\xi} - \sqrt{\mathcal{A}}}\right| \right),
    \label{eq:slip_rice_dimensionless}
\end{equation}
with $\mathcal{K}$ given by Eq.~\eqref{eq:K_II_rice}, which reduces to Eq.~\eqref{eq:slip_rice_uncoupled_dimensionless}, when $\beta = 0$.

\section{Conclusions}

We have presented a boundary integral formulation for steadily propagating semi-infinite plane strain fractures in linear poroelastic media. Using fundamental solutions of plane strain poroelasticity for an instantaneous fluid source and an edge dislocation, including both normal and slip modes, and applying the principle of temporal superposition, we analytically obtained the distributions of pore pressure and stress tensor components induced by a unit steadily moving fluid source and edge dislocation. Subsequent spatial superposition yielded boundary integral equations for semi-infinite tensile (Mode I) and shear (Mode II) crack problems, relating the normal and shear stresses, as well as the pore fluid pressure, all defined on the fracture surfaces, to the distributions of dislocation density and the derivative of the fluid displacement function. The resulting formulation should be interpreted as a boundary operator for steadily moving semi-infinite cracks under the hydraulic boundary conditions embedded in the adopted fundamental solutions.

We have developed a numerical methodology for solving the system of boundary integral equations with respect to the fracture opening (for a tensile crack), slip (for a shear crack), and the fluid displacement function, assuming that the traction and pore fluid pressure profiles are prescribed. The approach is based on representing the unknown functions on each spatial segment as a superposition of two power-law basis functions whose forms are chosen to reproduce the leading-order asymptotic behavior of the solution in the near-field and far-field, respectively. Thus, the method relies on problem-specific asymptotes and on convergence checks with respect to the computational interval and the number of collocation points.

The proposed boundary integral formulation was systematically verified against a series of benchmark problems, including a steadily propagating tensile fracture subjected to exponential normal loading with either impermeable surfaces or permeable surfaces on which zero pore fluid pressure is prescribed; a stress-free tensile fracture with permeable surfaces subjected to an imposed exponential pore fluid pressure; and a steadily propagating shear fracture with permeable surfaces, on which zero pore fluid pressure is prescribed, subjected to uniform shear loading over a finite region. In all cases, the numerical solutions exhibited excellent agreement with available analytical and semi-analytical results, confirming the correctness of the boundary integral equations and the accuracy of the proposed numerical methodology.

The developed formulation provides an accurate and efficient method, written directly in the moving-tip coordinate system, for analyzing semi-infinite steadily propagating cracks in poroelastic media. The system of equations derived here in the context of poroelasticity may also be adapted to related elasto-diffusive problems, such as thermoelasticity, through appropriate modification of the governing physical parameters. Furthermore, the numerical procedure presented in this work provides a basis for future extensions in which additional physical processes are incorporated, such as lubrication flow within a steadily propagating tensile crack or frictional resistance in a steadily propagating shear fracture under plane strain poroelasticity. Finally, the hydraulic boundary conditions adopted here could also be generalized to account for a leaky interface in shear fracture problems. These developments define useful directions for future work.


\section*{Data availability}
The numerical implementation and notebooks reproducing the verification tests are available in the GitHub repository: \url{https://github.com/evgenii-kanin/semi-infinite-fracture-poroelasticity}.

\section*{CRediT authorship contribution statement}
\textbf{Evgenii Kanin:} Conceptualization, Formal analysis, Investigation, Methodology, Software, Validation, Visualization, Writing - original draft. \textbf{Andreas Möri:} Conceptualization, Formal analysis, Investigation, Methodology, Software, Validation, Visualization, Writing - original draft. \textbf{Dmitry Garagash:} Conceptualization, Investigation, Methodology, Supervision, Writing - review \& editing. \textbf{Brice Lecampion:} Conceptualization, Funding acquisition, Investigation, Methodology, Project administration, Supervision, Writing - review \& editing.

\section*{Acknowledgments}
This work was partly funded by the Swiss National Science Foundation under grant \#192237.

\appendix

\section{Detailed derivation of the analytical expressions for steadily moving fundamental solutions}
\label{sec:derivations_moving_loadings}

In this appendix, we present detailed derivations of the pore pressure and stress fields generated by steadily moving loadings, namely a fluid source and an edge dislocation. All solutions are obtained in the moving coordinate system $(x, y)$ attached to the propagating loading. The following auxiliary variables are introduced and used throughout this appendix:
\begin{equation*}
    r = \sqrt{x^2 + y^2}, ~~~~ s = \frac{Vu}{r}, ~~~~ \xi = \frac{Vx}{2c}.
\end{equation*}
When deriving the pore pressure and stresses along the $x-$axis, the definition of $s$ is modified to $s = V u / |x|$.
 
\subsection{Steadily moving fluid source}
\label{sec:derivation_fluid_source}

\subsubsection{Pore pressure $p^{sm}$}
\label{sec:derivation_fluid_source_pressure}

Substituting Eq.~\eqref{eq:p_si} into Eq.~\eqref{eq:field_moving_loading} yields:
\begin{equation*}
    p^{sm} (x, y) = \frac{1}{4\pi\kappa}\int_{0}^{\infty} \frac{\diff u}{u} \exp{\left[-\frac{\left(x - Vu\right)^2 + y^2}{4 c u}\right]}.
\end{equation*}

The argument of the exponent can be rewritten as:
\begin{equation*}
    -\frac{\left(x - Vu\right)^2 + y^2}{4 c u} = -\frac{r^2}{4cu} - \frac{u V^2}{4c} + \xi = -\frac{V r}{4c} \left(s + \frac{1}{s}\right) + \xi.
\end{equation*}

Applying the change of integration variable $u \to s$, we rewrite the integral as follows:
\begin{equation*}
    p^{sm} (x, y) = \frac{e^\xi}{4\pi\kappa}\int_{0}^{\infty} \frac{\diff s}{s} \exp{\left[-\frac{V r}{4c} \left(s + \frac{1}{s}\right)\right]}.
\end{equation*}

Using the integral representation of the modified Bessel function of the second kind \citep{abramowitz1948handbook}:
\begin{equation}
    \mathrm{K}_a (z) = \frac{1}{2} \int_0^\infty s^{a-1} \exp{\left[-\frac{z}{2} \left(s + \frac{1}{s}\right)\right]}, ~~~~ z > 0,
    \label{eq:bessel_k}
\end{equation}
we write the final expression for the pore pressure distribution:
\begin{equation*}
    p^{sm} (x, y) = \frac{e^\xi}{2\pi\kappa} \mathrm{K}_0 \left(\frac{Vr}{2c}\right).
\end{equation*}

The expression for the pressure field along the $x-$axis is given by:
\begin{equation*}
    p^{sm} (x, 0) = \frac{e^\xi}{2\pi\kappa} \mathrm{K}_0(|\xi|).
\end{equation*}

\subsubsection{Stress component $\sigma_{yy}^{sm}$}
\label{sec:derivation_fluid_source_sigma_yy}

Substituting Eq.~\eqref{eq:sigma_si} into Eq.~\eqref{eq:field_moving_loading} yields:
\begin{flalign*}
    & \sigma^{sm}_{yy}(x, y) = \frac{\eta}{\pi S} \int_0^{\infty} \frac{\diff u}{(x - Vu)^2 + y^2} \bigg\{\left(1 - \frac{2 y^2}{(x - Vu)^2 + y^2}\right)\left(1 - \exp{\left[-\frac{\left(x - Vu\right)^2 + y^2}{4 c u}\right]}\right) - \\
    & -\frac{(x - Vu)^2}{2 c u} \exp{\left[-\frac{\left(x - Vu\right)^2 + y^2}{4 c u}\right]}\bigg\}.
\end{flalign*}

Further, we focus on the derivation of the stress distribution along the $x-$axis (where $y=0$):
\begin{equation*}
    \sigma^{sm}_{yy}(x, 0) = \frac{\eta}{\pi S} \int_0^{\infty} \frac{\diff u}{(x - Vu)^2} \bigg(\underbrace{1 - \exp{\left[-\frac{\left(x - Vu\right)^2}{4 c u}\right]}}_{J_1} - \underbrace{\frac{(x - Vu)^2}{2 c u} \exp{\left[-\frac{\left(x - Vu\right)^2}{4 c u}\right]}}_{J_2}\bigg).
\end{equation*}

Using the results of Appendix~\ref{sec:derivation_fluid_source_pressure}, we write the expression for the integral $J_2$:
\begin{equation*}
    J_2 = -\frac{\eta}{2 \pi \kappa}\int_0^{\infty} \frac{\diff u}{u} \exp{\left[-\frac{\left(x - Vu\right)^2}{4 c u}\right]} = -\frac{\eta}{\pi \kappa} e^\xi \mathrm{K}_0 (|\xi|).
\end{equation*}

By applying the change of integration variable $u \to s$, the integral $J_1$ can be rewritten in the following form:
\begin{equation*}
    J_1 = \frac{\eta}{\pi S}\int_0^{\infty} \frac{\diff u}{(x - Vu)^2} \bigg(1 - \exp{\left[-\frac{\left(x - Vu\right)^2}{4 c u}\right]} \bigg) = \frac{\eta}{\pi S V}\frac{1}{|x|}\int_0^{\infty} \frac{\diff s}{(s\mp1)^2} \left(1 - \exp{\left[-\frac{|\xi|\left(s\mp1\right)^2}{2s}\right]}\right),
\end{equation*}
where the upper signs correspond to $x > 0$, and the lower signs to $x < 0$; the same convention is adopted throughout the derivations below. Integration by parts further yields:
\begin{equation*}
    J_1 = -\frac{\eta}{\pi S V}\frac{1}{x}\left\{1 - \frac{e^\xi\xi}{2} \int_0^{\infty} \left(\frac{1}{s} \pm \frac{1}{s^2}\right)\exp{\left[- \frac{|\xi|}{2}\left(s + \frac{1}{s}\right)\right]}\diff s\right\}.
\end{equation*}

Using Eq.~\eqref{eq:bessel_k} and the symmetry property \citep{abramowitz1948handbook}:
\begin{equation}
    \mathrm{K}_a(z) = \mathrm{K}_{-a}(z),
    \label{eq:symmetry}
\end{equation}
we obtain:
\begin{equation}
     J_1 = -\frac{\eta}{\pi S V}\frac{1}{x}\bigg(1 - e^{\xi} \xi \bigg[\mathrm{K}_0(|\xi|) \pm  \mathrm{K}_1(|\xi|)\bigg]\bigg).
     \label{eq:integral_J_1}
\end{equation}

Upon combining $J_1$ and $J_2$, the stress component takes the following form:
\begin{equation*}
    \sigma^{sm}_{yy}(x, 0) = -\frac{\eta}{\pi S V} \frac{1}{x} + \frac{\eta}{2 \pi \kappa}e^{\xi}\bigg[\pm \mathrm{K}_1(|\xi|) - \mathrm{K}_0(|\xi|)\bigg] = -\frac{\eta}{2\pi\kappa}\bigg(\frac{1}{\xi} + e^{\xi}\bigg[\mathrm{K}_0(|\xi|) \mp \mathrm{K}_1(|\xi|)\bigg]\bigg).
\end{equation*}

\subsubsection{Stress component $\sigma_{xx}^{sm}$}
\label{sec:derivation_fluid_source_sigma_xx}

Substituting Eq.~\eqref{eq:sigma_si} into Eq.~\eqref{eq:field_moving_loading} yields:
\begin{flalign*}
    & \sigma^{sm}_{xx}(x, y) = \frac{\eta}{\pi S} \int_0^{\infty} \frac{\diff u}{(x - Vu)^2 + y^2} \bigg\{\left(1 - \frac{2 (x - Vu)^2}{(x - Vu)^2 + y^2}\right)\left(1 - \exp{\left[-\frac{\left(x - Vu\right)^2 + y^2}{4 c u}\right]}\right) - \\
    & -\frac{y^2}{2 c u} \exp{\left[-\frac{\left(x - Vu\right)^2 + y^2}{4 c u}\right]}\bigg\}.
\end{flalign*}

We consider the stress distribution along the $x-$axis (where $y=0$):
\begin{equation*}
    \sigma^{sm}_{xx}(x, 0) = -\frac{\eta}{\pi S} \int_0^{\infty} \frac{\diff u}{(x - Vu)^2} \left(1 - \exp{\left[-\frac{\left(x - Vu\right)^2}{4 c u}\right]}\right).
\end{equation*}

Using Eq.~\eqref{eq:integral_J_1}, we write the final expression for the analyzed stress component:
\begin{equation*}
    \sigma^{sm}_{xx}(x, 0) = \frac{\eta}{\pi S V} \frac{1}{x}-\frac{\eta}{2 \pi \kappa }e^\xi \bigg[\mathrm{K}_0(|\xi|) \pm  \mathrm{K}_1(|\xi|)\bigg] = -\frac{\eta}{2 \pi \kappa} \bigg(-\frac{1}{\xi} + e^\xi \left[\mathrm{K}_0(|\xi|) \pm  \mathrm{K}_1(|\xi|)\right] \bigg).
\end{equation*}

\subsubsection{Stress component $\sigma_{xy}^{sm}$}
\label{sec:derivation_fluid_source_sigma_xy}

Substituting Eq.~\eqref{eq:sigma_si} into Eq.~\eqref{eq:field_moving_loading} yields:
\begin{flalign*}
    & \sigma^{sm}_{xy}(x, y) = -\frac{\eta}{\pi S} \int_0^{\infty} \frac{\diff u}{(x - Vu)^2 + y^2} \bigg\{\frac{2 (x - Vu)y}{(x - Vu)^2 + y^2}\left(1 - \exp{\left[-\frac{\left(x - Vu\right)^2 + y^2}{4 c u}\right]}\right) - \\
    & -\frac{(x - Vu)y}{2 c u} \exp{\left[-\frac{\left(x - Vu\right)^2 + y^2}{4 c u}\right]}\bigg\}.
\end{flalign*}

From the expression above, it follows that $\sigma^{sm}_{xy}(x, 0) = 0$.

\subsection{Steadily moving edge dislocation}
\label{sec:derivation_edge_dislocation}

\subsubsection{Pore pressure $p^{em}$}
\label{sec:derivation_edge_dislocation_pressure}

Substituting Eq.~\eqref{eq:p_ei} into Eq.~\eqref{eq:field_moving_loading} yields:
\begin{itemize}
    \item slip edge dislocation:
    \begin{flalign*}
        & p^{em}_1(x, y) = (p^{em}_1)^0(x, y) + \Delta p^{em}_1(x, y), \\
        & (p^{em}_1)^0(x, y) = \frac{\eta}{\pi S} \frac{y}{r^2}, ~~~~ \Delta p^{em}_1(x, y) = - \frac{\eta}{4 \pi \kappa} \int_0^{\infty} \frac{y}{u^2} \exp{\left[-\frac{\left(x - Vu\right)^2 + y^2}{4 c u}\right]} \diff u.
    \end{flalign*}

    \item normal edge dislocation:
    \begin{flalign*}
        & p^{em}_2(x, y) = (p^{em}_2)^0(x, y) + \Delta p^{em}_2(x, y), \\
        & (p^{em}_2)^0(x, y) = -\frac{\eta}{\pi S} \frac{x}{r^2}, ~~~~ \Delta p^{em}_2(x, y) = \frac{\eta}{4 \pi \kappa} \int_0^{\infty} \frac{x - Vu}{u^2} \exp{\left[-\frac{\left(x - Vu\right)^2 + y^2}{4 c u}\right]} \diff u.
    \end{flalign*}
\end{itemize}

First, we evaluate the integral appearing in the term $\Delta p^{em}_1(x, y)$. We introduce the change of integration variable $u \to s$ and apply Eqs.~\eqref{eq:bessel_k}, \eqref{eq:symmetry}:
\begin{equation*}
    \Delta p^{em}_1(x, y) = - \frac{\eta V}{4 \pi \kappa} \frac{y}{r} e^\xi \int_0^{\infty} \frac{1}{s^2} \exp{\left[-\frac{V r}{4c} \left(s + \frac{1}{s}\right)\right]} \diff s = - \frac{\eta V}{2 \pi \kappa} \frac{y}{r} e^\xi\mathrm{K}_1\left(\frac{V r}{2 c}\right).
\end{equation*}
As a result, the expression for the pressure field corresponding to the slip edge dislocation takes the form:
\begin{equation*}
    p^{em}_1(x, y) = \frac{\eta}{\pi S} \frac{y}{r^2} - \frac{\eta V}{2 \pi \kappa} \frac{y}{r} e^\xi\mathrm{K}_1\left(\frac{V r}{2 c}\right), 
\end{equation*}
providing $p_1^{em}(x, 0) = 0$.

Next, we consider the normal edge dislocation and evaluate the term $\Delta p^{em}_2(x, y)$. The change of integration variable $u \to s$ together with Eqs.~\eqref{eq:bessel_k} and \eqref{eq:symmetry} yields: 
\begin{equation*}
    \Delta p^{em}_2(x, y) = \frac{\eta V}{4\pi\kappa} \frac{e^\xi}{r}\int_{0}^{\infty} \left(\frac{x}{s^2} - \frac{r}{s} \right)\exp{\left[-\frac{V r}{4c} \left(s + \frac{1}{s}\right)\right]} \diff s = \frac{\eta V}{2\pi\kappa} e^\xi \left[\frac{x}{r} \mathrm{K}_1\left(\frac{V r}{2c}\right) - \mathrm{K}_0\left(\frac{V r}{2c}\right)\right].
\end{equation*}
The pressure field corresponding to the normal edge dislocation is given by:
\begin{equation*}
    p^{em}_2 (x, y) = -\frac{\eta}{\pi S} \frac{x}{r^2} + \frac{\eta V}{2\pi\kappa} e^\xi \left[\frac{x}{r} \mathrm{K}_1\left(\frac{V r}{2c}\right) - \mathrm{K}_0\left(\frac{V r}{2c}\right)\right],
\end{equation*}
which, along the $x-$axis, reduces to the following expression:
\begin{equation*}
    p^{em}_2 (x, 0) = -\frac{\eta}{\pi S} \frac{1}{x} + \frac{\eta V}{2 \pi \kappa} e^\xi \bigg[\pm \mathrm{K}_1(|\xi|) - \mathrm{K}_0(|\xi|)\bigg] = -\frac{\eta V}{2 \pi \kappa} \left(\frac{1}{\xi} + e^\xi \bigg[\mathrm{K}_0(|\xi|)\mp \mathrm{K}_1(|\xi|)\bigg]\right).
\end{equation*}

\subsubsection{Stress component $\sigma_{yy}^{em}$}
\label{sec:derivation_edge_dislocation_sigma_yy}

Substituting Eq.~\eqref{eq:sigma_ei} into Eq.~\eqref{eq:field_moving_loading} yields:

\begin{itemize}
    \item slip edge dislocation:
    \begin{flalign*}
        & \sigma^{em}_{yy1}(x, y) = (\sigma^{em}_{yy1})^0(x, y) + \Delta \sigma^{em}_{yy1}(x, y), \\
        & (\sigma^{em}_{yy1})^0(x, y) = \frac{G}{2\pi (1-\nu_u)} \frac{y}{r^2} \left(1 - \frac{2y^2}{r^2} \right), \\
        & \Delta \sigma^{em}_{yy1}(x, y) = -\frac{4c\eta^2}{\pi S} \int_0^{\infty} \frac{y}{((x - Vu)^2 + y^2)^2} \bigg\{\left(3 - 4 \frac{y^2}{(x - Vu)^2 + y^2}\right) \times \\
        & \times \left(1 - \left(1+\frac{(x - Vu)^2 + y^2}{4c u}\right) \exp{\left[-\frac{\left(x - Vu\right)^2 + y^2}{4 c u}\right]}\right) - \\
        & - \frac{(x - Vu)^2 ((x - Vu)^2 + y^2)}{8 c^2 u^2} \exp{\left[-\frac{\left(x - Vu\right)^2 + y^2}{4 c u}\right]} \bigg\} \diff u.
    \end{flalign*}

    \item normal edge dislocation:
    \begin{flalign*}
        & \sigma^{em}_{yy2}(x, y) = (\sigma^{em}_{yy2})^0(x, y) + \Delta \sigma^{em}_{yy2}(x, y), \\
        & (\sigma^{em}_{yy2})^0(x, y) = \frac{G}{2\pi (1-\nu_u)} \frac{x}{r^2} \left( 1 + \frac{2 y^2}{r^2}\right), \\
        & \Delta \sigma^{em}_{yy2}(x, y) = \frac{4c\eta^2}{\pi S} \int_0^{\infty} \frac{x - Vu}{((x - Vu)^2 + y^2)^2} \bigg\{\left(1 - 4 \frac{y^2}{(x - Vu)^2 + y^2} \right) \times \\
        &\times \left(1 - \left(1+\frac{(x - Vu)^2 + y^2}{4c u}\right) \exp{\left[-\frac{\left(x - Vu\right)^2 + y^2}{4 c u}\right]}\right) - \\
        & - \frac{(x - Vu)^2((x - Vu)^2 + y^2)}{8 c^2 u^2} \exp{\left[-\frac{\left(x - Vu\right)^2 + y^2}{4 c u}\right]} \bigg\} \diff u.
    \end{flalign*}
\end{itemize}

We consider the stress distribution along the $x-$axis (where $y=0$). For a slip edge dislocation, $\sigma^{em}_{yy1}(x, 0) = 0$, while for a normal edge dislocation:
\begin{flalign*}
    & \sigma^{em}_{yy2}(x, 0) = (\sigma^{em}_{yy2})^0(x, 0) + \Delta \sigma^{em}_{yy2}(x, 0), \\
    & (\sigma^{em}_{yy2})^0(x, 0) = \frac{G}{2\pi (1-\nu_u)} \frac{1}{x}, \\
    & \Delta \sigma^{em}_{yy2}(x, 0) = \frac{4c\eta^2}{\pi S} \int_{0}^{\infty} \frac{\diff u}{(x - V u)^3} \left\{1 - \left(1 + \frac{(x - V u)^2}{4cu} +  \frac{(x - V u)^4}{8 c^2 u^2}\right)\exp{\left[-\frac{\left(x - V u\right)^2}{4 c u}\right]}\right\}.
\end{flalign*}

We introduce a change of integration variable $u \to s$ in the term $\Delta \sigma^{em}_{yy2}(x, 0)$: 
\begin{equation*}
    \Delta \sigma^{em}_{yy2} (x, 0) = -\frac{4c\eta^2}{\pi S V} \frac{1}{x^2} \int_{0}^{\infty} \frac{\diff s}{(s\mp 1)^3} \bigg\{\underbrace{1 - \bigg(1 + \frac{|\xi|\left(s \mp 1\right)^2}{2s}}_{J_1} + \underbrace{\frac{\xi^2\left(s \mp 1\right)^4}{2s^2}}_{J_2}\bigg) \exp{\left[-\frac{|\xi|\left(s\mp1\right)^2}{2s}\right]} \bigg\}.
\end{equation*}

The expression for the integral $J_2$ can be written, using Eqs.~\eqref{eq:bessel_k} and \eqref{eq:symmetry}:
\begin{equation*}
    J_2 = \frac{\eta^2 V}{2 \pi \kappa} \int_{0}^{\infty} \left(\frac{1}{s} \mp \frac{1}{s^2}\right) \exp{\left[-\frac{|\xi|\left(s\mp1\right)^2}{2s}\right]} \diff s  = \frac{\eta^2 V}{\pi \kappa} e^\xi\bigg[\mathrm{K}_0(|\xi|) \mp  \mathrm{K}_1(|\xi|)\bigg].
\end{equation*}

Applying integration by parts, we obtain the following expression for the integral $J_1$: 
\begin{flalign}
    & J_1 = -\frac{4c\eta^2}{\pi S V} \frac{1}{x^2} \int_{0}^{\infty} \frac{\diff s}{(s \mp 1)^3} \bigg\{1 - \bigg(1 + \frac{|\xi|\left(s \mp 1\right)^2}{2s}\bigg) \exp{\left[-\frac{|\xi|\left(s\mp1\right)^2}{2 s}\right]} \bigg\} = \nonumber \\
    & = -\frac{2c\eta^2}{\pi S V} \frac{1}{x^2}\left\{1+\frac{\xi^2}{4}\int_{0}^{\infty} \left(\frac{1}{s} - \frac{1}{s^3}\right)\exp{\left[-\frac{|\xi|\left(s\mp1\right)^2}{2s}\right]} \diff s\right\} = -\frac{2c\eta^2}{\pi S V} \frac{1}{x^2}\bigg[1 - |\xi|e^\xi \mathrm{K}_1(|\xi|)\bigg], \label{eq:integral_J_1_2}
\end{flalign}
where we utilize Eq.~\eqref{eq:bessel_k} and the recurrence relation:
\begin{equation}
    \mathrm{K}_2(z) = \mathrm{K}_0(z) + \frac{2}{z}\mathrm{K}_1(z), ~~ z > 0.
    \label{eq:bessel_k_recurrence}
\end{equation}

Combining $(\sigma^{em}_{yy2})^0$ and integrals $J_1$, $J_2$, we write the final expression for the stress component:
\begin{equation*}
    \sigma^{em}_{yy2}(x, 0) = \frac{G}{2\pi (1-\nu_u)} \frac{1}{x}\bigg\{1 - \frac{\nu_u - \nu}{1-\nu}\left(\frac{1}{\xi} \mp e^\xi \mathrm{K}_1(|\xi|) - 2 \xi e^\xi \bigg[\mathrm{K}_0(|\xi|) \mp  \mathrm{K}_1(|\xi|)\bigg]\right)\bigg\}.
\end{equation*}

\subsubsection{Stress component $\sigma_{xx}^{em}$}
\label{sec:derivation_edge_dislocation_sigma_xx}

Substituting Eq.~\eqref{eq:sigma_ei} into Eq.~\eqref{eq:field_moving_loading} yields:
\begin{itemize}
    \item slip edge dislocation:
    \begin{flalign*}
        & \sigma^{em}_{xx1}(x, y) = (\sigma^{em}_{xx1})^0(x, y) + \Delta \sigma^{em}_{xx1}(x, y), \\
        & (\sigma^{em}_{xx1})^0(x, y) = -\frac{G}{2\pi (1-\nu_u)} \frac{y}{r^2} \left(1 + \frac{2 x^2}{r^2} \right), \\
        & \Delta \sigma^{em}_{xx1}(x, y) = -\frac{4c\eta^2}{\pi S} \int_0^{\infty} \frac{y}{((x - Vu)^2 + y^2)^2} \bigg\{\left(1 - 4 \frac{(x - Vu)^2}{(x - Vu)^2 + y^2}\right) \times \\
        & \times \left(1 - \left(1+\frac{(x - Vu)^2 + y^2}{4c u}\right) \exp{\left[-\frac{\left(x - Vu\right)^2 + y^2}{4 c u}\right]}\right) - \\
        & - \frac{y^2((x - Vu)^2 + y^2)}{8 c^2 u^2} \exp{\left[-\frac{\left(x - Vu\right)^2 + y^2}{4 c u}\right]} \bigg\} \diff u.
    \end{flalign*}
    \item normal edge dislocation:
    \begin{flalign*}
        & \sigma^{em}_{xx2}(x, y) = (\sigma^{em}_{xx2})^0(x, y) + \Delta \sigma^{em}_{xx2}(x, y), \\
        & (\sigma^{em}_{xx2})^0(x, y) = -\frac{G}{2\pi (1-\nu_u)}\frac{x}{r^2} \left(1 - \frac{2 x^2}{r^2}\right), \\
        & \Delta \sigma^{em}_{xx2}(x, y) = \frac{4c\eta^2}{\pi S} \int_0^{\infty} \frac{x - Vu}{((x - Vu)^2 + y^2)^2} \bigg\{\left(3 - 4 \frac{(x - Vu)^2}{(x - Vu)^2 + y^2} \right) \times \\
        & \times \left(1 - \left(1+\frac{(x - Vu)^2 + y^2}{4c u}\right) \exp{\left[-\frac{\left(x - Vu\right)^2 + y^2}{4 c u}\right]}\right) - \\
        & - \frac{y^2((x - Vu)^2 + y^2)}{8 c^2 u^2} \exp{\left[-\frac{\left(x - Vu\right)^2 + y^2}{4 c u}\right]} \bigg\} \diff u.
    \end{flalign*}
\end{itemize}

We derive the stress distribution along the $x-$axis (where $y=0$). For a slip edge dislocation, $\sigma^{em}_{xx1}(x, 0) = 0$, while for a normal edge dislocation:
\begin{flalign*}
    & \sigma^{em}_{xx2}(x, 0) = (\sigma^{em}_{xx2})^0(x, 0) + \Delta \sigma^{em}_{xx2}(x, 0), \\
    & (\sigma^{em}_{xx2})^0(x, 0) = \frac{G}{2\pi (1-\nu_u)}\frac{1}{x}, \\
    &  \Delta \sigma^{em}_{xx2} (x, 0) = -\frac{4c\eta^2}{\pi S} \int_{0}^{\infty} \frac{\diff u}{(x - V u)^3} \left\{1 - \left(1 + \frac{(x - V u)^2}{4cu}\right)\exp{\left[-\frac{\left(x - V u\right)^2}{4 c u}\right]}\right\}.
\end{flalign*}

A change of integration variable $u \to s$ in the term $\Delta \sigma^{em}_{xx2}(x, 0)$ yields: 
\begin{equation*}
    \Delta \sigma^{em}_{xx2} (x, 0) = \frac{4c\eta^2}{\pi S V} \frac{1}{x^2} \int_{0}^{\infty} \frac{\diff s}{(s \mp 1)^3} \left\{1 - \bigg(1 + \frac{|\xi|\left(s \mp 1\right)^2}{2s}\bigg)\exp{\left[-\frac{|\xi|\left(s\mp1\right)^2}{2s}\right]} \right\}.
\end{equation*}

Next, we use Eq.~\eqref{eq:integral_J_1_2} and write the final expression for stress component:
\begin{equation*}
    \sigma^{em}_{xx2} (x, 0) = \frac{G}{2\pi (1-\nu_u)} \frac{1}{x}\bigg(1 + \frac{\nu_u - \nu}{1-\nu}\left[\frac{1}{\xi} \mp e^\xi \mathrm{K}_1(|\xi|)\right]\bigg).
\end{equation*}

\subsubsection{Stress component $\sigma_{xy}^{em}$}
\label{sec:derivation_edge_dislocation_sigma_xy}

Substituting Eq.~\eqref{eq:sigma_ei} into Eq.~\eqref{eq:field_moving_loading} yields:
\begin{itemize}
    \item slip edge dislocation:
    \begin{flalign*}
        & \sigma^{em}_{xy1}(x, y) = (\sigma^{em}_{xy1})^0(x, y) + \Delta \sigma^{em}_{xy1}(x, y), \\
        & (\sigma^{em}_{xy1})^0(x, y) = \frac{G}{2\pi (1-\nu_u)} \frac{x}{r^2} \left(1 - \frac{2 y^2}{r^2}\right), \\
        & \Delta \sigma^{em}_{xy1}(x, y) = -\frac{4c\eta^2}{\pi S} \int_0^{\infty} \frac{x - Vu}{((x - Vu)^2 + y^2)^2} \bigg\{\left(1 - 4 \frac{y^2}{(x - Vu)^2 + y^2} \right) \times \\
        & \times \left(1 - \left(1+\frac{(x - Vu)^2 + y^2}{4c u}\right) \exp{\left[-\frac{\left(x - Vu\right)^2 + y^2}{4 c u}\right]}\right) + \\
        & + \frac{y^2((x - Vu)^2 + y^2)}{8 c^2 u^2} \exp{\left[-\frac{\left(x - Vu\right)^2 + y^2}{4 c u}\right]} \bigg\} \diff u. 
    \end{flalign*}
    \item normal edge dislocation:
    \begin{flalign*}
        & \sigma^{em}_{xy2}(x, y) = (\sigma^{em}_{xy2})^0(x, y) + \Delta \sigma^{em}_{xy2}(x, y), \\
        & (\sigma^{em}_{xy2})^0(x, y) = -\frac{G}{2\pi (1-\nu_u)} \frac{y}{r^2} \left(1 - \frac{2 x^2}{r^2} \right), \\
        & \Delta \sigma^{em}_{xy2}(x, y) = \frac{4c\eta^2}{\pi S} \int_0^{\infty} \frac{y}{((x - Vu)^2 + y^2)^2} \bigg\{\left(1 - 4 \frac{(x - Vu)^2}{(x - Vu)^2 + y^2} \right) \times \\
        & \times \left(1 - \left(1+\frac{(x - Vu)^2 + y^2}{4c u}\right) \exp{\left[-\frac{\left(x - Vu\right)^2 + y^2}{4 c u}\right]}\right) + \\ 
        & + \frac{(x - Vu)^2((x - Vu)^2 + y^2)}{8 c^2 u^2} \exp{\left[-\frac{\left(x - Vu\right)^2 + y^2}{4 c u}\right]} \bigg\} \diff u. 
    \end{flalign*}
\end{itemize}

We focus on the stress distribution along the $x-$axis (where $y=0$). For a normal edge dislocation, $\sigma^{em}_{xy2}(x, 0) = 0$, while for a slip edge dislocation:
\begin{flalign*}
    & \sigma^{em}_{xy1}(x, 0) = (\sigma^{em}_{xy1})^0(x, 0) + \Delta \sigma^{em}_{xy1} (x, 0), \\
    & (\sigma^{em}_{xy1})^0(x, 0) = \frac{G}{2\pi (1-\nu_u)} \frac{1}{x}, \\
    &  \Delta \sigma^{em}_{xy1} (x, 0) = -\frac{4c\eta^2}{\pi S} \int_{0}^{\infty} \frac{\diff u}{(x - V u)^3} \left\{1 - \left(1 + \frac{(x - V u)^2}{4cu}\right)\exp{\left[-\frac{\left(x - V u\right)^2}{4 c u}\right]}\right\}.
\end{flalign*}

We apply a change of integration variable $u \to s$ in the term $\Delta \sigma^{em}_{xy1}(x, 0)$: 
\begin{equation*}
    \Delta \sigma^{em}_{xy1} (x, 0) = \frac{4c\eta^2}{\pi S V} \frac{1}{x^2} \int_{0}^{\infty} \frac{\diff s}{(s \mp 1)^3} \left\{1 - \bigg(1 + \frac{|\xi|\left(s \mp 1\right)^2}{2s}\bigg)\exp{\left[-\frac{|\xi|\left(s\mp1\right)^2}{2s}\right]} \right\}.
\end{equation*}

Using Eq.~\eqref{eq:integral_J_1_2}, we can write the final expression for stress component $\sigma^{em}_{xy1}$ as:
\begin{equation*}
    \sigma^{em}_{xy1} (x, 0) = \frac{G}{2\pi (1-\nu_u)} \frac{1}{x}\bigg(1 + \frac{\nu_u - \nu}{1-\nu}\left[\frac{1}{\xi} \mp e^{\xi} \mathrm{K}_1(|\xi|)\right]\bigg).
\end{equation*}

\bibliographystyle{cas-model2-names}
\bibliography{Bibliography}

\end{document}